\documentclass[11pt,a4paper]{article}
\usepackage{amsmath}
\usepackage[font=footnotesize,labelfont=bf]{caption}
\usepackage[font=footnotesize,labelfont=rm,singlelinecheck=on,justification=raggedright]{subcaption}
\usepackage{overpic}
\usepackage{bm}

\newcommand{\beq}{\begin{equation}}
\newcommand{\eeq}{\end{equation}}
\newcommand{\bea}{\begin{eqnarray}}
\newcommand{\eea}{\end{eqnarray}}
\newcommand{\bean}{\begin{eqnarray*}}
\newcommand{\eean}{\end{eqnarray*}}
\newcommand{\E}{\delta {\cal E}}
\newcommand{\B}{\delta {\cal B}}
\newcommand{\wo}{{\widetilde \omega}}

\begin{document}

\title{\textbf{A renormalisation group equation for transport co-efficients
   in $\mathbf{(2+1)}$-dimensions derived from the AdS/CMT correspondence}}

\author{Brian P. Dolan\footnote{email: bdolan@thphys.nuim.ie} \\ \\
  \textit{Department of Theoretical Physics, Maynooth University}\\
  \textit{Maynooth, Co.~Kildare, Ireland}\\
  and \\ 
  \textit{School of Theoretical Physics}\\
  \textit{Dublin Institute for Advanced Studies}\\
  \textit{10 Burlington Rd., Dublin, Ireland}}

\maketitle

\vspace{-12cm}
\rightline{DIAS-STP-20-05}
\vspace{12cm}

\begin{abstract}
  Within the framework of the AdS/CMT correspondence
  asymptotically anti-de Sitter black holes in four space-time dimensions
  can be used to 
analyse transport properties in two space dimensions.
A non-linear renormalisation group equation for the conductivity in two dimensions is derived in this model and, as an example of its application,
both the Ohmic and Hall DC and AC conductivities are
studied in the presence of a magnetic field, using a bulk dyonic solution of the Einstein-Maxwell equations in asymptotically AdS$_4$ space-time.  The  ${\cal Q}$-factor of the cyclotron resonance is shown to decrease as the temperature is increased and increase as the charge density is increased in a fixed magnetic field.
Likewise the dissipative Ohmic conductivity at resonance increases as the temperature is decreased and as the charge density is increased.   The analysis also involves a discussion of the piezoelectric effect in the context of the AdS/CMT framework.
\end{abstract}

\newpage

\section{Introduction}

The holographic AdS/CFT or, more generally AdS/CMT, approach to condensed matter systems at strong coupling has been an active area of investigation for a number of years now, for a review see \cite{HLS}. These ideas are still speculative, the strongest evidence for the validity of the AdS/CFT program comes from supersymmetric
 $SU(N)$ gauge theories at large $N$ in higher dimensions, but a na{\" i}ve application of similar ideas to low dimensional non-relativistic problems,
while not as well supported mathematically, often gives qualitative results that are at least intriguing and merit further investigation.   

In \cite{Hartnoll+Kovtun} and \cite{Hartnoll+Herzog}
the properties of transport functions in two spatial dimensions at finite temperature in the presence of
a magnetic field were studied using this method.  The general idea is to take a charged black-hole in a 4-dimensional asymptotically AdS space-time, with a conformal
field theory on the $(2+1)$-dimensional asymptotic boundary. In the framework
of the AdS/CFT formulation a classical solution of Maxwell-Einstein gravity in the bulk corresponds to a strongly coupled quantum system on the boundary.
The black-hole event horizon acts as a one-way membrane that gives rise to dissipative phenomena on the boundary.
The Hawking temperature of the black-hole provides a thermal background and
its electric charge generates a chemical potential.  A static magnetic field normal to the two space directions on the boundary can be incorporated by giving the black-hole a magnetic charge, promoting it to the status of a dyon in $(3+1)$-dimensions.  This setup
gives a promising framework in which to study the quantum Hall effect and there has been a number of papers examining this possibility \cite{SKMS}-\cite{Alejo+Nastase}.

In \cite{Hartnoll+Kovtun,Hartnoll+Herzog} a dyonic black hole with a planar event horizon was used to study transport properties, thermal and electrical conductivity and the thermoelectric 
co-efficient were investigated.  The basic idea is that information on transport co-efficients can be obtained by studying the second variation of the action on-shell, when the first variation vanishes.
In the present work a first order differential equation that the conductivity must satisfy, as a function of the bulk radial co-ordinate, is derived in this setting, giving a non-linear renormalisation group equation.
The equation is singular at the event horizon in that the co-efficient of the derivative term vanishes there, reducing it to an algebraic equation, and the RG equation then dictates its own boundary conditions in the infra-red (this is essentially the attractor  mechanism).
In general the equation must be solved numerically to obtain both the Ohmic and the Hall AC and DC conductivities as functions of the radial coordinate, but a particular truncation of the equation
has a simple solution, which is an approximation studied in \cite{Hartnoll+Herzog}. In this approximation the conductivity has a pole in the complex frequency plane and the numerical solutions indicate that this pole persists in the full solution albeit with renormalised residue and position in the complex frequency plane.
The infra-red and ultra-violet cases are studied in the full numerical solution,
for various values of the electric and magnetic charges, and real and imaginary parts of both the Ohmic and the Hall conductivities are extracted at finite frequency.
The ${\cal Q}$-factor of the cyclotron resonance is studied as a function of the electric and magnetic charges and it is
shown to decrease as the temperature increases and increase as the ratio of the charge density to the magnetic field increases. 

The layout of the paper is as follows: in \S\ref{sec:GKPW} we briefly summarise the Gubser-Klebanov-Polyakov-Witten (GKPW) hypothesis and how it can be used to determine response functions on the boundary of the bulk theory; in \S\ref{sec:RG-equation} the formalism is applied to a dyonic black brane
in 4-dimensional asymptotically AdS Einstein-Maxwell theory and the non-linear RG equation for the conductivity on the $2+1$-dimensional boundary is derived; solutions of the RG equation are studied, both numerically and in certain
analytic approximations, in \S\ref{sec:solutions} and presented in a number of figures; finally the results are summarised in
\S\ref{sec:Conclusions}. Two appendices contain a discussion of the piezoelectric effect and technical details about Riccati equations.

\section{Response functions and the GKPW proposal\label{sec:GKPW}}

The GKPW proposal \cite{GKP,Witten-GKP} relates a conformal field theory on
the boundary of an asymptotically AdS space-time to a gravitational theory in the bulk.  More specifically it states that the generating functional for the
quantum field theory on the boundary, with sources $h_i$ for operators 
${\cal O}^i$, is equal to the partition 
function in the bulk when bulk fields $\varphi_i$ take the boundary value $h_i$.
In the Euclidean version, with
\[ Z_{CFT}[h] = \left< e^{- \int_{\partial {\cal M}}h_i {\cal O}^i} \right> \]
and 
\beq Z_{Grav}[h] = \int_{\{\varphi_i|_{\partial {\cal M}}=h_i\}} 
\kern -5pt {\cal D}\varphi_i \,e^{-S_{Grav}[\varphi]/\hbar}\label{eq:Z_Grav}\eeq 
where $\varphi_i$ represents all the bulk fields, including the metric, and $\partial {\cal M}$ is the boundary,  GKPW proposed that
\[ Z_{QFT}[h] =  Z_{Grav}[h].\] 
The effective action in the bulk, $S_{eff}[\phi]$, is defined as
the Legendre transform of the
generating functional $W[J,h]$, 
\[S_{eff}[\phi,h] = W[J,h] - \phi.J\]
where
\[e^{-W[J,h]/\hbar} = \int_{\{\varphi_i|_{\partial {\cal M}}=h_i\}}
\kern -5pt {\cal D}\varphi_i \,e^{-\frac{1}{\hbar}(S_{Grav}[\varphi]  + J.\varphi)}\]
with
\beq
\left.\frac{\delta S_{eff}[\phi,h]}{\delta \phi_i}\right|_h = -J^i\label{eq:J-def}
\eeq
so, when external sources vanish,
\beq
\left.\frac{\delta S_{eff}[\phi,h]}{\delta \phi_i}\right|_h = 0.\label{eq:EoM}\eeq

When quantum corrections in the bulk are negligible, as will be assumed here, the effective action equals the classical action $S_{eff}[\phi,h]=S_{Grav}[\phi,h]$ and (\ref{eq:EoM}) is
equivalent to the classical equations of motion with fixed boundary conditions.
Assuming the boundary conditions pick out a unique classical solution, $\phi_h$, the bulk partition function (\ref{eq:Z_Grav}) can be approximated by 
\beq Z_{Grav}[h] = e^{-S_{Grav}[\phi_h,h]/\hbar}\label{eq:Z-classical-bulk}\eeq
(if the boundary conditions do not pick out a unique solution, we would need to sum over all solutions $\phi_h$ compatible with the boundary conditions).

Although the variation of the action is zero for a classical solution, the variation of the Lagrangian ${\cal L}(\phi_i,\partial_\mu \phi_i)$ need not be --- in
general it will be a surface term.  For a classical solution
\[ \delta \bigl\{{\cal L}(\phi,\partial_\mu \phi) \sqrt{-g}\,\bigr\}
  = \partial_\mu \theta^\mu(\phi,\delta \phi) \quad \hbox{modulo equations of motion}.\]
$\theta^\mu(\phi,\delta \phi)$ can be non-zero in the bulk but
when $\phi_i|_{\partial {\cal M}} = h_i$ is fixed and $\delta \phi_i|_{\partial {\cal M}}=0$ it must vanishes on the boundary.
However if the boundary conditions are varied then $\theta$ can, and in general will,
be non-zero on the boundary. 
Define 
\[\Theta[\phi_h,\delta h] = \int_{\partial {\cal M}} \theta^\mu(\phi_h,\delta h)
d^3 \Sigma_\mu \]
 where $d^3 \Sigma_\mu$ is the volume element on the boundary.
Then
\[\frac{\delta S_{Grav}[\phi_h,h]}{\delta h_i(x)}  = \frac{\delta \Theta[\phi_h,h]}{\delta h_i(x)}.\]

This can now be related to the boundary CFT response functions.
These are defined by
\[ \chi^{i j}(x,y;h)= \left< {\cal O}^i(x) {\cal O}^j(y)\right> -\left< {\cal O}^i(x)\right>\left< {\cal O}^j(y)  \right> =\frac{\delta \ln Z_{QFT}[h]}{\delta h_i(x) \delta h_j(y)}.\]
When quantum corrections in the bulk are negligible we can now use (\ref{eq:Z-classical-bulk}) 
together with the GKPW hypothesis  to equate
\beq\chi^{i j}(x,y;h)  = -\frac{\delta^2 S_{Grav}[\phi_h,h]}{\delta h_i(x) \delta h_j(y)}
=-\frac{\delta^2 \Theta[\phi_h,h]}{\delta h_i(x)\delta h_j(y)}.
\eeq
This is how response functions can be calculated in the AdS/CFT framework,
\beq
\boxed{\vline height 22pt depth 14pt width 0pt \quad
  \chi^{i j}(x,y;h) =-\frac{\delta^2 \Theta[\phi_h,h]}{\delta h_i(x)\delta h_j(y)}.\quad\label{eq:dS=dTheta}}
\eeq

In the next section this formalism will be used to analyze condensed matter systems
in two spatial dimensions in the presence of a background magnetic field.

\section{The RG equation for the conductivity\label{sec:RG-equation}}

The starting point is that of \cite{Hartnoll+Kovtun,Hartnoll+Herzog}, a bulk theory which is 4-dimensional Einstein-Maxwell with a negative cosmological constant $\Lambda = -\frac{3}{L^2}$,
\beq S = \int \left( \frac{1}{2 \kappa^2} R + \frac{6}{L^2} - \frac{1}{4} F_{\mu\nu} F^{\mu \nu}\right) \sqrt{-g}\, d x^4\label{eq:Action}\eeq
where $\kappa^2 = 8 \pi G$ and $c=1$, and we shall also use units with $\frac{e^2}{h}=1$. 

Using co-ordinates $t$, $r$, $x$ and $y$ let $z=\frac{L}{r}$ be a dimensionless radial co-ordinate, $z \rightarrow 0$ being asymptotic infinity.
There is a Reissner-N\"ordstrom AdS black-hole solution of (\ref{eq:Action}) with a flat event horizon and constant radial electric and magnetic fields
\beq  F_{z t} = \tilde q\, L, \qquad F_{x y} = \tilde m \label{eq:F-background}\eeq
(the factor of $L$ ensures that $\tilde q$ and $\tilde m$ have the same dimensions, $mass^{1/2}\times length^{-3/2}$).
The line element is
\[ d s^2 = \frac{1}{z^2}\left(- f(z) d t^2 + \frac{ L^2 d z^2}{f(z)} + d x^2 + d y^2\right),\] 
with
\beq f(z)= 1 -  \frac{\kappa^2 M z^3 }{4 \pi L} + \frac{\kappa^2 L^2 (\tilde q^2 + \tilde m^2) z^4}{2 }
\label{eq:g-background}\eeq
where $M$ is the black-hole mass.

The event horizon could be either an infinite plane or a flat torus
with a finite area (we shall use the latter to keep global quantities finite with non-zero densities)
and it lies at the largest root, $z_h$, of
\[ \frac{\kappa^2  M z^3}{4 \pi L}  
=  1 + \frac{\kappa^2 L^2 (\tilde q^2+\tilde m^2) z^4}{2}.\] 
The event horizon has area ${\cal A}=\frac{1}{z_h^2} \int_{Torus} d x d y$
and the total charge  is
\[ Q=-\frac{\tilde q}{z_h^2}\int_{Torus} d x d y =  -\tilde q  {\cal A}.\]
The charge density at a general value of $z\le z_h$ is
\beq \rho(z) = -\frac{z^2}{z_h^2} \tilde q\label{eq:rho}\eeq
(the minus sign is because the outward normal is defined as the direction of
decreasing $z$).

It will be convenient to re-scale $t$, $z$, $x$ and $y$ by a factor of $z_h$ to 
\[ t \rightarrow \frac{t}{z_h}, \quad  
x \rightarrow \frac{x}{z_h}, \quad  
y \rightarrow \frac{y}{z_h}\]
and define 
\[u =\frac{z}{z_h}\]
in terms of which
\[ d s^2 = \frac{1}{u^2} \left(- f(u) d t^2 + \frac{L^2 d u^2}{f(u)} + d x^2 + d y^2\right)\]
where
\[ f(u)= (1-u)\bigl\{ 1 + u + u^2  - (q^2 + m^2) u^3\bigr\}\]
with
$0\le u \le 1$ outside the horizon and $q^2 =\kappa^2 \tilde q^2 L^2 z_h^4/2$ and 
$m^2 =\kappa^2 L^2\tilde m^2 z_h^4/2 $ dimensionless.
The Hawking temperature is
\beq T_H = \frac{(3-q^2 - m^2)\hbar }{4 \pi z_h L},\label{eq:T_H}\eeq
so  $q^2 + m^2$ is constrained to be less than or equal to $3$.

To study transport properties we perturb the solution by introducing oscillating transverse electric and magnetic fields which depend on $u$ but are independent of $x$ and $y$.
These can be derived from a vector potential
\beq \delta A_\alpha(u,t) = e^{-i\omega t} \delta  \widetilde A_\alpha(u)\label{eq:delta-A}\eeq
with $\alpha= x,y$,
\[ \delta  E_\alpha = -\delta  F_{t \alpha}=-\delta \dot A_\alpha= i \omega \delta A_\alpha , 
\qquad \delta B^\alpha =-\frac{\epsilon^{t \alpha u \beta}}{\sqrt{-g}} \delta F_{u \beta}= -\frac{u^4}{L} \epsilon^{\alpha \beta} \delta A'_\beta  \]
where $\epsilon^{t x u y} =\epsilon^{x y}=1$,
$\delta \dot A_\beta = \partial_t (\delta A_\beta)$ and $\delta A'_\beta = \partial_u (\delta A_\beta)$.
For static fields we can use
\[\delta  A_\alpha(u,t) = \frac{e^{-i\omega t}}{ i\omega} \delta  E_\alpha + \delta  A_\alpha(u),\]
with
\[\delta E_\alpha = -\lim_{\omega \rightarrow 0} \delta \dot  A_\alpha(u,t),  
\qquad \delta  B^\alpha =  -\frac{u^4}{L} \epsilon^{\alpha \beta} \delta A'_\beta(u),
\] 
provided $\delta E_\alpha$ is independent of $u$.

The analysis is initially identical to that of \cite{Hartnoll+Kovtun,Hartnoll+Herzog}.
We require not only that $\delta E_\alpha$ and $\delta B^\alpha$ satisfy Maxwell's equations in the background metric but also take into account the back reaction on the metric to first order in the variation, \textit{i.e.} $\delta A_\alpha(u,t)$ must satisfy the equation of motion.
The transverse electromagnetic field then back reacts on the metric producing
metric variations encoded into two functions of $t$ and $u$ defined by
\beq\delta G_\alpha(t,u) = e^{-i\omega t}  \delta \widetilde G(u)= u^2 \delta g_{t \alpha}.\label{eq:delta-g}\eeq
This variation of the metric components leaves the orthonormal 1-forms
\[ e^0 = \frac{\sqrt{f}}{u} d t, \qquad e^3 = \frac{1}{u \sqrt{f}}d u \]
unchanged while the transverse orthonormal 1-forms do change 
\[ e^1 \rightarrow e^1 + \delta e^1 = \frac{1}{u}(d x + \delta G_x d t), \qquad
e^2 \rightarrow e^2 + \delta e^2 = \frac{1}{u}(d y + \delta G_y d t),\]
where we ignore terms of $O(\delta G_\alpha^2)$.
The vierbein matrix and its inverse for the new metric can be chosen as
\[ e^a{}_\mu = \frac{1}{u}\begin{pmatrix}
\sqrt{f} & 0 & 0 & 0 \\
\delta G_x & 1 & 0 & 0 \\
\delta G_y & 0 & 1 & 0 \\
0 & 0 & 0 & \frac{1}{\sqrt{f}}
\end{pmatrix}, \qquad (e^{-1})^\mu{}_a = u\begin{pmatrix}
\frac{1}{\sqrt{f}} & 0 & 0 & 0 \\
-\frac{\delta G_x}{\sqrt{f}} & 1 & 0 & 0 \\
-\frac{\delta G_y}{\sqrt{f}} & 0 & 1 & 0 \\
0 & 0 & 0 & \sqrt{f}
\end{pmatrix}.\]
This change then feeds in to the electric field seen by an inertial observer,
\begin{align*}
  \delta E_i &
               = - \delta F_{0 i} \\
  & =
\delta F_{\mu \nu}(e^{-1})^\mu{}_0 (e^{-1})^\nu{}_i  + F_{\mu \nu} (\delta e^{-1})^\mu{}_0 (e^{-1})^\nu{}_i + F_{\mu \nu} (e^{-1})^\mu{}_0 (\delta e^{-1})^\nu{}_i + O(\delta^2),\end{align*}
where $0$ and $i=1,2$ are orthonormal indices.
In an orthonormal basis the variation in the transverse electric field is
\[\delta E_1=\frac{u^2}{\sqrt{f}} (\delta E_x -\tilde m \delta G_y),
\qquad
\delta E_2 = \frac{u^2}{\sqrt{f}} (\delta E_y +\tilde m \delta G_x),\]
while the transverse magnetic field merely acquires a multiplicative factor,
\[\delta B^1 = \frac{\sqrt{f}}{u^2} \delta B^x, 
\quad \delta B^2 = \frac{\sqrt{f}}{u^2} \delta B^y.\]
The $-\tilde m \epsilon_\alpha{}^\beta \delta G_\alpha$  contributions to $\delta E_i$ are a kinematic effect.  The metric is not invariant under Galilean transformations,
rather $t\rightarrow t$, $x \rightarrow x -\delta v_x t$, $y\rightarrow y-\delta v_y t$, with  $\delta v_\alpha$ small, changes the metric components, 
$\delta g_{t \alpha} = - \frac{\delta v_\alpha}{u^2} = \frac{1}{u^2} \delta G_\alpha$,
with $\delta g_{t t}\sim O(\delta v^2)$, so
\[ \delta E_\alpha - \tilde m \epsilon^{\alpha \beta} \delta G_\beta = 
\delta E_\alpha + \tilde m \epsilon^{\alpha \beta} \delta v_\beta \]
and an inertial observer sees a contribution to the electric field arising from the combination $v^\alpha$ with the radial magnetic field.

Following \cite{Hartnoll+Herzog} define
\[
\E_x = \delta E_x -\tilde m \delta G_y,\qquad
\E_y = \delta E_y +\tilde m \delta G_x,\]
then the perturbation $\E_\alpha$ will generate a current
\[ \delta J^\alpha = \sigma^{\alpha \beta} \E_\beta\]
with $\sigma^{\alpha \beta}$ the conductivity tensor.
Assuming the transverse space is isotropic this
will satisfy
\[ \sigma^{x y} = - \sigma^{y x}, \qquad \sigma^{x x} = \sigma^{y y}.\]
In complex co-ordinates $x\pm i y$
\[\sigma_\pm = (\sigma^{x y} \pm i \sigma^{x x}),
\qquad J_\pm =J^x \pm i J^y, \qquad
\E_\pm = \E_x \pm \E_y\]
in terms of which
\beq \delta J_\pm =\mp i\sigma_{\pm}(u)\E_\pm\qquad \Rightarrow \qquad
\sigma_\pm = \pm i \frac{\delta J_\pm}{\E_\pm}.
\label{eq:sigma-pm}\eeq
Note that $\E_-$ is not the complex conjugate of $\E_+$ in general,
since $\delta E_\alpha$ itself is complex for AC fields.

The authors in \cite{Hartnoll+Kovtun,Hartnoll+Herzog} further define 
\[\B^x = \frac{f}{u^4}\delta B^x = -\frac{f}{L}  \delta A'_y,\qquad
\B^y = \frac{f}{u^4}\delta B^y = \frac{f}{L} \delta A'_x\]
and show that Einstein's equations can be used to eliminate 
$\delta G_\alpha$ from Maxwell's equations resulting in
\begin{align}
\widetilde \omega 
\left(\mp f \E_\pm' + \widetilde \omega \B_\pm\right) 
&= 4 m u^2 (q  \E_\pm +  m \B_\pm) \label{eq:deltaE-B-I}\\
f\left(q\E_\pm' + m \B_\pm'\right) 
&= \mp \widetilde w (m  \E_\pm -  q \B_\pm), \label{eq:deltaE-B-II}
\end{align}
where $\widetilde \omega = \omega L$. These can be re-arranged to obtain
\beq
\pm\, \widetilde \omega \begin{pmatrix} \E_\pm' \\ \B_\pm' \end{pmatrix}
=  \begin{pmatrix} 
- 4 u^2 q m &  - 4 u^2 m^2 + \frac{\widetilde \omega^2}{f}  \\ 
4 u^2 q^2  - \frac{\widetilde \omega^2}{f} & {4 u^2 q m }
\end{pmatrix}
\begin{pmatrix} \E_\pm \\ \B_\pm \end{pmatrix}.
\label{eq:E'-B'}\eeq
From these we can derive an RG  equation for the electrical conductivity,
describing how it changes as $u$ is varied.

The two complex functions $\sigma_\pm$ have complete information about both the real (dissipative) and imaginary (refractive) parts of $\sigma^{x x}$ as well as the real and imaginary parts of $\sigma^{x y}$.
 In the notation of the previous section, for this solution and this specific form of variation, 
\[\Theta(A_\mu,\delta A_\mu, g_{\mu\nu},\delta g_{\mu\nu}) 
= -\frac{1}{2} \int_{\partial {\cal M}} \delta A_\nu F^{\mu \nu}\sqrt{-g}\,  d^3 \Sigma_\mu + O(\delta^2) 
\]
(with the variations specified in (\ref{eq:delta-A}) and (\ref{eq:delta-g}) the Einstein scalar itself gives no contribution to the variation at first order).
Choose space-time to be bounded by the event-horizon, $u=1$, an outer cylinder at some value of $u<1$ together with two constant $t$ space-like hypersurfaces $\Sigma_{t_\pm}$ at some future time $t_+$ and past time $t_-$. If we choose $t_+ - t_-$ to be an integer multiple of $\frac{2 \pi}{\omega}$ the contributions from $\Sigma_{t_\pm}$ cancel and
\begin{align} 
\Theta =-\frac{1}{2} \int_{\cal M} \partial_\mu (\delta A_\nu F^{\mu \nu}\sqrt{-g})  d^4 x
& = -\frac{1}{2} \int d t d x d y \left[\delta A_\alpha F^{u \alpha} \sqrt{-g} \,\right]_1^u\nonumber \\
& = -\frac{1}{2} \int d t d x d y \left( \delta A_\alpha  F^{u \alpha}\sqrt{-g}  \right)_u,\label{eq:Theta-dM}\end{align} 
where the contribution from $u=1$ has been discarded
since $F^{u \alpha}= \frac{u^4 f(u)}{L^2}F_{u \alpha}$ vanishes there.
With (\ref{eq:delta-A}) and (\ref{eq:delta-g}) 
\[\delta F^{u \alpha} = \frac{u^4 f}{L^2} \delta A_\alpha' - \frac{u^4 \tilde q}{L} \delta G_\alpha\]
and $\delta (\sqrt{-g})=0$ with $\sqrt{-g}=\frac{L}{u^4}$, so
\[ \delta \Theta(u) = -\frac{1}{2} \int d t d x d y \left( \frac{f}{L} \delta A_\alpha 
 \delta A_\alpha' - \tilde q \, \delta A_\alpha \delta G_\alpha   \right)_u. \]
The current is
\begin{align}
  \delta J^\alpha(u) &= -\frac{\delta S_{Class}}{\delta A_\alpha}
                       = -\frac{\delta \Theta}{\delta A_\alpha}\nonumber \\
                     &  = \frac{f}{L} \delta A_\alpha'(u)-\tilde q \,\delta G_\alpha
                       = \frac{f}{u^4} \epsilon^\alpha{}_\beta \delta B^\beta
                      -\tilde q \, \delta G_\alpha.\label{eq:J-delta-G}
\end{align}
The contribution of $\delta G_\alpha$ to the current is a manifestation of the piezoelectric effect (see appendix \ref{app:piezoelectric})
but the conductivity is defined as the response to a variation of the electric field, not the metric ---  while the
$\tilde q \, \delta G_\alpha$ term contributes to the current it
does not contribute to the conductivity.  In linear response
the conductivity (\ref{eq:sigma-pm}) is therefore\footnote{It is not the case that $\delta G_\alpha=0$ in the linearised equations of motion, it will not be zero in the bulk as it is related to $\E_\alpha$ and $\B^\alpha$ (though it can be set to zero at the boundary as we can always add a constant to $\delta G_\alpha$). Equation (\ref{eq:sigma_B_over_E}) is taken as the definition of the conductivity for a general value of
  $u<1$.}
\beq \sigma_\pm(u) = \left.\frac{\delta J_\pm}{\E_\pm}\right|_{\delta G=0}
= \left.\frac{\B_\pm}{\E_\pm}\right|_{\delta G=0},\label{eq:sigma_B_over_E}\eeq
as in  \cite{Hartnoll+Herzog}.
Equations (\ref{eq:E'-B'}) and (\ref{eq:sigma_B_over_E}) can now be used to derive an RG equation for $\sigma_\pm(u)$.
Differentiating (\ref{eq:sigma_B_over_E}) with respect to $u$ and using (\ref{eq:E'-B'}) gives
\begin{align*}
\left( \frac{\B_\pm}{\E_\pm} \right)'
&= \frac{\B_\pm' \E_\pm - \B_\pm \E_\pm'}{\E_\pm^2}\\
&=\pm \frac{1}{\widetilde \omega  f}
\left\{ (4 u^2 f  q^2 -\widetilde  \omega^2) 
+ 8 u^2 f  q m \left( \frac{\B_\pm}{\E_\pm} \right) 
+\left(4 u^2 f m^2 -\widetilde \omega^2\right) \left( \frac{\B_\pm}{\E_\pm} \right)^2 \right\}.\end{align*}
This can be re-arranged to write the RG equation for the conductivity as a Riccati equation
\beq \boxed{\vline height 22pt depth 12pt width 0pt \quad 
\pm \widetilde \omega f(u) \sigma_\pm'
= 4 u^2 f(u) ( q + m \sigma_\pm)^2 - \widetilde \omega^2 (1+ \sigma_\pm^2)
\label{eq:dsigma-du} \quad
}\eeq
(the relevance of Riccati equations to holographic 2-point functions was observed in \cite{PT,LPTV}) and matrix version of coupled Riccati RG equations was presented in \cite{GTWW1}-\cite{GTWW4}.
The RG equation is also related to the Hamilton-Jacobi equation involving the effective action \cite{deBoer-V2}.

\section{Solutions of the conductivity RG equation\label{sec:solutions}}

To study solutions of (\ref{eq:dsigma-du}) first note that
it is singular at the event horizon, since
$f(u) \rightarrow 0$ as $u\rightarrow 1$ and, if $\wo\ne0$, we must necessarily use boundary conditions $\sigma_\pm^2 = -1$ at $u=1$.
Next, as observed in \cite{Hartnoll+Herzog},
\[\sigma_-(m)=-\sigma_+(-m) \]
which allows us to extract
\begin{align}
Re(\sigma^{x y})&=\frac{1}{2} Re\bigl( \sigma_+(m) -\sigma_+(-m) \bigr), \qquad
Im(\sigma^{x y})=\frac{1}{2} Im\bigl( \sigma_+(m) -\sigma_+(-m) \bigr),\label{eq:sigmaxy} \\
Re(\sigma^{x x})&=\frac{1}{2} Im\bigl(\sigma_+(m) +\sigma_+(-m) \bigr), \qquad
Im(\sigma^{x x})=-\frac{1}{2} Re\bigl(\sigma_+(m) +\sigma_+(-m)\bigr).\label{eq:sigmaxx}
\end{align}
At the event horizon $\sigma_\pm^2=-1$ is independent of $m$ so
$\sigma^{x y} = 0$ and $\sigma_\pm = \pm i\sigma^{x x}$.
Since $Re(\sigma^{x x})$ is necessarily positive $\sigma^{x x}=1$ at the event horizon 
and we must use boundary conditions
\[ \sigma_\pm|_{u=1} =\pm i.\] 
These boundary conditions correspond to matter falling into the event horizon
and, as emphasised in \cite{HLS}, this is the
source of dissipation in holography.  
In particular these boundary conditions imply that
it is not the case that the solutions of (\ref{eq:dsigma-du})
are related by 
\[ \sigma_+(-\wo)=\sigma_-(\wo)\]
but rather
\beq \sigma_+(-\wo) = \sigma^*_-(\wo).\label{eq:sigma-minus-omega}\eeq
Since $Re(\sigma^{x x}) = \frac{1}{2}Im(\sigma_+ - \sigma_-)$ changing the sign of $\wo$
corresponds to sending $Re(\sigma^{x x}) \rightarrow - Re(\sigma^{x x})$, as one expects from time reversal. The sign of $\wo$ should always be chosen so that the dissipative Ohmic conductivity, $Re(\sigma^{x x})$, is non-negative.

Equation (\ref{eq:dsigma-du}) is also invariant under electromagnetic duality in the bulk,
\[ q \rightarrow m,\quad m \rightarrow -q,\quad \sigma_\pm \rightarrow -\frac{1}{\sigma_\pm},\] 
a property of the conductivity noted in \cite{Hartnoll+Kovtun}.

Of course the real and imaginary parts of (\ref{eq:sigmaxy}) and (\ref{eq:sigmaxx}) are not independent, they are related
by the Kramers-Kronig relations
\begin{align*}
Re(\sigma^{x x}(\wo) &=\frac{1}{\pi} P \int_{-\infty}^\infty \frac{Im\bigl(\sigma^{x x}(\wo\,')\bigr)}{(\wo\,' -\wo)} d\, \wo\,',\\
Im(\sigma^{x x}(\wo) &=-\frac{1}{\pi} P \int_{-\infty}^\infty \frac{Re\bigl(\sigma^{x x}(\wo\,')\bigr)}{(\wo\,' -\wo)} d\, \wo\,',\\
\end{align*} 
and similarly for $\sigma^{x y}(\wo)$.

Some consequences of equation (\ref{eq:dsigma-du}) in certain limits are immediate:

\subsection{AC conductivity}

\subsubsection{$u\rightarrow 1$}

At the event horizon, as $u\rightarrow 1$, $f\rightarrow 0$ and, if $\omega\ne 0$,
\beq \sigma_\pm = \pm i \qquad \Rightarrow \qquad \sigma^{x y}=0,\  \sigma^{x x}=1.\label{sigma-horizon}\eeq
This is the attractor mechanism.

\subsubsection{Large $\wo$ or small $u$}
 When $\wo$ is large or $u$ is small the first term on the right-hand side of (\ref{eq:dsigma-du}) can be ignored and 
\beq  f(u) \sigma_\pm' = \mp \wo (1+\sigma_\pm^2).\eeq
As $\wo \rightarrow \infty$ (or as $u\rightarrow 1$) the solution becomes a constant $\sigma_+ = i$ and
\beq
\sigma^{x x} \rightarrow 1, \qquad  \sigma^{x y} \rightarrow 0.
\label{eq:omega->infinity-conductivities}\eeq
On the other hand, as $u\rightarrow 0$ for large but finite $\wo$, $f\rightarrow 1$ and
\beq \sigma_\pm' = \mp \wo (1+\sigma_\pm^2)+O(u^2).\label{eq:asymptotic-RG}\eeq
This is independent of $m$ and hence, from equations (\ref{eq:sigmaxy}) and (\ref{eq:sigmaxx}),
\[\sigma^{x y} \approx \ O(u^3)
  \qquad \Rightarrow \qquad \sigma_+ \ = \ i \sigma^{x x} +O(u^3) \]
so
\begin{align*}
  \sigma^{x x}(u,\wo)
  = & \\
  & \kern - 45pt \sigma^{x x}(0,\wo)
+ i\bigl\{1-\sigma^{x x}(0,\wo)\bigr\}\wo u
    + \sigma^{x x}(0,\wo)\bigl\{ 1-\sigma^{x x}(0,\wo)\bigr\}(\wo u)^2 + O(u^3),
    \end{align*}
where $\sigma^{x x}(0,\wo)$ must be determined by numerically
integrating (\ref{eq:dsigma-du}) from $u=1$ to $u=0$ with the boundary condition $\sigma_\pm=\pm i$ at $u=1$.  The classical Drude form of the conductivity, allowing for the cyclotron resonance, would be
\beq \sigma^{x x}(0,\wo) = \frac{\sigma_0}{1-i(\omega-\omega_0)\tau}\label{eq:Drude}\eeq
with cyclotron frequency $\wo_0$ and relaxation time $\tau$. 

\subsubsection {Cyclotron resonance} 

The presence of a resonance is straightforward to show when the second term on the
right hand side of (\ref{eq:dsigma-du}) is small relative to the first and can be ignored. 
We then have
\beq
\pm \widetilde \omega \sigma_+'= 4 u^2 ( q + m \sigma_+)^2. \label{eq:dsigma-du-truncated}
\eeq
With the boundary condition $\sigma_+|_{u=1} = i$
the solution is
\beq
\sigma_+ = \frac{3 i \wo - 4 q (q+i m)(1-u^3)}{3 \wo + 4 m (q+ i m) (1-u^3)}.\label{eq:sigma-Pade-u}
\eeq
In the complex $\sigma_+$-plane, plotting $Im(\sigma_+)$ against $Re(\sigma_+)$ using $\wo$ as a parameter, this
is a perfect circle lying entirely in the upper-half complex plane and tangent to the real axis.
In terms of the filling factor $\nu=\frac{q}{m}$
the radius is $\frac{1+\nu^2}{2 }$
and the centre is at $Re(\sigma_+)=-\nu$, $Im(\sigma_+)=\frac{1+\nu^2}{2}$.  
The approximate form (\ref{eq:sigma-Pade-u}) for $\sigma_+$ was found in \cite{Hartnoll+Kovtun}
for small $q^2$ and $m^2$, both of the order of $\wo$, and indeed it works best in this range.

When $u<1$ there is a pole in the complex $\wo$-plane at
\[ \wo_* = -\frac 4 3 m (q + i m)(1-u^3)\]
and near $\wo_*$ equation (\ref{eq:sigma-Pade-u}) has the form of a resonance
\beq \sigma_+ = \frac{i\sigma_0 \Gamma}{\wo-(\wo_0 - i \Gamma)}\label{eq:pole-form-of-sigma} \eeq
with resonance frequency, width and amplitude at resonance
\beq \wo_0 = -\frac 4 3 q m(1-u^3),\qquad \Gamma = \frac 4 3 m^2(1-u^3),
\qquad \sigma_0 =\frac{q}{m^2} (-2m +i q).\label{eq:omega_0-linear}\eeq

In this approximation the corresponding ${\cal Q}$-factor
\beq {\cal Q} = \frac{|\wo_0|}{2 \Gamma}= \frac{1}{2}\left| \frac q m \right|.\label{eq:Q-factor-approx}\eeq
is one-half the filling factor, independent of $u$, and the residue is
\beq \sigma_0\, \Gamma =   
\frac{ 4 q}{3} \left(  - 2 m 
+ i q\right)(1-u^3)\label{eq:residues}\eeq
The resonance disappears at $u=1$ and is strongest at $u=0$ where
\beq \wo_0 = - \frac{4}{3} q m.\label{eq:linear-omega0}\eeq

When $\sigma_+(m)$ and $\sigma_+(-m)$ are combined to construct the Ohmic and Hall conductivities
in (\ref{eq:sigmaxy}) and (\ref{eq:sigmaxx}) one might na{\rm \"i}vely be led,
at $u=0$, to
\beq
\frac{1}{2 i} \bigl( \sigma_0(m) + \sigma_0(-m) \bigr) = \frac{q^2}{m^2}
\quad \hbox{and} \quad
\left| \frac{1}{2} \bigl( \sigma_0(m) - \sigma_0(-m)\bigr)\right|=\left|\frac q m \right|,
\label{eq:naive-sigma_0}
\eeq 
but these are not the same as the maximum values of $Re(\sigma^{x x})$
and $\bigl|Im(\sigma^{x y})\bigr|$.  
Using  (\ref{eq:sigma-Pade-u}), and setting $u=0$, gives
\begin{align}
\sigma^{xx}& = \frac{3 \,\wo\bigl\{3 \,\wo+ 4 i (q^2 + m^2)  \bigr\} }{(3\wo + 4 m(q + i m))(3\wo - 4 m(q - i m) )}\label{eq:sigma_xx_pole_approx}\\
\sigma^{x y}& = \frac{8 q m\bigl\{2 (q^2 + m^2) - 3 i \wo \bigr\}}{(3\wo + 4 m(q + i m))(3\wo - 4 m(q - i m) )}.\label{eq:sigma_xy_pole_approx}
\end{align}
In these expressions the maximum of $Re(\sigma^{x x})$
occurs when
\beq
(\wo^{x x}_0)^2={\frac { 16\,m^2(q^2 + m^2)^{\frac 3 2}  \left( 2 |q|+\sqrt{q^2 + m^2} \right) 
}{9\,(3\,{q}^{2}-{m}^{2})}} 
\label{eq:omega_0-xx-approx}\eeq
where 
\beq \sigma^{x x}_{Max}:=Re\bigl( \sigma^{x x}(\wo^{x x}_0)\bigr)= 
\frac{ (q^2+ m^2) \left( 2\,|q| +\sqrt{q^2+ m^2} \right) }{4 \,|q| \left( {m}^{2}-{q}^{
2} + |q|\,\sqrt{q^2+ m^2}\right)}.\label{eq:Pade-sigma-xx}\eeq
Since $\wo^{x x}_0$ is only real for $m^2 < 3 \,q^2$, we would expect that $m$ must be restricted to this range
and  $\sigma^{x x}_{Max}=1$ when the bound is saturated and $\wo^{x x}_0 \rightarrow \infty$.
However although $\wo^{x x}_0$ diverges when $m^2 = 3 q^2$, and becomes imaginary for 
$m^2> 3 q^2$, $\sigma^{x x}_{Max}$ is well defined for all $0<m^2< 3-q^2$
and indeed the
approximation (\ref{eq:Pade-sigma-xx}) is remarkably close to the numerical
result shown in Fig.\,{\ref{fig:peaks}(a) when $q=0.1$ for the whole of the allowed range of $m$.

The full numerical solutions presented below indicate that near the cyclotron frequency, and for a significant band on either side of it, 
$\bigl|Im\bigl( \sigma^{x y}(\wo)\bigr)\bigr|$  behaves very like the dissipative part of the Ohmic conductivity while the real part displays the properties of a refractive conductivity
($\sigma^{x y}$ changes sign when either $m$ or $q$ change sign).  
The maximum of $\bigl|Im\bigl( \sigma^{x y}(\wo)\bigr)\bigr|$ is at 
\beq
(\wo^{x y}_0)^2=
\frac {16 m^2\Bigl( \,{m}^{2}-\,{q}^{2}+2\,\sqrt { q^4+ q^2 m^2 + m^4  }\Bigr)}{9}
\eeq
where $\sigma^{x y}_{Max}=\Bigl|Im\bigl(\sigma^{x y}(\wo^{x y}_0)\bigr)\Bigr|$ is
\beq \sigma^{x y}_{Max}
=
{\frac {|q| \left| {m}^{2}-{q}^{2}+2\,\sqrt {{m}^{4}+{m}^{2}{q}^{2}+{q}^{4}} \right|^{3/2}}
{4\Bigl|{m}^{4}+{q}^{4}+ 
\left( {m}^{2}-{q}^{2} \right) \sqrt {{m}^{4}+{m}^{2}{q}^{2}+{q}^{4}}\Bigr|}}.\label{eq:Pade-sigma-xy}
\eeq

These expressions can be more succinctly written in terms of the filling factor $\nu = \frac{q}{m}$ 
\begin{align*}
\sigma^{x x}_{Max}
= \frac{(1+\nu^2)\Bigl(2|\nu| + \sqrt{1+\nu^2}\Bigr)}
 {4\,|\nu|\Bigl(1-\nu^2 + |\nu|\sqrt{1+\nu^2}\Bigr)}\\
\wo^{x x}_0=\frac{4 m^2  (1+\nu^2)^{\frac 3 4}}{3}
\sqrt{ \frac{\bigl( 2|\nu|+ \sqrt{ 1+\nu^2}\bigr) }{ 3\nu^2 -1}}\ ,\\
\sigma^{x y}_{Max}
 = \frac{|\nu|\Bigl| 1 - \nu^2 + 2\sqrt{1 + \nu^2 + \nu^4}\Bigr|^{\frac 3 2} }
{4\Bigl| 1 + \nu^4 +(1-\nu^2)\sqrt{1 + \nu^2 + \nu^4}\Bigr| }\\
\wo^{x y}_0  = \frac{4 m^2}{3} \sqrt{ 1 - \nu^2 + 2\sqrt{1 + \nu^2 + \nu^4}}\ .
\end{align*}

It is not immediately obvious
what the widths of the peaks are in (\ref{eq:sigma_xx_pole_approx}) and (\ref{eq:sigma_xy_pole_approx}).  We shall define the widths by assuming the resonance form (\ref{eq:pole-form-of-sigma}), {\it e.g.} for $\sigma^{x x}(\wo)$
\[\bigl( \sigma^{x x}(\wo)\bigr)^{-1} = \frac{\wo - \wo^{x x}_0 +i \Gamma^{x x}}{i\sigma^{x x}_{Max}\Gamma^{x x}},\]
for which the width would be given by
\beq \bigl(\Gamma^{x x}\bigr)^{-1} 
=- Re\bigl(\sigma^{x x}(\wo_0)\bigr){Im \left[\frac{d (\sigma^{x x})^{-1}}{d \,\wo}\right]}_{\wo = \wo^{x x}_0}
%\vline width 0.5pt height 17pt depth  7pt }
\ .\label{eq:Gamma_xx_def}\eeq
For any $\sigma^{x x}(\wo)$ with a maximum of $Re \bigl(\sigma^{x x}(\wo)\bigr)$ at $\wo_0$, 
where $Re\bigl(\sigma^{x x}(\wo^{x x}_0)\Bigr)=\sigma^{ x x}_{Max}$, equation (\ref{eq:Gamma_xx_def}) 
 will be used as the definition of $\Gamma^{x x}$.
Similarly
\[ \bigl(\Gamma^{x y}\bigr)^{-1}  = \Bigl|Im\bigl(\sigma^{x y}(\wo_0)\bigr)\Bigr|
{ Re \left[\frac{d (\sigma^{x y})^{-1}}{d \,\wo}\right]}_{\wo = \wo^{x y}_0}.\]

With these definitions the widths arising from the analytic approximations
(\ref{eq:sigma_xx_pole_approx}) and (\ref{eq:sigma_xy_pole_approx}) are not very illuminating but are given here for completeness, 
\begin{align*}
\Gamma^{x x}&=
{\frac { 8\,m^2 \nu\sqrt {1+{\nu}^{2}}
\left( 1-{\nu}^{2} +|\nu|\sqrt {1+{\nu}^{2}} \right) \left(2+ 3\,|\nu|\sqrt {1+{\nu}^{2}} \right) ^{2}  }
{ 3\,\left( 3\,{\nu}^{2}-1 \right)  
\left( 1+3\,{\nu}^{2}+6\,{\nu}^{4} + 3\,|\nu| \left( {\nu}^{2}+1 \right) ^{3/2}\right) }},\\
\Gamma^{x y}&= 
\left|{\frac { 8 m^2\bigl( 1+{\nu}^{4}+(1-\nu^2)\sqrt {{\nu}^{4}+{\nu}^{2}+1} \bigr)  
\bigl( 5-2\,{\nu}^{2}+{\nu}^{4}+8\,\sqrt {{\nu}^{4}+{\nu}^{2}+1} \bigr) ^{2}}
{3 \left( 1-{\nu}^{2}+2\,\sqrt {{\nu}^{4}+{\nu}^{2}+1} \right) ^{2} \left( 1+{\nu}^{2}\right) ^{3}}}\right|.
\end{align*}
These expressions are compared to the full numerical solutions graphically below.

\subsubsection{$m=0$}

When $m=0$ equation (\ref{eq:dsigma-du}) can also be solved
analytically in the approximation that $\wo$ is small.
It is a Riccati equation which can easily be recast as a second order, linear, homogeneous equation. 
Details are given in appendix \ref{app:2nd-order-RG} and in this section we examine the case $m=0$.
Define the function $X(u)$ via
\beq f X' -  \wo \sigma_+ X=0,  \label{eq:X-def}\eeq 
Then, from (\ref{eq:D2X}) with $m=0$, equation (\ref{eq:dsigma-du}) can be re-expressed as
\[
f^2 X''+ff' X'+\left(\wo^2-4 q^2 u^2 f \right) X  =0,\]
which, for $\wo=0$,  reduces to
\beq f X'' + f' X' = 4 q^2 u^2 X,\label{eq:D2Xm=0}\eeq
with corrections of $O(\wo^2)$. 
Since $f(1)=0$ and $f'(1)=-3+q^2$ a boundary condition on $X(u)$ at $u=1$ 
is
\[X'(1)=-\left( \frac{4 q^2 }{ 3- q^2 }\right) X(1),\]
Up to an overall normalisation equation (\ref{eq:D2Xm=0}) determines its own boundary conditions because $u=1$ is a singular point.
The solution is immediate
\[X(u)= c\bigl\{3(1+q^2)-4 q^2 u\bigr\}\]
with $c$ constant and we arrive at
\[\wo \sigma_+(u)= \frac{f X'}{X}
=
-\frac{4 q^2(1-u)\bigl(1+u+u^2 - q^2 u^3\bigr)}{3(1+q^2) - 4 q^2 u},\]
so asymptotically
\[ \sigma_+(0)=-\frac{4 q^2}{ 3(1+q^2)\wo}.\]

Hence, from (\ref{eq:sigmaxy}) and (\ref{eq:sigmaxx}), as $u\rightarrow 0$ the refractive Ohmic conductivity has a pole at $\wo=0$,
\beq 
Im(\sigma^{x x})
=\frac{4 q^2}{3(1+q^2)\wo},
\label{eq:m=0}\eeq
with zero width and the ${\cal Q}$-factor diverges.
This is a better approximation than (\ref{eq:sigma-Pade-u}), with $m=0$ and $u=0$, because (\ref{eq:sigma-Pade-u}) is only valid for $q^2\sim O(\wo)$ when $\wo\ll 1$.

\subsection{DC conductivity}
 
\subsubsection{$0<u<1$}

 Away from the event horizon $f\ne 0$ and, when $\omega=0$, 
\[ \sigma_\pm = -\frac{q}{m} \qquad \Rightarrow \qquad \sigma_{x y} 
= -\frac{q}{m},\quad  \sigma_{x x}=0,\]
the Ohmic conductivity vanishes and the Hall conductivity is $-\frac{q}{m}$,
as found in \cite{Hartnoll+Herzog}.

\subsubsection{$u \rightarrow 1$}
 
We can go all the way to $u=1$ for the DC conductivity with the solution
\beq \sigma_\pm(u) = -\frac{q}{m} + \left( \frac{q}{m}  \pm i\right)H(u)\label{eq:omega=0-conductivities}\eeq
where 
\[ H(u) = \begin{cases} 0, & u<1;\\ 1, & u=1\end{cases}
\]
is the step function.
Thus again $\sigma^{x y}=0$ and $\sigma^{x x}=1$,
again the same value that one gets from infalling boundary conditions at the horizon for a 2-dimensional system \cite{HLS}.

As observed earlier for any $\omega$ the conductivity at $u=1$ is always $\sigma^{x y}=0$,
$\sigma^{x x}=1$, an attractive fixed point in the infra-red. 
A universal critical point at $\sigma^{x x}=1$ was predicted in \cite{FGG,CFGWY}.

\subsubsection{$u\rightarrow 0$}

The $u\rightarrow 0$ and $\wo \rightarrow 0$ limits do not commute and we
cannot trust (\ref{eq:omega=0-conductivities}) at $u=0$.
Instead consider the conductivity (\ref{eq:Drude}) in the limit $\sigma_0\rightarrow \infty$, $\tau\rightarrow\infty$,
with the ratio $\frac{r_0}{L}=\frac{\sigma_0}{\tau}$ finite,
\[\sigma^{x x}(\wo) \rightarrow \frac{i \sigma_0}{\tau(\wo-\wo_0)}.\] 
The Kramers-Kronig relations
require that a pole in $Im\bigl(\sigma^{xx}(\omega)\bigr)$ is associated
with a $\delta$-function singularity\ in $Re\bigl(\sigma^{xx}(\omega)\bigr)$,
\beq\sigma^{xx}(\omega)=\frac{r_0}{L}\left( \frac{\tau }{1-i(\omega-\omega_0)\tau}\right)\quad 
\mathop{\longrightarrow}_{\tau\rightarrow \infty} \quad 
r_0\left(\pi\delta(\wo-\wo_0) +\frac{i}{(\wo-\wo_0)}\right).\label{eq:omega-residue}\eeq
$\wo_0=0$ when $m=0$ and a $\delta$-function in the DC conductivity is
a signal of a superconductor.
  Equation (\ref{eq:m=0}) gives  
\[ r_0 = \frac{4 q^2}{ 3(1+q^2)}. \]

\subsection{Numerical solutions}

In this section we shall present some conductivity profiles obtained by integrating (\ref{eq:dsigma-du}) numerically. Some conductivities obtained from 
numerical integration of asymptotically AdS solutions of a bulk theory consisting of Einstein-Maxwell with a negative cosmological constant have been calculated before, \cite{LPTV,Hartnoll+Herzog+Horowitz}.
A more thorough investigation
is made here as an example of the application  of equation (\ref{eq:dsigma-du}).

In Fig.\,\ref{fig:1-1-omega} the conductivities are plotted for $q=m=1$ as functions of $u$ for various values of $\wo$ . The approach to the analytic form 
(\ref{eq:omega=0-conductivities}) as $\wo$ approaches zero is evident.
The negative of the imaginary part of the Hall conductivity (purple curve)
is plotted to emphasize the similarity (and difference) with the real part of
the Ohmic conductivity (red curve) ---  the same flip in sign would be achieved by changing the sign of either $q$ or $m$.  

\begin{figure}
  %\centering
\vspace{-4cm}
\hspace{-1.5cm}
\begin{subfigure}[t]{0.45\textwidth}
%\centering
\includegraphics[scale=0.4]{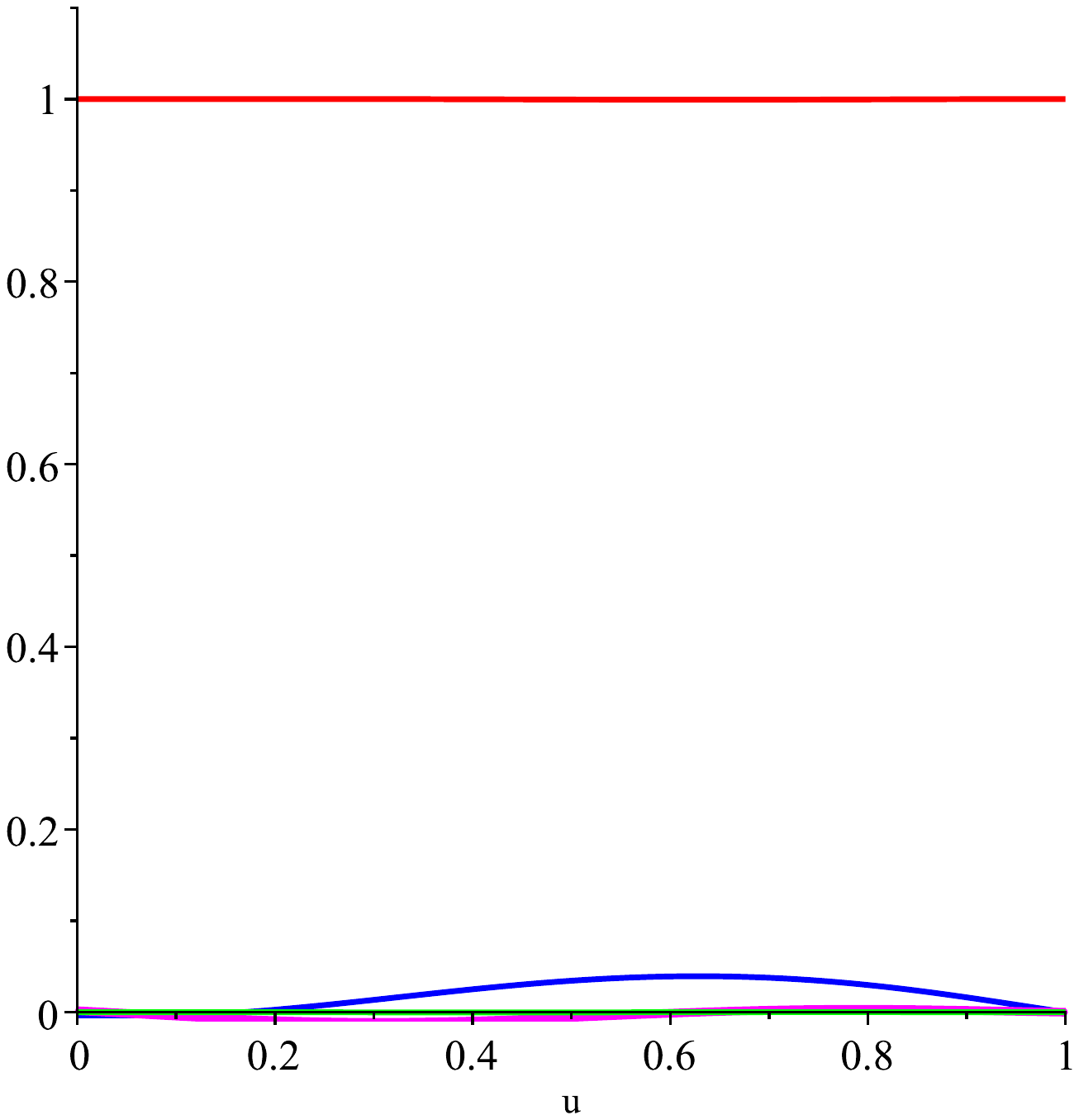}
\vspace{-5.5cm}
\subcaption{$\scriptstyle \wo =5$}
\end{subfigure}
\begin{subfigure}[t]{0.45\textwidth}
\centering
\includegraphics[scale=0.4]{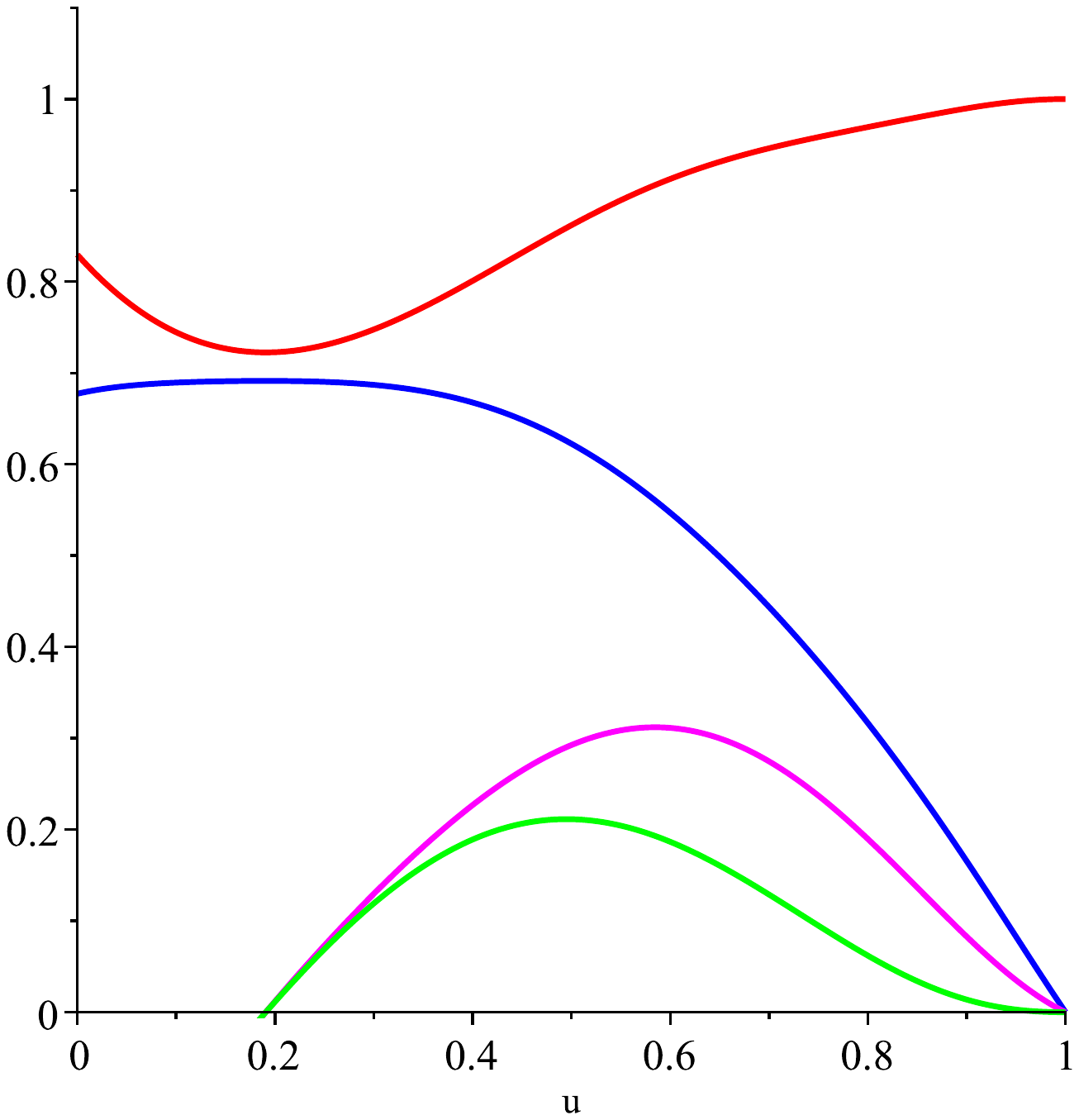}
\vspace{-5.5cm}
\subcaption{$\scriptstyle \wo =1.5$}
\end{subfigure}

\vspace{-5cm}
\hspace{-1.5cm}
\begin{subfigure}[t]{0.45\textwidth}
\centering
\includegraphics[scale=0.4]{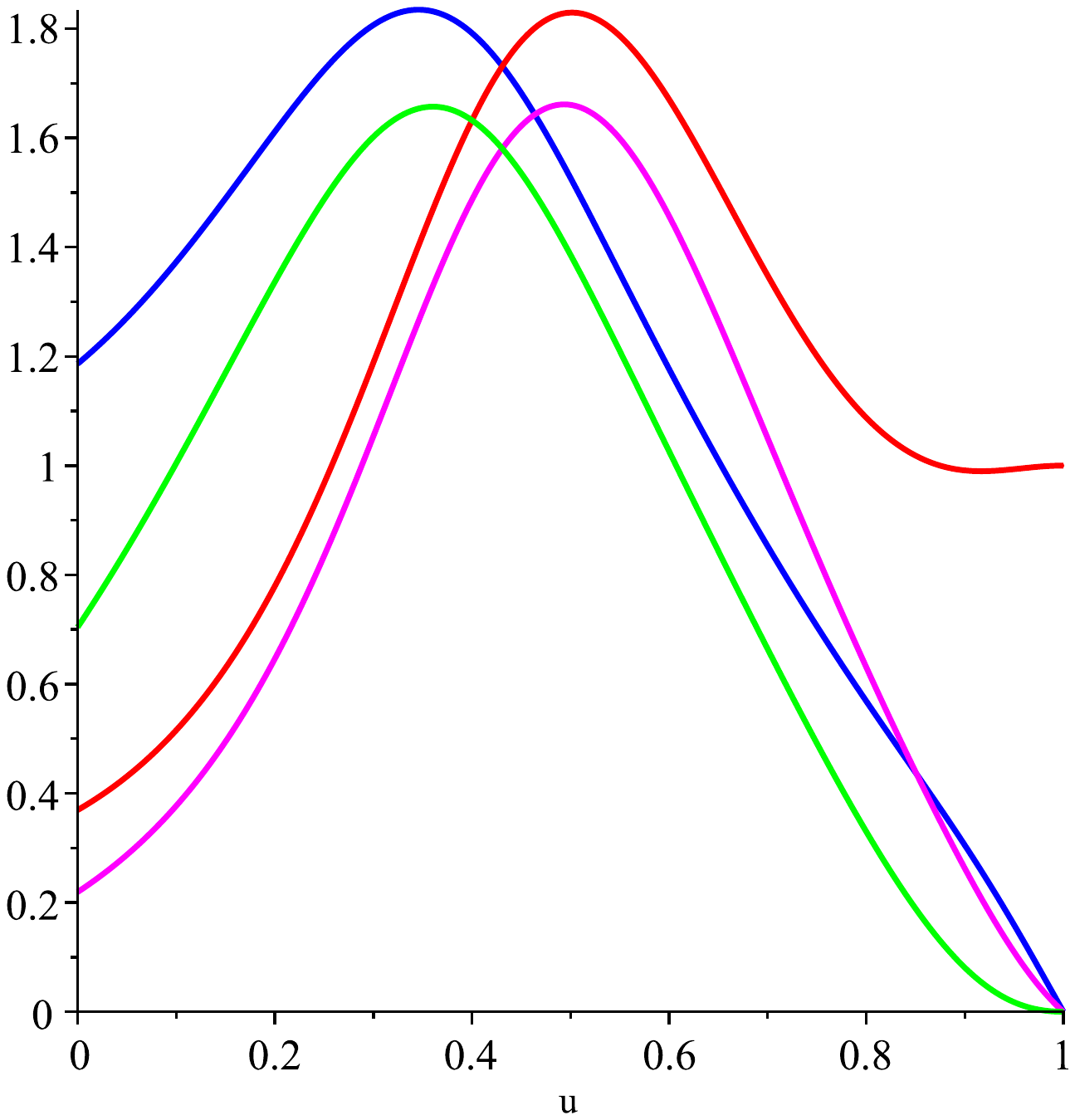}
\vspace{-5.5cm}
\subcaption{$\scriptstyle \wo =1$}
\end{subfigure}
\begin{subfigure}[t]{0.45\textwidth}
\centering
\includegraphics[scale=0.4]{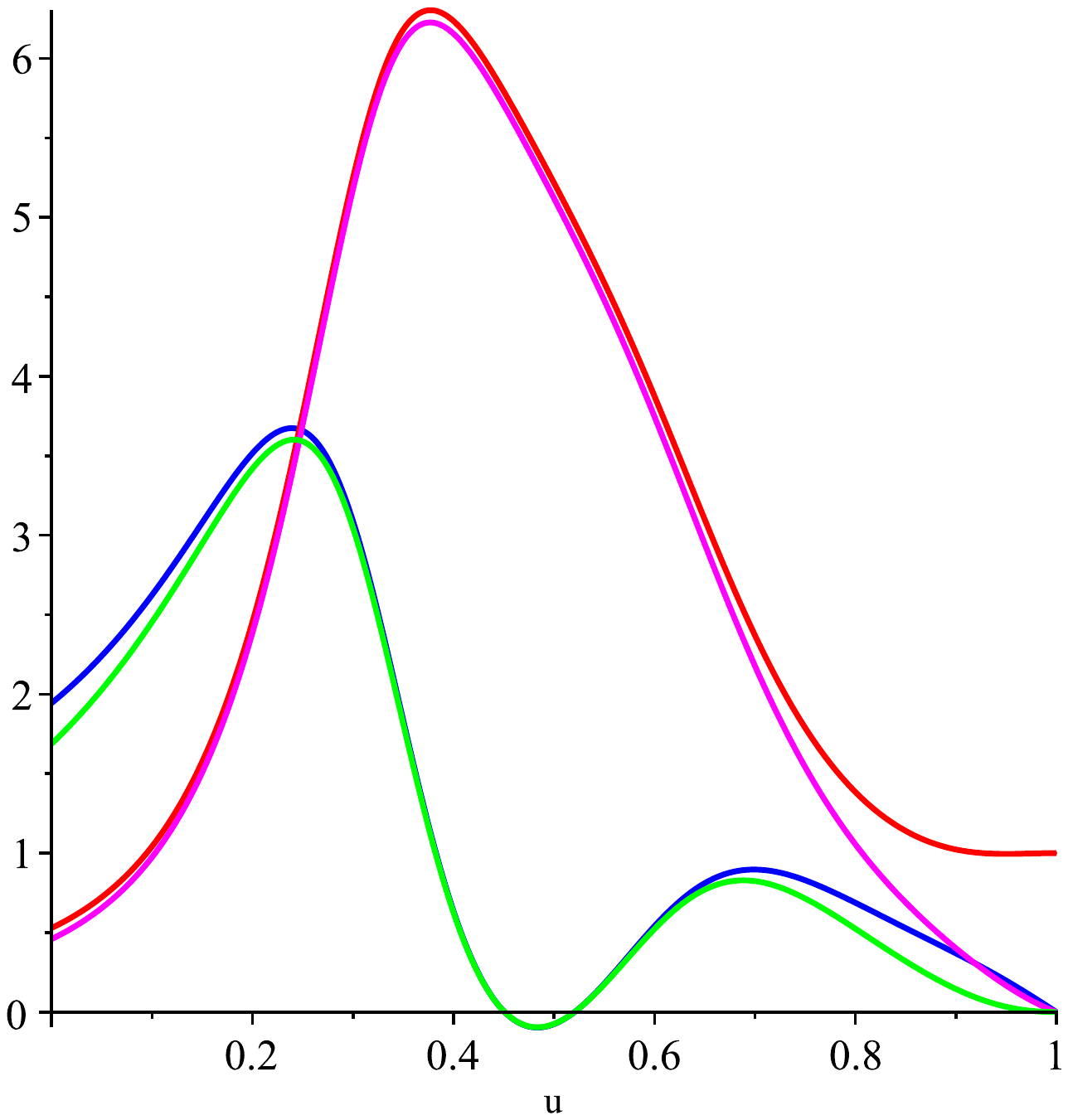}
\vspace{-5.5cm}
\subcaption{$\scriptstyle \wo =0.85$}
\end{subfigure}

\vspace{-5.5cm}
\hspace{-1.5cm}
\begin{subfigure}[t]{0.45\textwidth}
\centering
\includegraphics[scale=0.4]{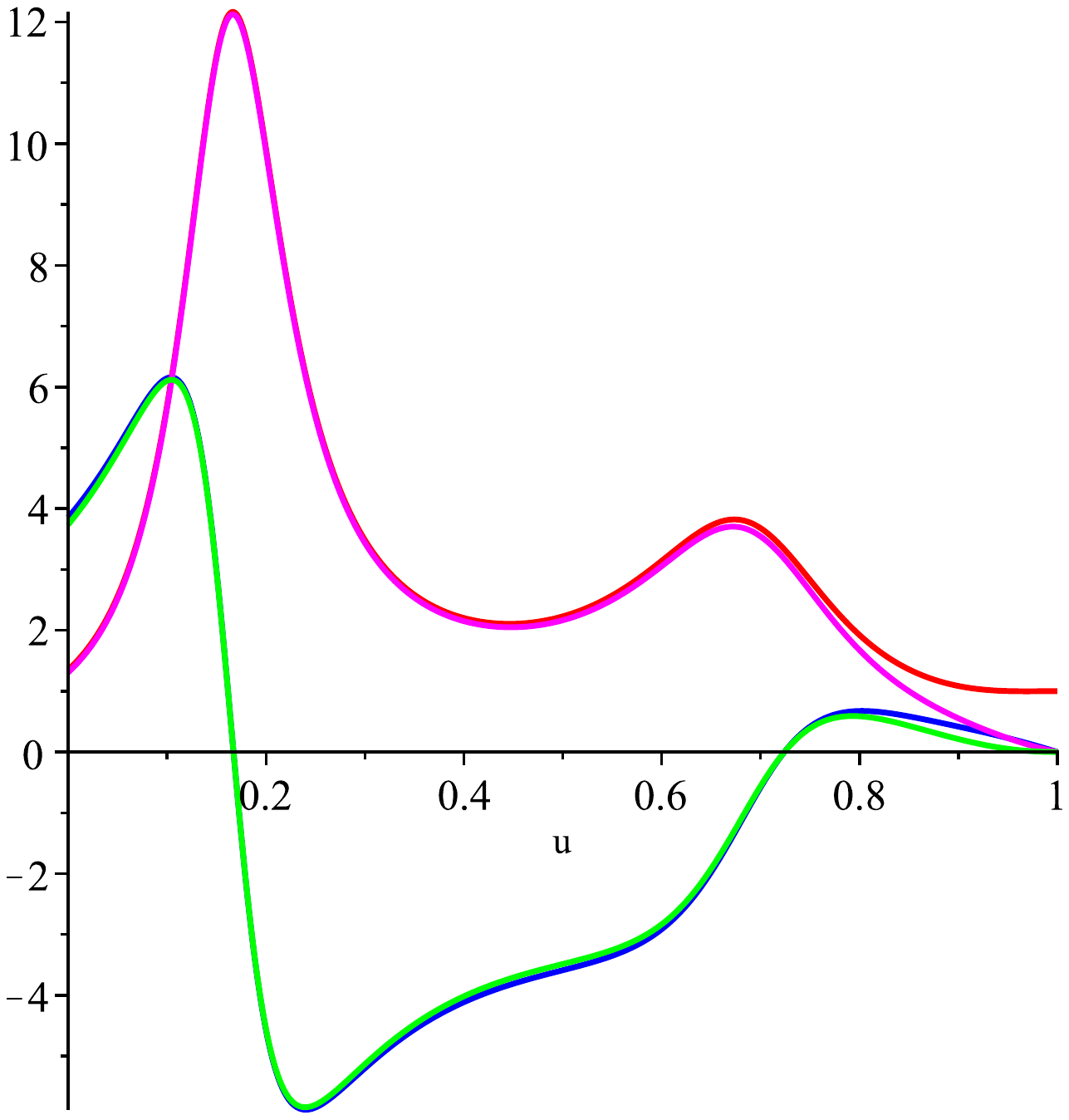}
\vspace{-5.6cm}
\subcaption{$\scriptstyle \wo =0.75$}
\end{subfigure}
\begin{subfigure}[t]{0.45\textwidth}
\centering
\includegraphics[scale=0.4]{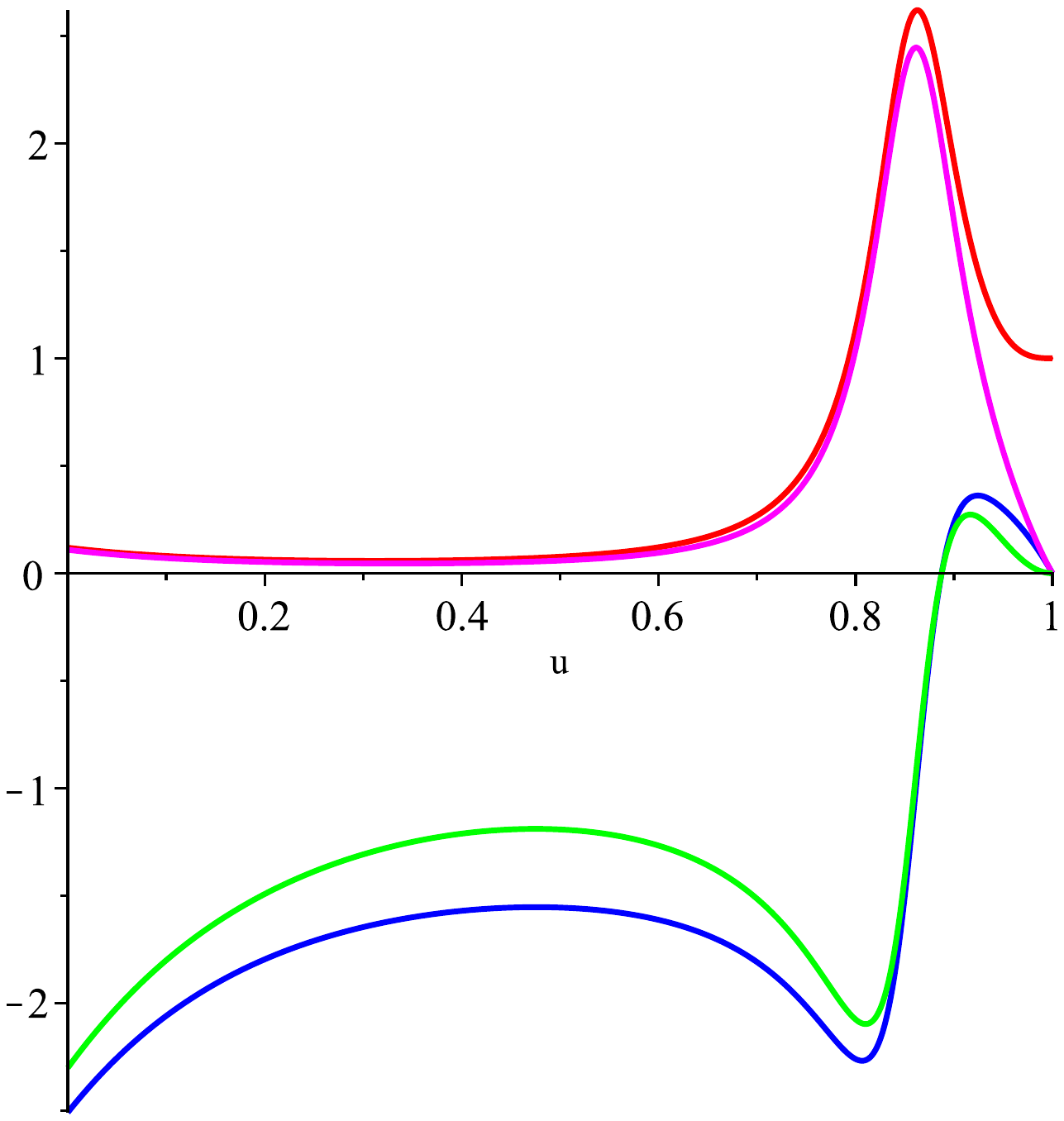}
\vspace{-5.6cm}
\subcaption{$\scriptstyle \wo =0.5$}
\end{subfigure}

\vspace{-5.7cm}
\hspace{-1.6cm}
\begin{subfigure}[t]{0.45\textwidth}
\centering
\includegraphics[scale=0.4]{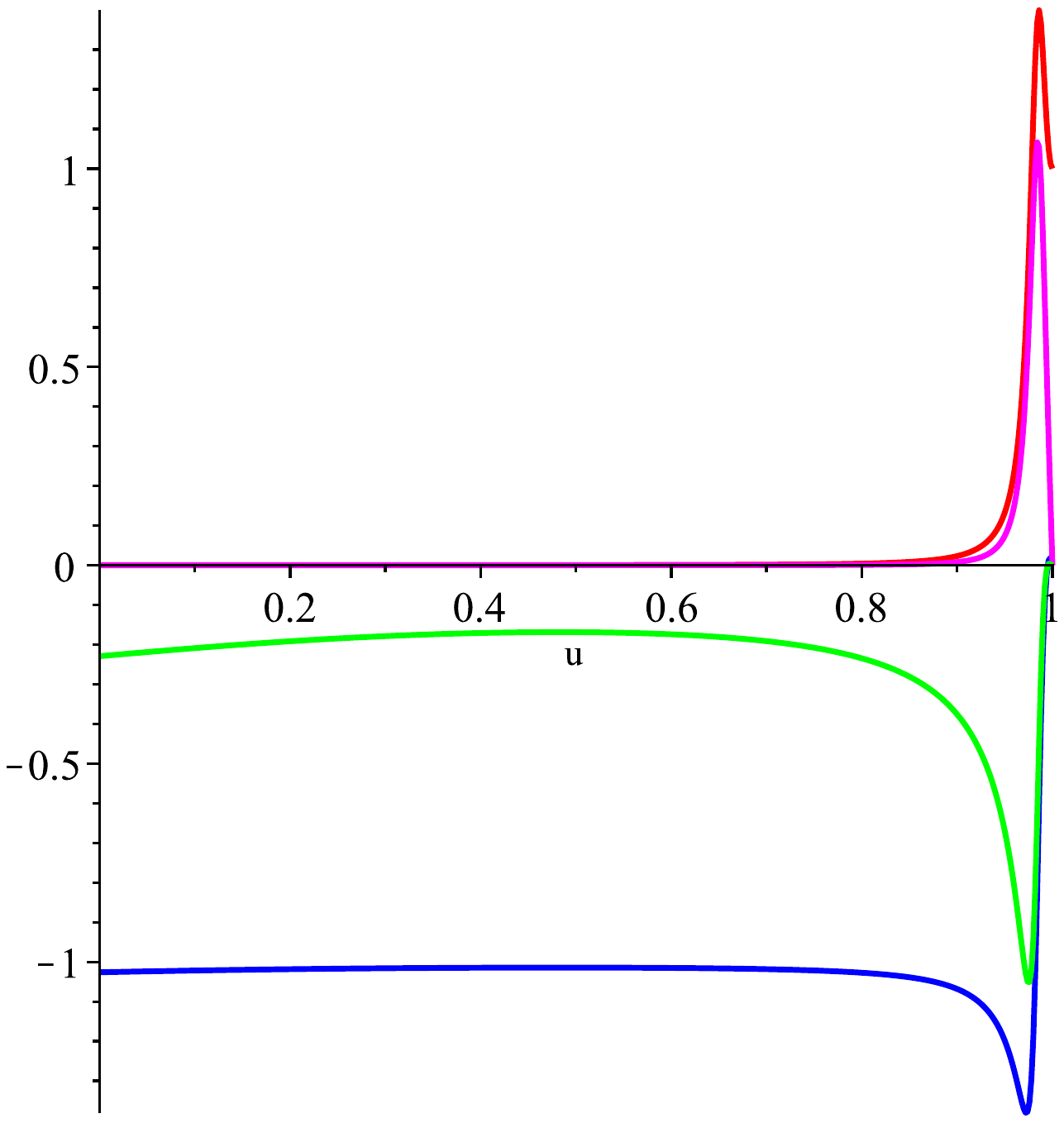}
\vspace{-5.6cm}
\subcaption{$\scriptstyle \wo =0.1$}
\end{subfigure}
\begin{subfigure}[t]{0.45\textwidth}
\centering
\includegraphics[scale=0.4]{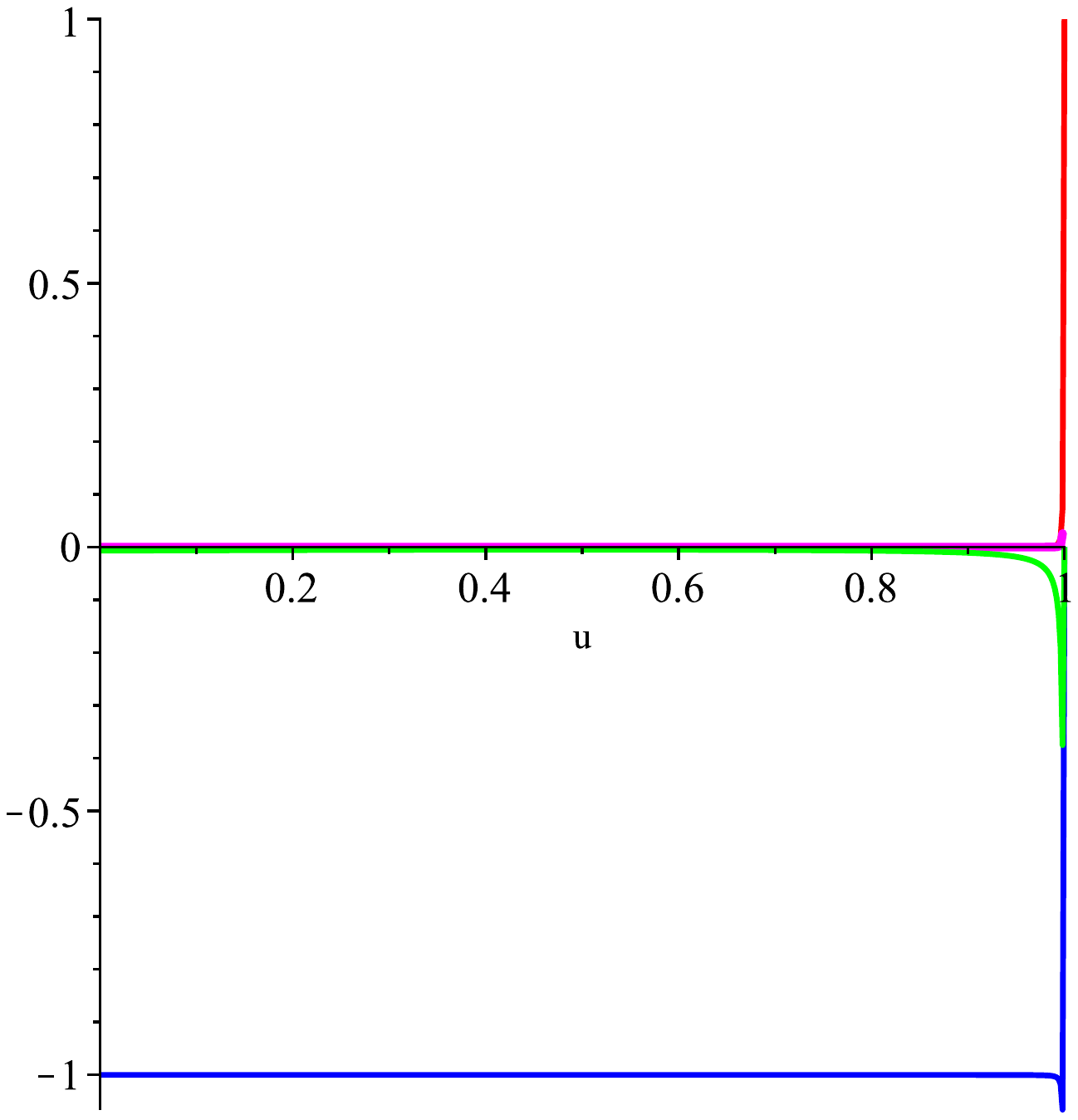}
\vspace{-5.6cm}
\subcaption{$\scriptstyle \wo =0.003$}
\end{subfigure}
\vspace{-4.8cm}
\caption{\scriptsize The conductivities as a function of $u$ for $q=m=1$ and various values of $\wo$.
The real part of the Ohmic conductivity is red,
the imaginary part is green. The real part of the Hall conductivity is blue
and the negative of the imaginary part is purple.
The approach to the analytic form (\ref{eq:omega->infinity-conductivities}) when
$\wo$ is large is evident in (a) and to the form
(\ref{eq:omega=0-conductivities}) as $\wo\rightarrow 0$ is evident in (h).}
\label{fig:1-1-omega}
\end{figure}

A general numerical analysis is rather involved as we are studying
four functions, $Re\bigl( \sigma^{x x}(u)\bigr)$, $Im\bigl( \sigma^{x x}(u)\bigr)$, $Re\bigl( \sigma^{x y}(u)\bigr)$ and $Im\bigl( \sigma^{x y}(u)\bigr)$,
in a 3-dimensional parameter space $(\wo,q,m)$.
Since $u\rightarrow 1$ is fixed by regularity conditions we shall simplify the analysis by focusing on the UV limit, $u \rightarrow 0$, and consider  $Re\bigl( \sigma^{x x}(\wo)\bigr)$, $Im\bigl( \sigma^{x x}(\wo)\bigr)$, $Re\bigl( \sigma^{x y}(\wo)\bigr)$ and $Im\bigl( \sigma^{x y}(\wo)\bigr)$ in a 2-parameter space  $(q,m)$.

The $u\rightarrow 0$ conductivity in zero magnetic field was studied in the
context of a holographic superconductor in \cite{Hartnoll+Herzog+Horowitz}.
In the presence of a magnetic field  a pole in $\sigma_+(\wo)$ in the complex $\wo$ plane was found in \cite{Hartnoll+Herzog}, associated with a cyclotron resonance.  Near such a pole with resonance frequency $\wo_0$,
width $\Gamma$ and
resonance value $\sigma_0$ the conductivity is of the form (\ref{eq:pole-form-of-sigma}),
\beq \sigma(\wo)= \sigma_0 \Gamma 
\left[ \frac{\Gamma+i(\wo - \wo_0)}{(\wo - \wo_0)^2 + \Gamma^2}\right]
=\frac{i \sigma_0 \Gamma}{\wo -(\wo_0 -i \Gamma)}.\label{eq:sigma-Drude}
\eeq 
Normally $Re\bigl(\sigma(\wo)\bigr)$ is the dissipative conductivity, with a peak at $\wo_0$, while $Im\bigl(\sigma(\wo)\bigr)$ is the refractive conductivity, with a zero at $\wo_0$ when $\sigma_0$ is real. That interpretation is not valid for $\sigma_+$ as its real and imaginary parts contain
information about both the Ohmic and the Hall conductivities and these must be extracted separately.
For the Ohmic conductivity, assuming the resonance form (\ref{eq:sigma-Drude}) with $\sigma_0$ real and positive, $Im\bigl(\sigma^{x x}(\wo)\bigr)$ vanishes at the same frequency as that for which  $Re\bigl(\sigma^{x x}(\wo)\bigr)$ is maximised.
For the Hall conductivity it is the opposite way round, the peak is in 
$Im\bigl(\sigma^{x y}(\wo)\bigr)$ and the zero is in $Re\bigl(\sigma^{x y}(\wo)\bigr)$.  As noted above when $q$ and $m$ are both positive the Hall maximum peak is in
$-Im\bigl(\sigma^{x y}(\wo)\bigr)$ rather than $Im\bigl(\sigma^{x y}(\wo)\bigr)$, but the
sign could be changed by changing the relative sign of $q$ and $m$.

The three parameters $\sigma_0$, $\wo_0$ and $\Gamma$ will in general
depend on $q$ and $m$ and the numerical analysis below shows that,
at fixed $q$, they differ significantly as functions of $m$
for the Ohmic and Hall conductivities when $q<1$
but are remarkably similar when $q>1$.

The numerical results are displayed in a number of figures and we first summarise the figures before describing them in more detail.

\bigskip

{\it Summary of figures:}
\begin{itemize}
\item Fig.\,\ref{fig:q=1,m=0.5}: conductivities for $q=1$, $m=0.5$.
\item Figs.\,\ref{fig:q=1,various-m}, \ref{fig:q=0.5,various-m} and \ref{fig:q=0,various-m}: cyclotron resonances for various $m$ at $q=1$, $q=0.5$ and $q=0$.
\item Figs.\,\ref{fig:omega_0-versus-m} and \ref{fig:peaks}: cyclotron frequencies and conductivity maxima at fixed $q$, as a function of $m$. 
\item Fig.\,\ref{fig:Gamma}: the inverse resonance widths at fixed $q$, as a function of $m$.
\item Fig.\,\ref{fig:Q-h}: the ${\cal Q}$ factor for the cyclotron resonances at fixed $q$ as a function of $m$.
\item Fig.\,\ref{fig:small-qm} and Fig.\,\ref{fig:parametric}: further details of the resonances for a small value of $q$, ($q=0.1)$.
\item Fig.\,\ref{fig:large-q-small-m}: further details of the resonances for
  a large value of $q$, ($q=1.5).$
\item Fig.\,\ref{fig:Gammasigma0}: the residues at $q=0.5$, $q=1$ and $q=1.5$ as functions of $m$.
\item Fig.\,\ref{fig:QT}: some resonance properties as a function of temperature.
\item Fig.\,\ref{fig:T=0}: zero temperature conductivities and ${\cal Q}$-factors as functions of the filling factor $\nu$.
\end{itemize}

\noindent These figures are now described  in more detail.

\subsubsection{Conductivities for $q=1$, $m=0.5$.}

The asymptotic, $u=0$, conductivities are plotted in Fig.\,\ref{fig:q=1,m=0.5} 
for $q=1$ and $m=0.5$ as a function of $\wo$ (similar plots appeared in \cite{LPTV}).  
The resonance found numerically in \cite{Hartnoll+Herzog} is clearly visible.
For these values of $q$ and $m$ the resonances for $\sigma^{x x}$ and $\sigma^{x y}$ are very similar, and the plots only differ significantly near $\wo=0$ (though we
know from (\ref{eq:omega->infinity-conductivities}) that they also differ for large $\wo$, but this
is not clearly visible on the scale in the plots).
\begin{figure}
\begin{subfigure}[t]{0.45\textwidth}
\hspace{-2cm}\includegraphics[scale=0.5]{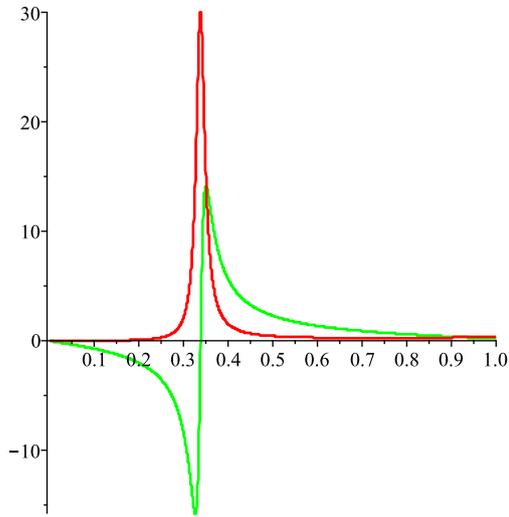}
\vspace{-6cm}
\subcaption{\scriptsize The real (red) and imaginary (green) Ohmic conductivities as a function of frequency.}
\end{subfigure}
\qquad
\begin{subfigure}[t]{0.45\textwidth}
\hspace{-1.8cm}\includegraphics[scale=0.5]{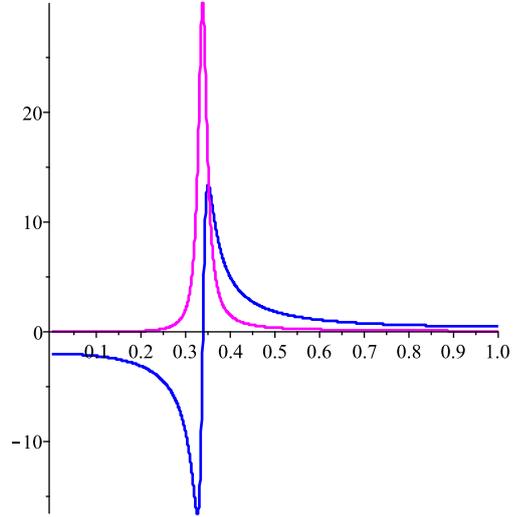}
\vspace{-6cm}
\subcaption{\scriptsize The real (blue) and minus the imaginary (magenta) Hall conductivities as a function of frequency.}
\end{subfigure}
\vspace{-4.5cm}

\centering
\begin{subfigure}[t]{1\textwidth}
\centering
{\includegraphics[scale=0.5]{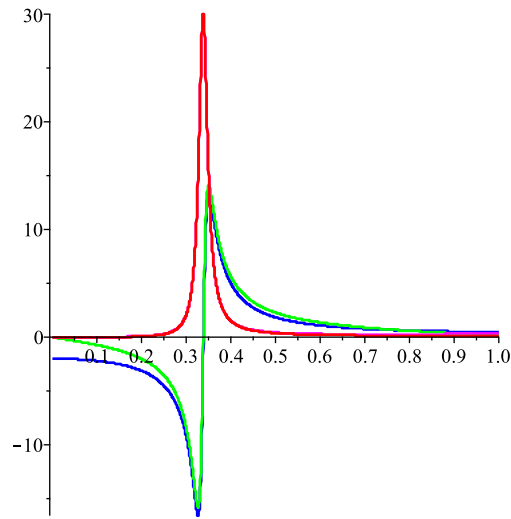}}
\vspace{-5.5cm}
  \subcaption{\scriptsize Ohmic and Hall
  conductivities overlaid as a function of frequency.}
\end{subfigure}
\vspace{-5cm}
\caption{\scriptsize Cyclotron resonance of the Ohmic conductivity (a), the Hall
conductivity (b) and both overlaid (c) as a function of $\scriptstyle \wo$, for $q=1$ and $m=0.5$.} \label{fig:q=1,m=0.5}
%\vspace{-0.9cm}
\end{figure}

\subsubsection{Cyclotron resonances for various $m$ at $q=1$, $q=0.5$ and $q=0$.}
In Figs.\,\ref{fig:q=1,various-m}, \ref{fig:q=0.5,various-m} and \ref{fig:q=0,various-m} a number of different resonances are overlaid for different $m$ at fixed $q$: $q=1$, $q=0.5$ and $q=0$
in Figs.\,\ref{fig:q=1,various-m}, \ref{fig:q=0.5,various-m} and \ref{fig:q=0,various-m} respectively. The resonance frequencies increase as $m$ increases and the
resonance peaks are strongest for larger $q$ and smaller $m$. $\sigma^{xy}$ vanishes when $q=0$ because in that case
$\sigma_+(m)=\sigma_+(-m)$.
\begin{figure}
\centering
\begin{subfigure}[t]{0.45\textwidth}
\hspace{-1.8cm}\includegraphics[scale=0.5]{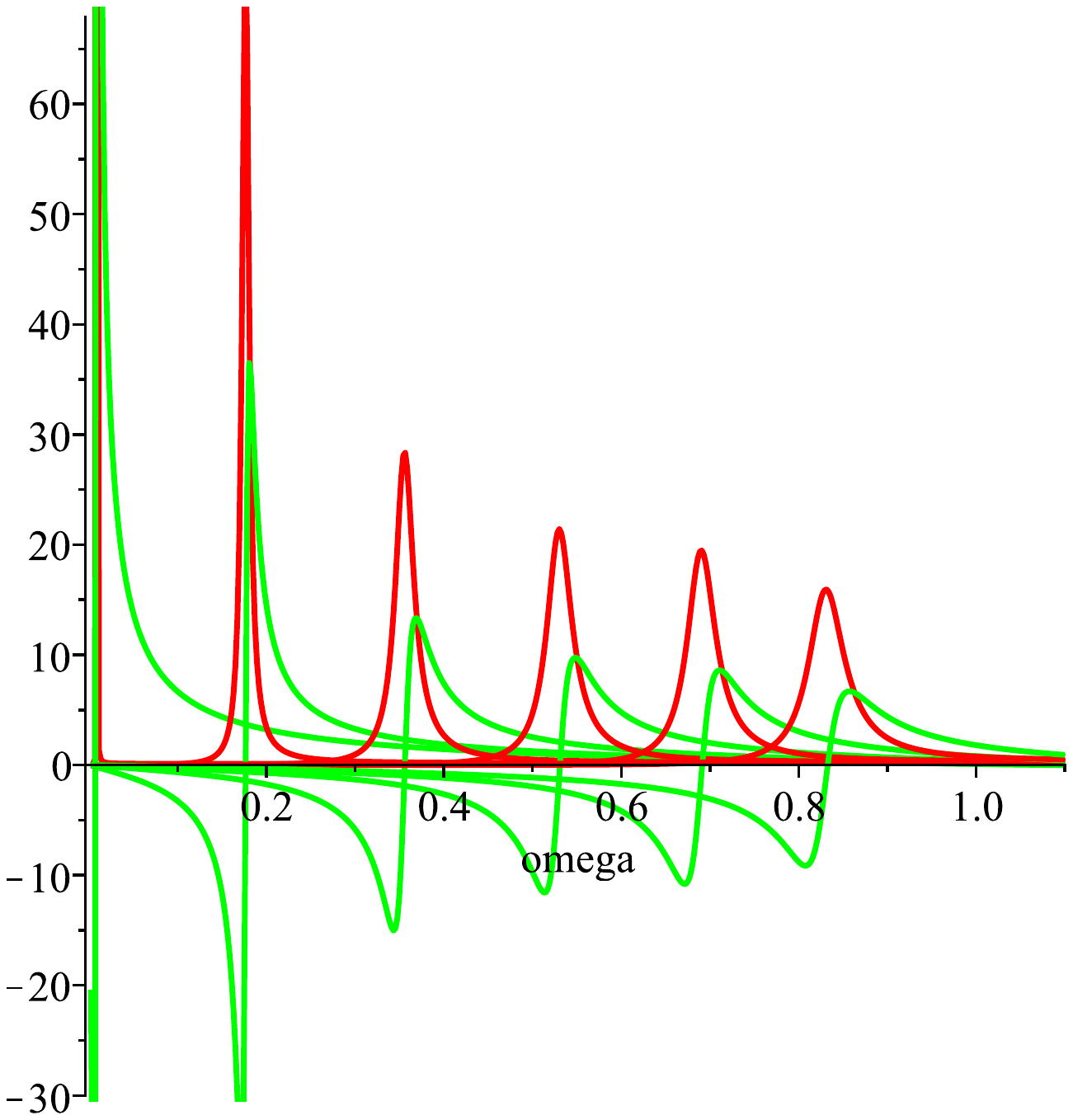}
\vspace{-6cm}
\subcaption{\scriptsize The real (red) and imaginary (green) Ohmic conductivities as a function of frequency for $\scriptstyle q=1$ and different $\scriptstyle m$.}
\end{subfigure}
\qquad
\begin{subfigure}[t]{0.45\textwidth}
\hspace{-1.5cm}\includegraphics[scale=0.5]{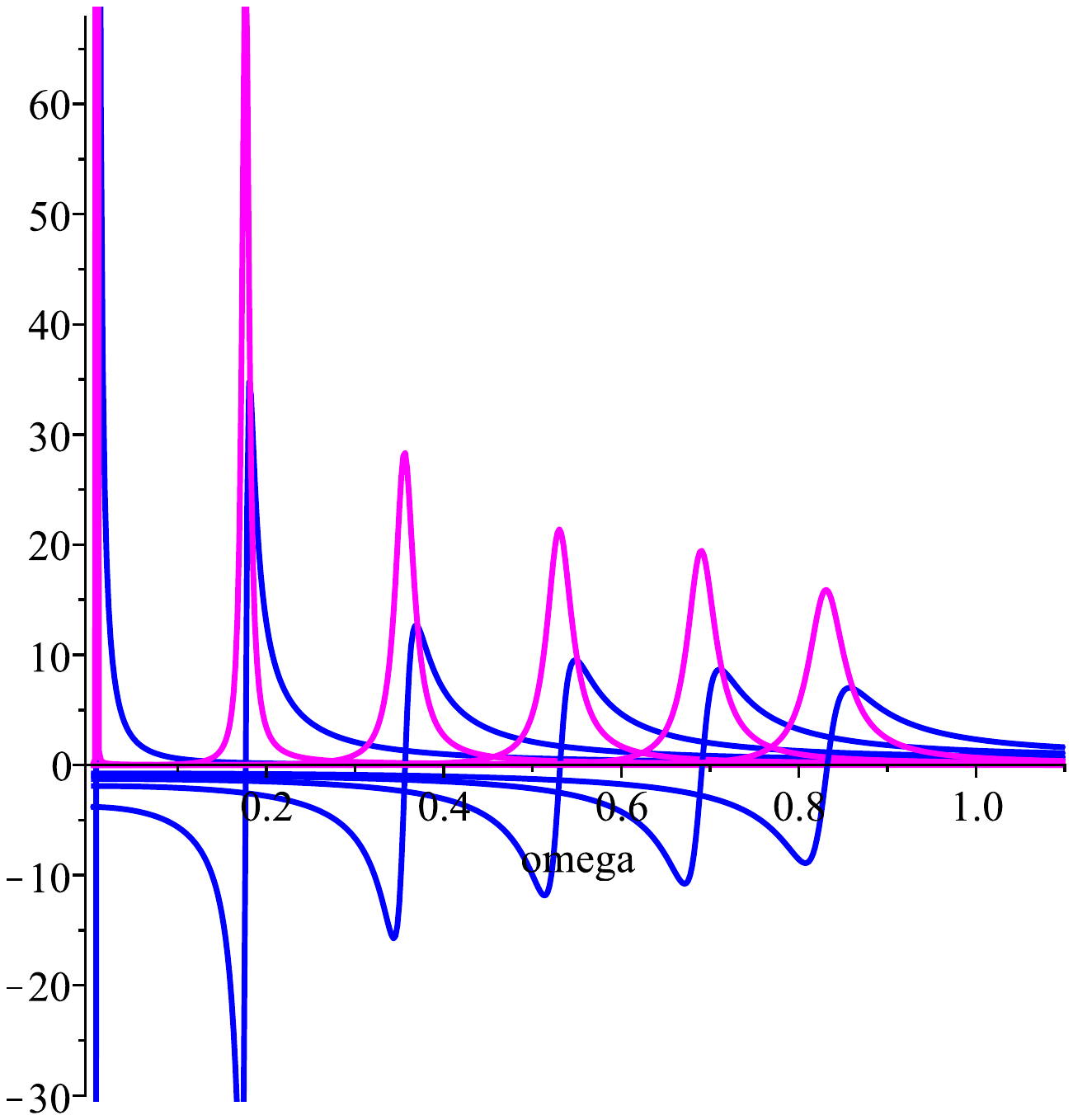}
\vspace{-6cm}
\subcaption{\scriptsize The real (blue) and imaginary (magenta) Hall conductivities as a function of frequency at fixed $\scriptstyle q=1$ and different $\scriptstyle m$.}
\end{subfigure}
\vspace{-4cm}
\caption{\scriptsize The Ohmic and Hall conductivities as a function of frequency 
for $q=1$ and a number of different $m$. The resonance frequencies increase as $m$ increases.}
\label{fig:q=1,various-m}
\end{figure}

\begin{figure}
%\vspace{-1cm}
\centering
\begin{subfigure}[t]{0.45\textwidth}
\hspace{-1.8cm}\includegraphics[scale=0.5]{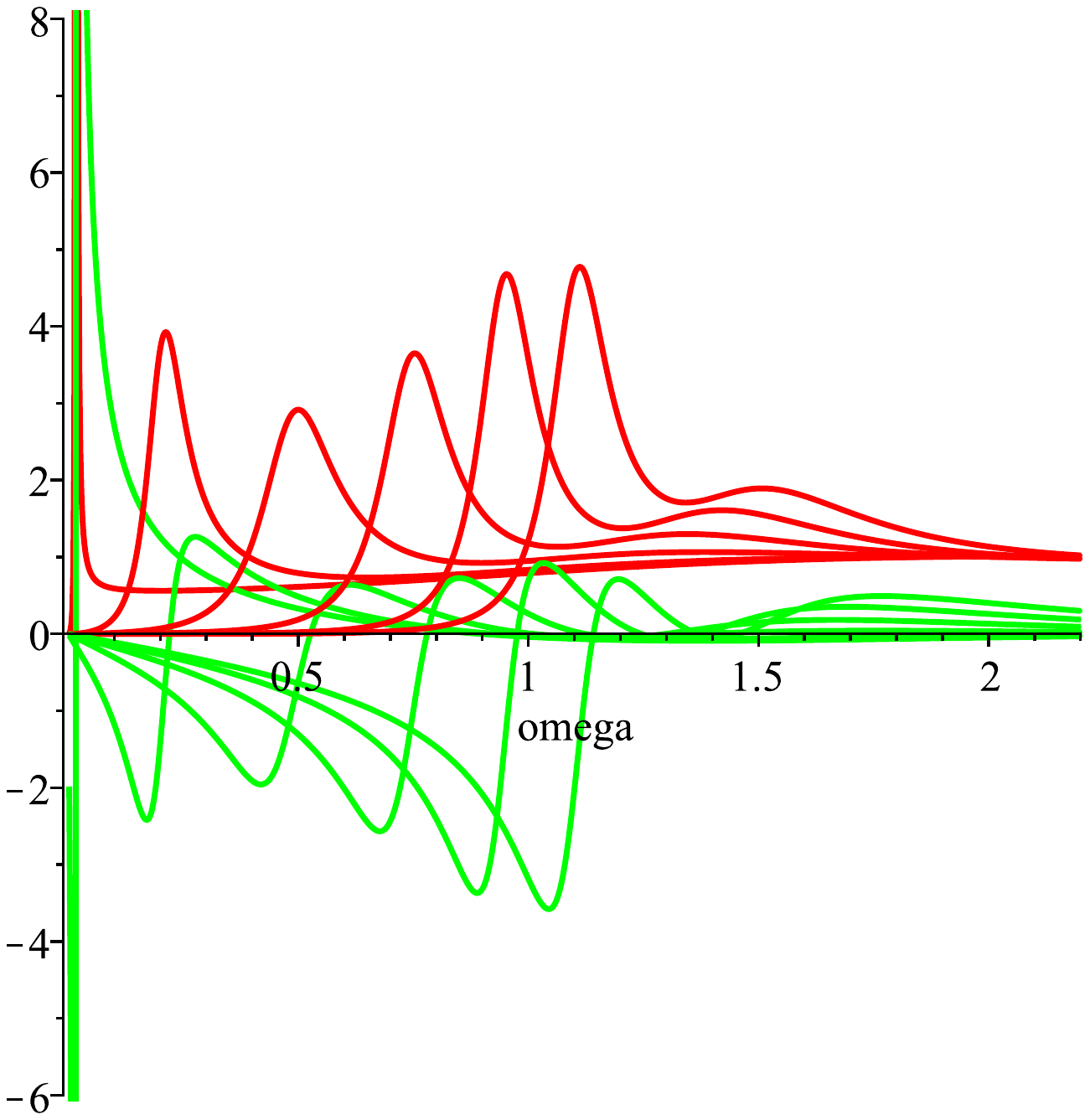}
\vspace{-6cm}
\subcaption{\scriptsize The real (red) and imaginary (green) Ohmic conductivities as a function of frequency for $\scriptstyle q=0.5$ and various $\scriptstyle m$.}
\end{subfigure}
\qquad
\begin{subfigure}[t]{0.45\textwidth}
\hspace{-1.5cm}\includegraphics[scale=0.5]{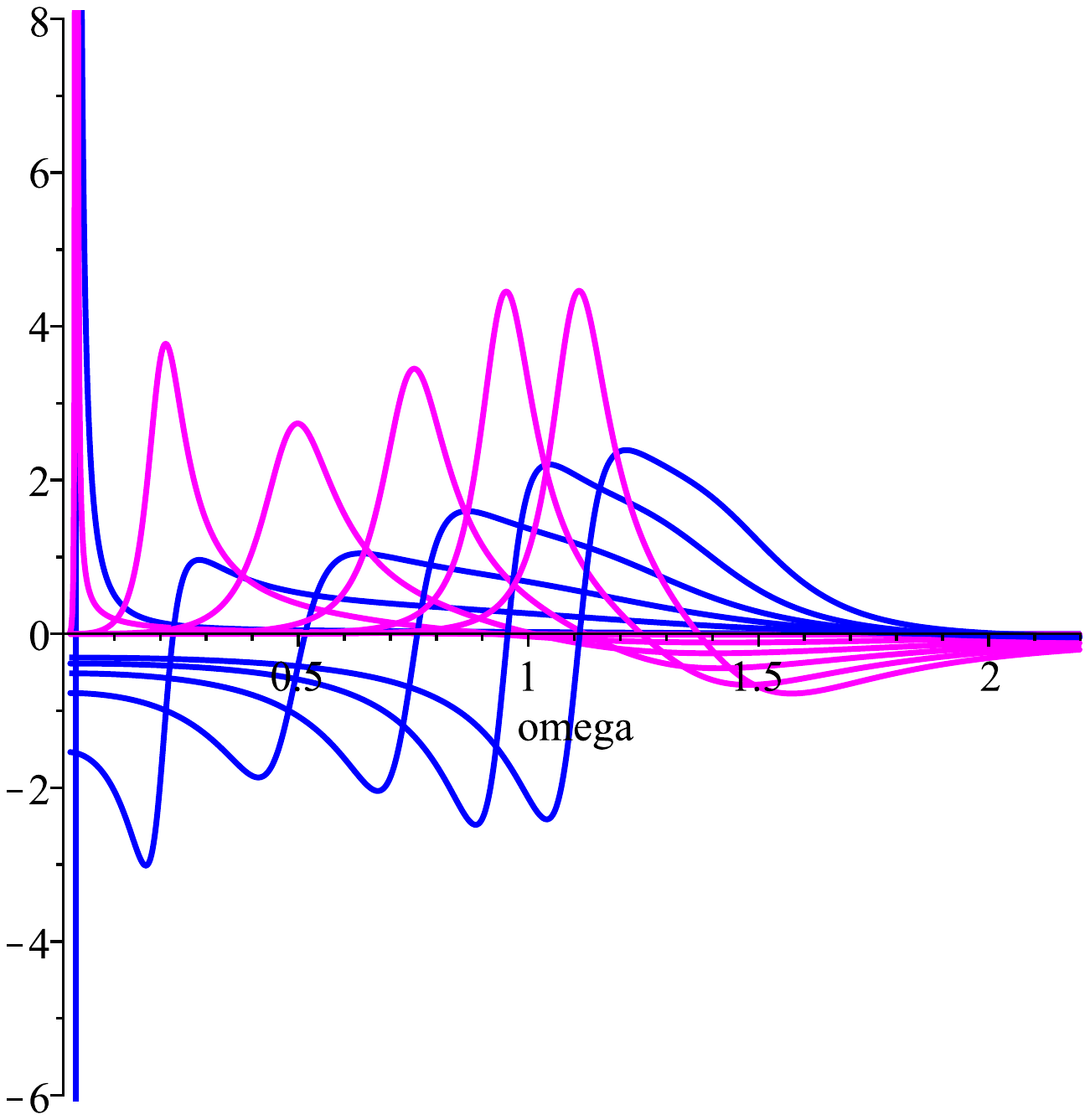}
\vspace{-6cm}
\subcaption{\scriptsize The real (blue) and minus the imaginary (magenta) Hall conductivities as a function of frequency for $\scriptstyle q=0.5$ and various $\scriptstyle m$.}
\end{subfigure}
\vspace{-4cm}
\caption{\scriptsize The Ohmic and Hall conductivities as a function of frequency 
for $q=0.5$ and a number of different $m$.}
\label{fig:q=0.5,various-m}
\end{figure}

\begin{figure}
  \centering
  \includegraphics[scale=0.5]{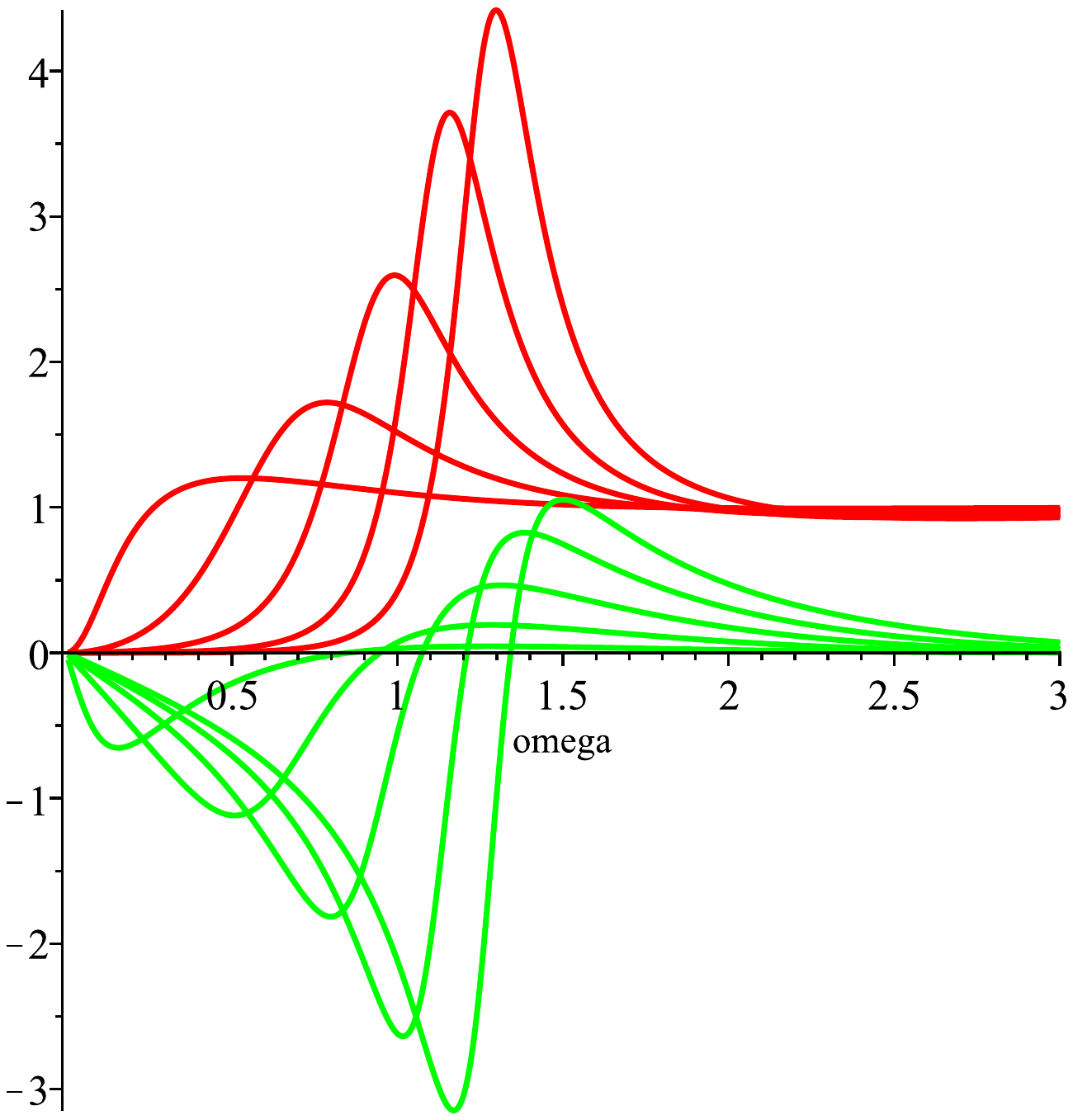}
%data/q=0-v3.7/
\vspace{-6cm}
\caption{\scriptsize The real (red) and imaginary (green) Ohmic conductivities as a function of frequency for $q=0$ and different $m$ (the Hall conductivity vanishes when $q=0$). As $m$ is reduced the resonance peak disappears.}
\label{fig:q=0,various-m}
\end{figure}

\subsubsection{Cyclotron frequencies and conductivity maxima at fixed $q$ as  a function of $m$.\label{sec:cyclotron_frequency}}
The resonance frequencies and resonance peaks of the conductivities
are shown in Figs.\,\ref{fig:omega_0-versus-m} and \ref{fig:peaks}, as a function of $m$ at 
$q=0.1$, $0.5$, $1$ and $1.5$.
For a sharp resonance it can be 
difficult to obtain an accurate value for the maximum of the conductivity numerically, 
as to do so requires a very fine discretisation of the frequency.
In that case we use the fit (\ref{eq:sigma-Drude}) for a well-defined resonance,
and assume that, for the Ohmic conductivity, the peak in $Re\bigl(\sigma^{x x}(\wo)\bigr)$
occurs at the same frequency, $\wo_0$, as that at which $Im\bigl\{\bigl(\sigma^{x x}(\wo)\bigl)^{-1}\bigr\}$ vanishes
(and vice versa for the Hall conductivity)
while at the same time $\sigma_0 = \Bigl\{Re\bigl((\sigma^{x x}(\wo_0))^{-1}\bigr)\Bigr\}^{-1}$.
For a sharp peak it is much easier numerically to determine the zero of  $Im\bigl\{\bigl(\sigma^{x x}(\wo)\bigr)^{-1}\bigr\}$ than it is to determine the maximum of $Re\bigl(\sigma^{x x}(\wo)\bigr)$ and for this reason
when the ${\cal Q}$-factor is greater than 10 ({\it i.e.} $q=1$ and $1.5$ in the plots)
the zero of $Im\bigl\{\bigl(\sigma^{x x}(\wo)\bigr)\bigr\}$ is used to define
$\wo_0$ and  $\sigma_0^{x x}=\Bigl\{Re\bigl(\sigma^{xx}(\wo_0)\bigr)^{-1}\Bigr\}$  then gives
$Re(\sigma^{x x})_{Max}$.
If the ${\cal Q}$-factor is less than 10 (\ref{eq:sigma-Drude}) is not a good fit,
but the peak is broad so it easier to ascertain the height and
resonance frequency just by examining $Re\bigl\{\bigl(\sigma^{x x}(\wo)\bigr)\bigr\}$ directly and determining where its maximum lies, and this is the method used for analysing for
$q=0.1$ and $0.5$.

It is somewhat surprising that the analytic approximations (\ref{eq:Pade-sigma-xx}) is such a good fit for the Ohmic conductivity in Fig.\,\ref{fig:peaks}(a) ($q=0.1$) for the full range of $m$, as
the corresponding fit to the resonance
frequency (\ref{eq:omega_0-xx-approx})  (Fig.\,\ref
{fig:omega_0-versus-m}(a)) diverges when
$m= \sqrt{3}\, q= 0.173$ (red dotted line) and is imaginary for $m> \sqrt{3}$.
For larger values of $q$ the Ohmic and Hall conductivities are indistinguishable at resonance.
\begin{figure}
\centering
\begin{subfigure}[t]{0.45\textwidth}
\hspace{-2.5cm}\includegraphics[scale=0.5]{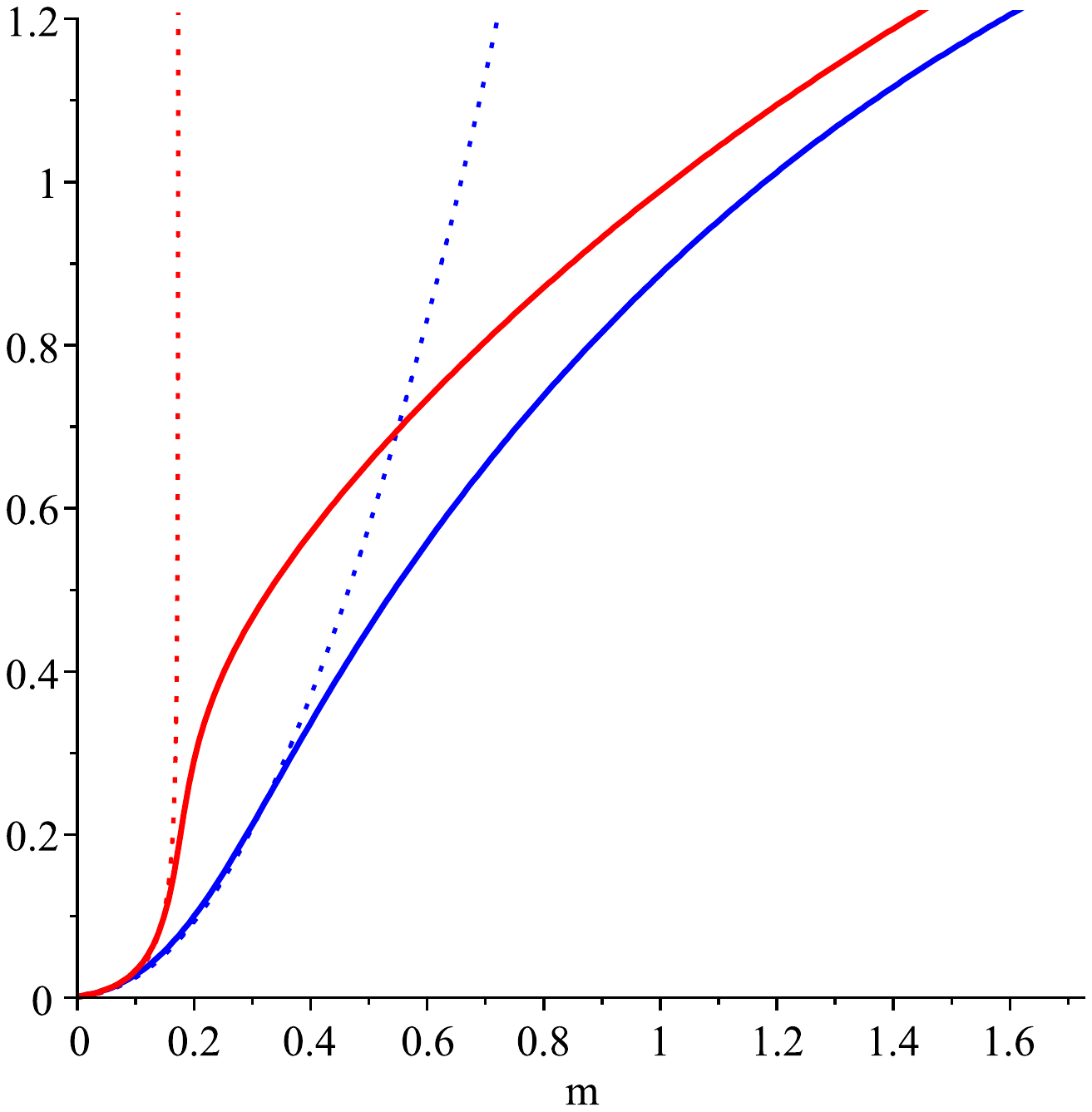} 
% AdS-CFT-sigma-equation-v4.6: data v4.40: OmegaRange=1.6,  NumB=200, nrange=2000;
\vspace{-6.5cm}
\subcaption{$\scriptstyle q=0.1$}
\end{subfigure}
\centering
\begin{subfigure}[t]{0.45\textwidth}
\hspace{-1.5cm}\includegraphics[scale=0.5]{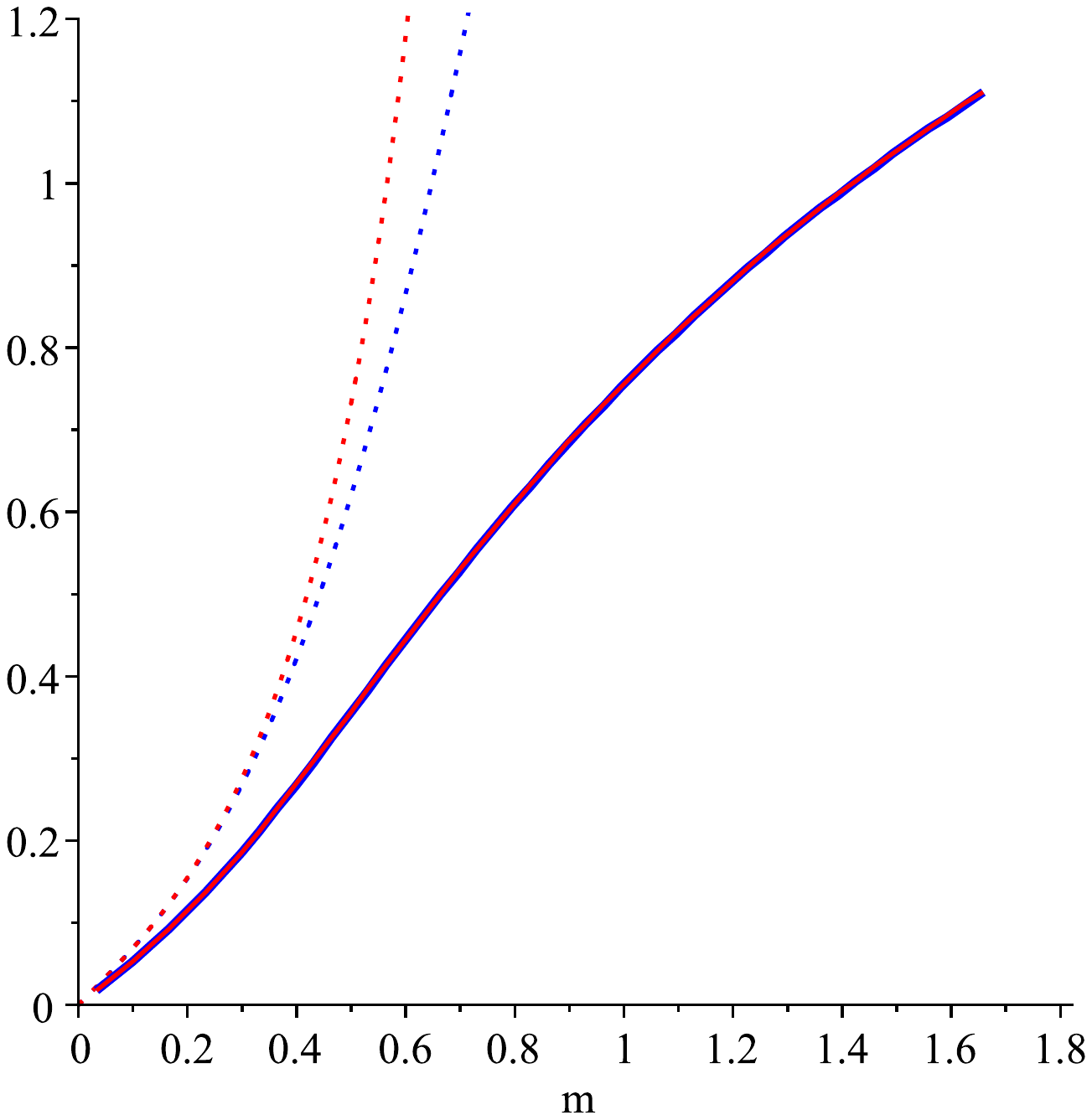}
%q=.5-v4.1-NumB=50-Blow=.33e-1-Bhigh=1.658-omega_low=0-omega_high=2.2-nrange=1000
\vspace{-6.5cm}
\subcaption{$\scriptstyle q=0.5$}
\end{subfigure}

\vspace{-5.5cm}

\centering
\begin{subfigure}[t]{0.45\textwidth}
\hspace{-2.5cm}\includegraphics[scale=0.5]{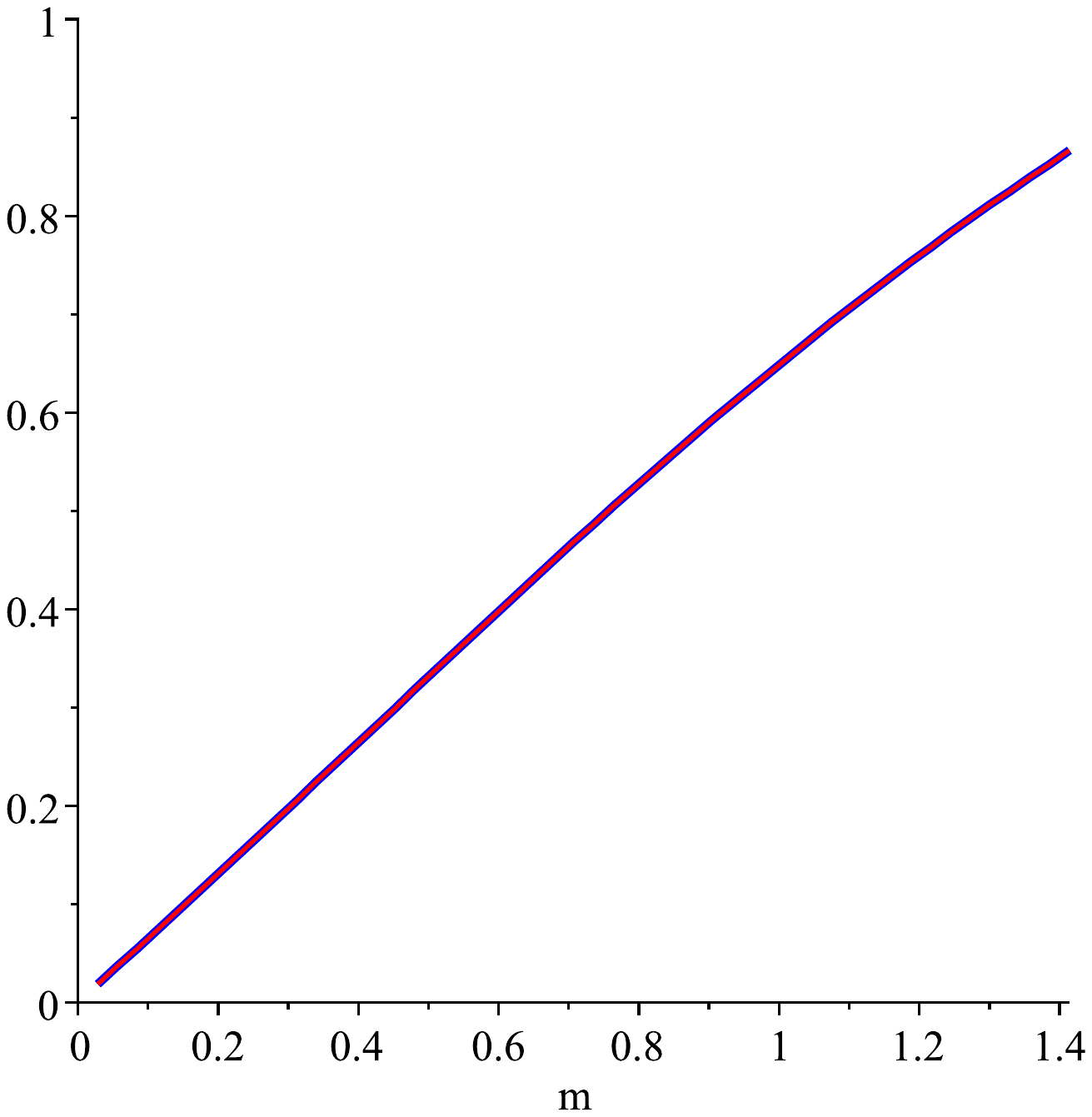}
% AdS-CFT-sigma-equation-v4.1: data v4.1 OmegaRange 1.1, NumB=50, nrange=1000; 
\vspace{-6.5cm}
\subcaption{$\scriptstyle q=1$}
\end{subfigure}
\vspace{-6cm}
\centering
\begin{subfigure}[t]{0.45\textwidth}
\hspace{-1.5cm}\includegraphics[scale=0.5]{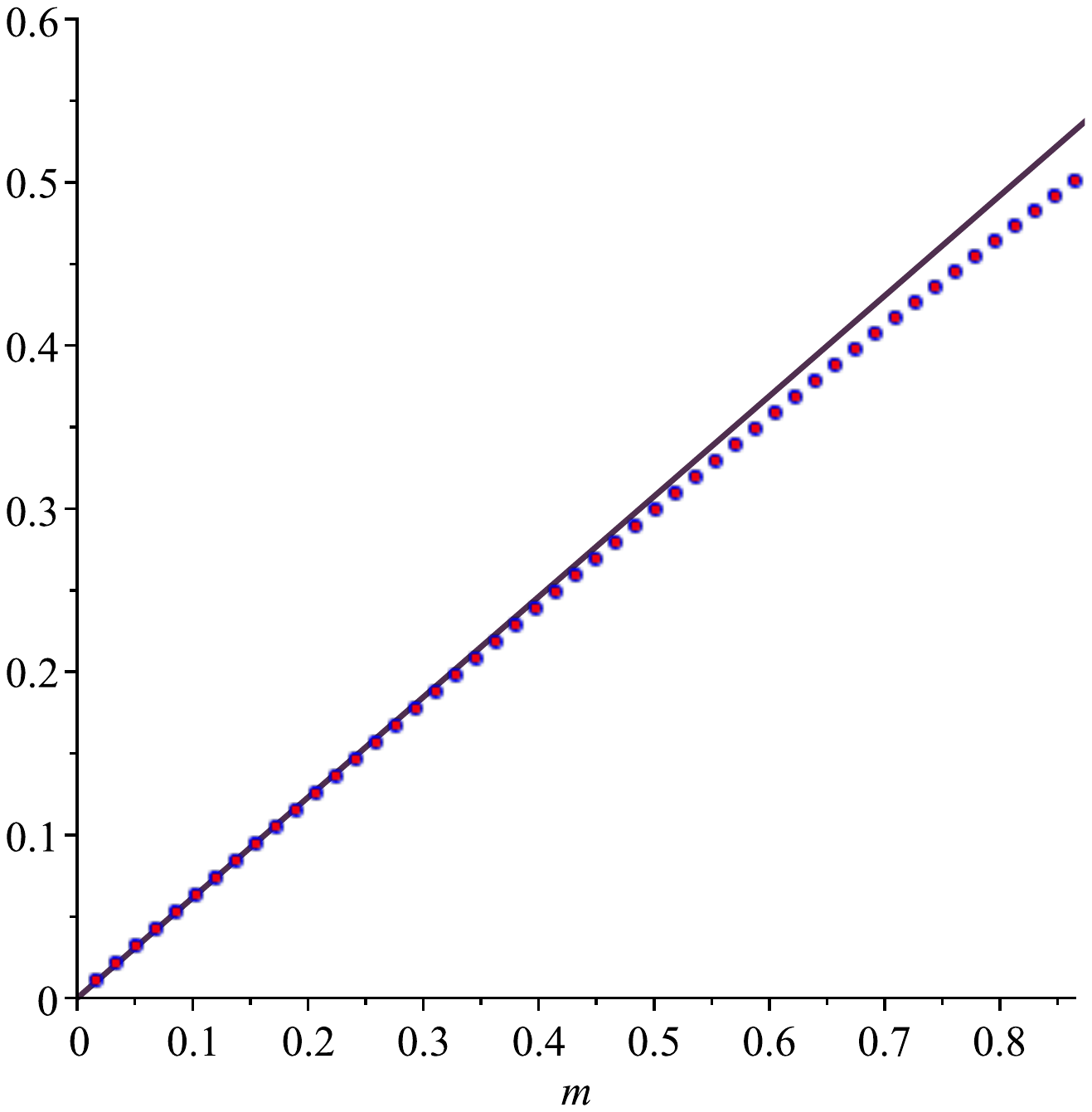}
%q=1.5-v4.1-NumB=50-Blow=.17e-1-Bhigh=.866-omega_low=0-omega_high=.6-nrange=1000
\vspace{-6.5cm}
\subcaption{\scriptsize $\scriptstyle q=1.5$ }
\end{subfigure}
\vspace{1cm}
\caption{\scriptsize The resonance frequency $\wo_0$ as a function of $m$.
$\wo_0$ for both the Ohmic conductivity (red) and the Hall conductivity (blue)
at fixed $q$ are superimposed (in (b), (c) and (d) they are indistinguishable on the scales shown).
As $q$ increases the frequency gets closer to the cyclotron form of exhibiting linearity in $m$.  The analytic approximations (\ref{eq:Pade-sigma-xx}) and (\ref{eq:Pade-sigma-xy}) are
  shown as dotted lines in (a) and (b). In (d) the numerical data are shown as dots and the linear fit $0.41 q m$ is shown in grey. A more detailed version of (a) at small $m$ is shown in Fig.\,\ref{fig:small-qm}(a)
and a more detailed version of (d) for small $m$ is shown in Fig.\,\ref{fig:large-q-small-m}(a).}
\label{fig:omega_0-versus-m}
\vspace{-0.3cm}
\end{figure}

\begin{figure}
%  \vspace{-1cm}
\centering
\begin{subfigure}[t]{0.45\textwidth}
\hspace{-2.5cm}\includegraphics[scale=0.5]{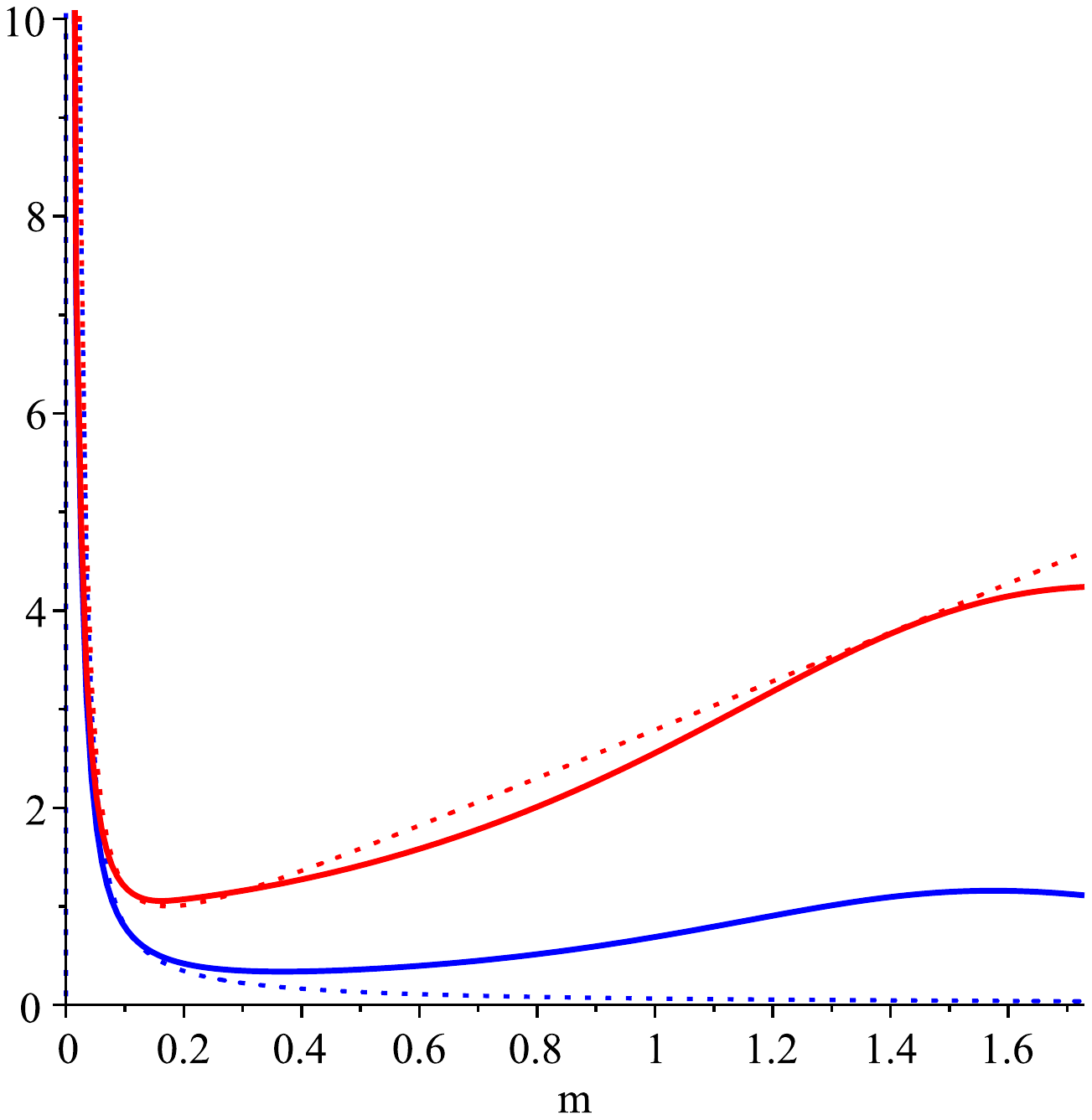} 
% AdS-CFT-sigma-equation-v4.6: "data/q=.1-v4.40-NumB=200-Bhigh=1.73-OmegaRange=1.6-nrange=2000"
\vspace{-6.5cm}
\subcaption{$\scriptstyle q=0.1$}
\end{subfigure}
\centering
\begin{subfigure}[t]{0.45\textwidth}
\hspace{-1cm}\includegraphics[scale=0.5]{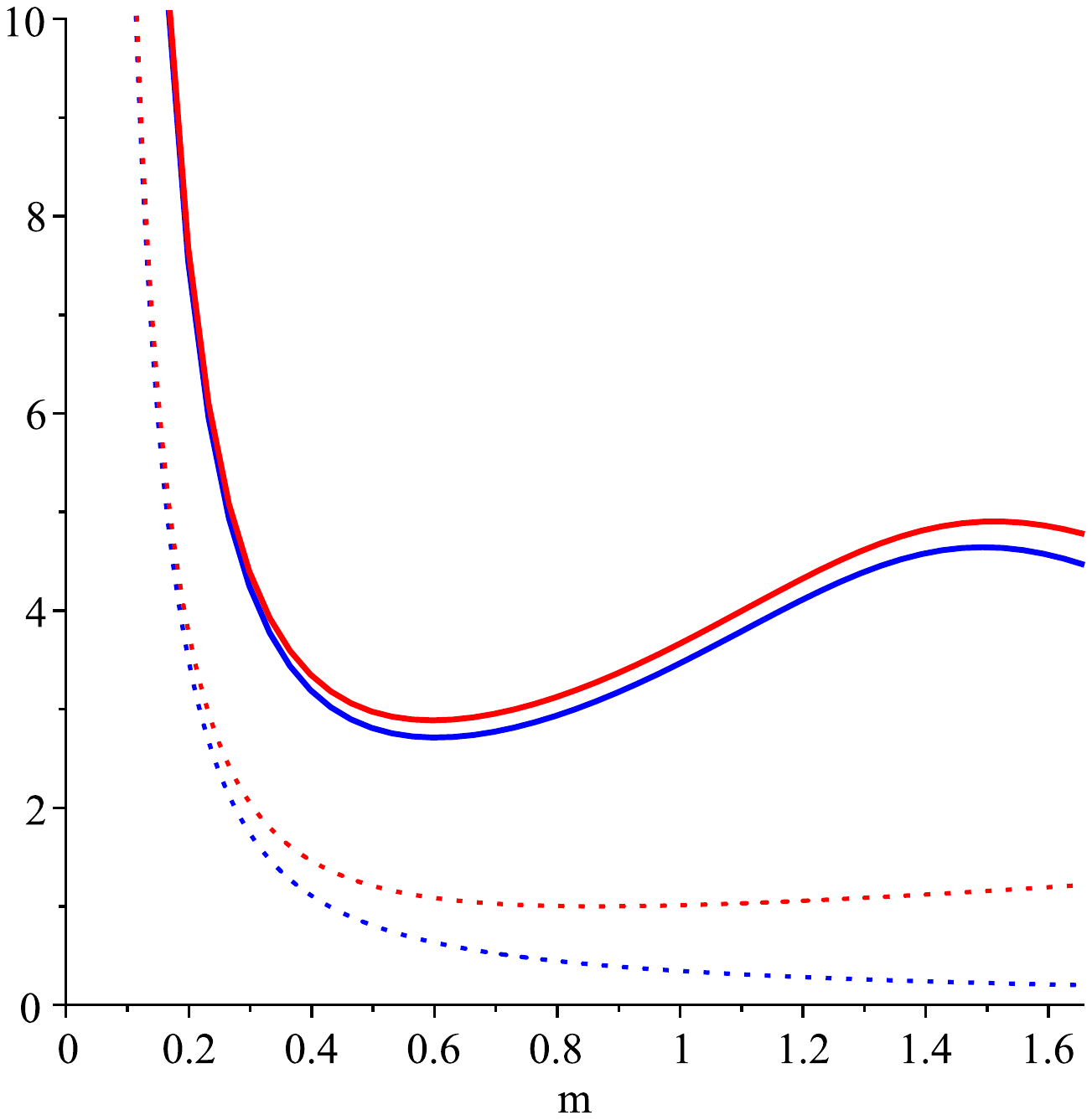}
%q=.5-v4.1-NumB=50-Blow=.33e-1-Bhigh=1.658-omega_low=0-omega_high=2.2-nrange=1000
\vspace{-6.5cm}
\subcaption{\scriptsize $\scriptstyle q=0.5$}
\end{subfigure}

\vspace{-5cm}

\begin{subfigure}[t]{0.5\textwidth}
\hspace{-2cm}\includegraphics[scale=0.5]{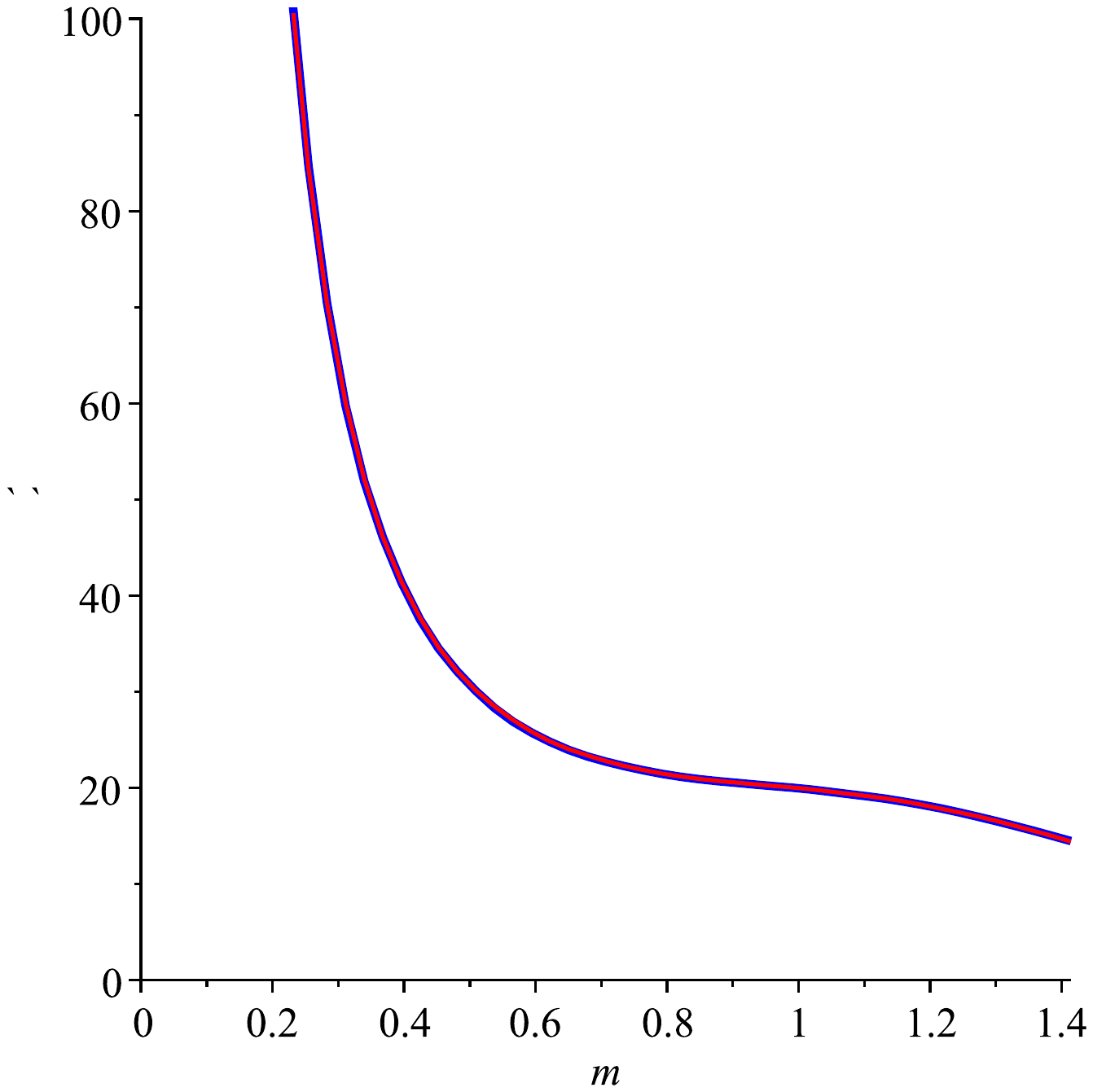}
% AdS-CFT-sigma-equation-v4.6: "data/q=1-v4.1/ OmegaRange=1.1,  NumB=50, nrange=1000, Bhigh=sqrt(3-q^2);
\vspace{-6.5cm}
\subcaption{$\scriptstyle q=1$}
\end{subfigure}
\begin{subfigure}[t]{0.45\textwidth}
\hspace{-1.5cm}\includegraphics[scale=0.5]{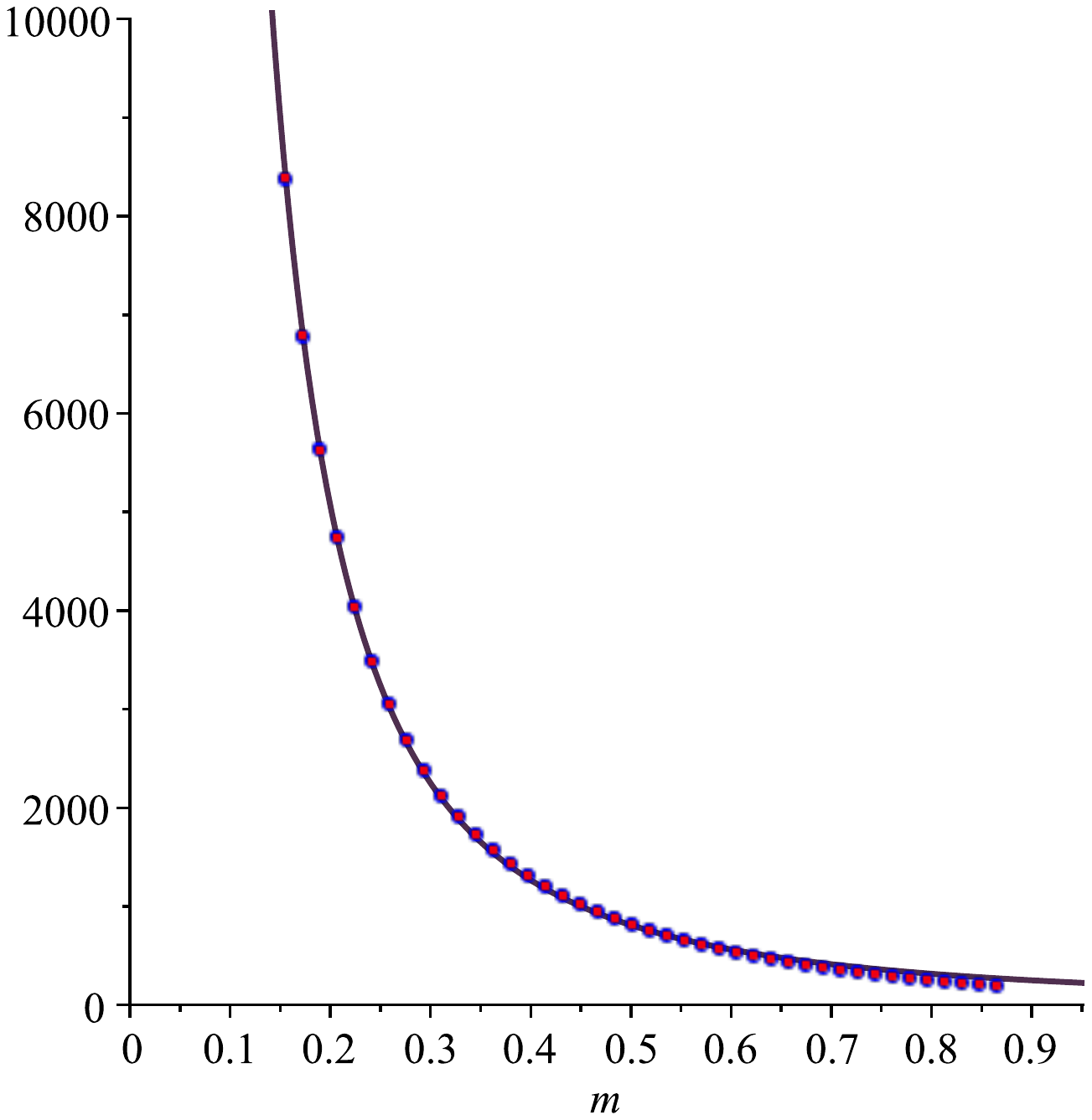}
%q=1.5-v4.1-NumB=50-Blow=.17e-1-Bhigh=.866-omega_low=0-omega_high=.6-nrange=1000
\vspace{-6.5cm}
\subcaption{\scriptsize $\scriptstyle  q=1.5$ }
\end{subfigure}
\vspace{-5cm}
\caption{\scriptsize The peaks for the resonances of the Ohmic conductivity (red) and Hall conductivity (blue) at fixed $q$ as a function of $m$. For clarity the data are represented as continuous lines in figures (a), (b) and (c) and as dots in figure (d).
  The analytic approximations (\ref{eq:Pade-sigma-xx}) and (\ref{eq:Pade-sigma-xy}) are
  shown as dotted lines in (a) and (b).  For $q=1.5$ the numerical data are
  almost indistinguishable on this scale from the analytic fit
  $\scriptstyle 90\frac{q^2}{m^2}$, shown in grey, except at the larger values of $m$.
A more detailed version of (a) at small $m$ is shown in Fig.\,\ref{fig:small-qm}(b)
and a more detailed version of (d) for small $m$ is shown in Fig.\,\ref{fig:large-q-small-m}(b).}
\label{fig:peaks}
\end{figure}

\subsubsection{The resonance widths at fixed $q$ as a function of $m$.}
In Fig.\,\ref{fig:Gamma} the  inverse widths $\Gamma^{-1}$ of the resonances at fixed $q$ as a function of $m$ are shown for a selection of values for $q$.  The widths are extracted from the numerical solutions by determining  the resonance frequencies and peak values
$\sigma_0^{x x}=Re(\bigl(\sigma^{x x}(\wo_0)\bigr)$
and $\sigma_0^{x y} = \Bigl|Im\bigl(\sigma^{x y}(\wo_0)\bigr)\Bigr|$ as described in
{\bf \ref{sec:cyclotron_frequency}} above
and assuming the forms 
\begin{align*}
 \bigl(\sigma^{x x}(\wo)\bigr)^{-1} & = 
\frac{\Gamma_{x x} -i(\wo - \wo_0)}{\sigma_0^{x x} \Gamma_{x x}}
 \qquad \Rightarrow \qquad
\Gamma_{x x}^{-1} = -\sigma_0^{x x} Im \left[ \frac{d(\sigma^{x x})^{-1}}{d\wo}\right]_{\wo=\wo_0},\\
 \bigl(\sigma^{x y}(\wo)\bigr)^{-1} & = \frac{i\Gamma_{x y} + (\wo - \wo_0)}{\sigma_0^{x y} \Gamma_{x y}}
\qquad \Rightarrow  \qquad \Gamma_{x y}^{-1} = \sigma_0^{x y} Re\left[ \frac{d(\sigma^{x y})^{-1}}{d\wo}\right]_{\wo=\wo_0}.\\
\end{align*}
The widths  $\Gamma_{x x}$ and $\Gamma_{x y}$ are then obtained by differentiating the
numerical solutions.  Again the widths for $\sigma^{x x}$ and $\sigma^{x y}$ are
indistinguishable on this scale for $q=1$ and $q=1.5$.

\begin{figure}
%  \vspace{-1cm}
\begin{subfigure}[t]{0.45\textwidth}
\hspace{-2.5cm}\includegraphics[scale=0.5]{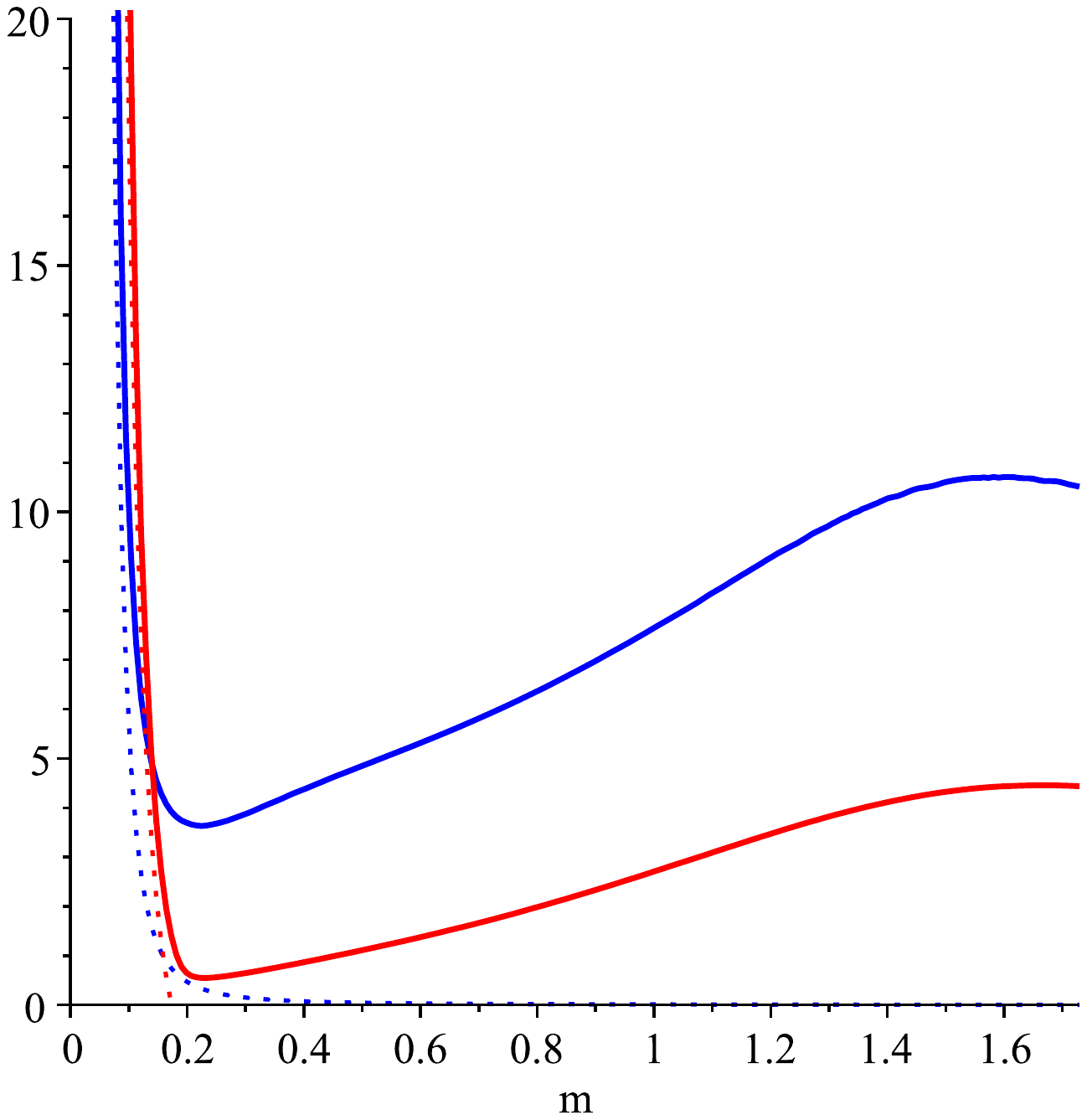}
% AdS-CFT-sigma-equation-v4.6.2: data/q=.1-v4.40-NumB=200-Bhigh=1.73-OmegaRange=1.6-nrange=2000
\vspace{-6.5cm}
\subcaption{\scriptsize $\scriptstyle q=0.1$}
\end{subfigure}
\begin{subfigure}[t]{0.45\textwidth}
\hspace{-1cm}\includegraphics[scale=0.5]{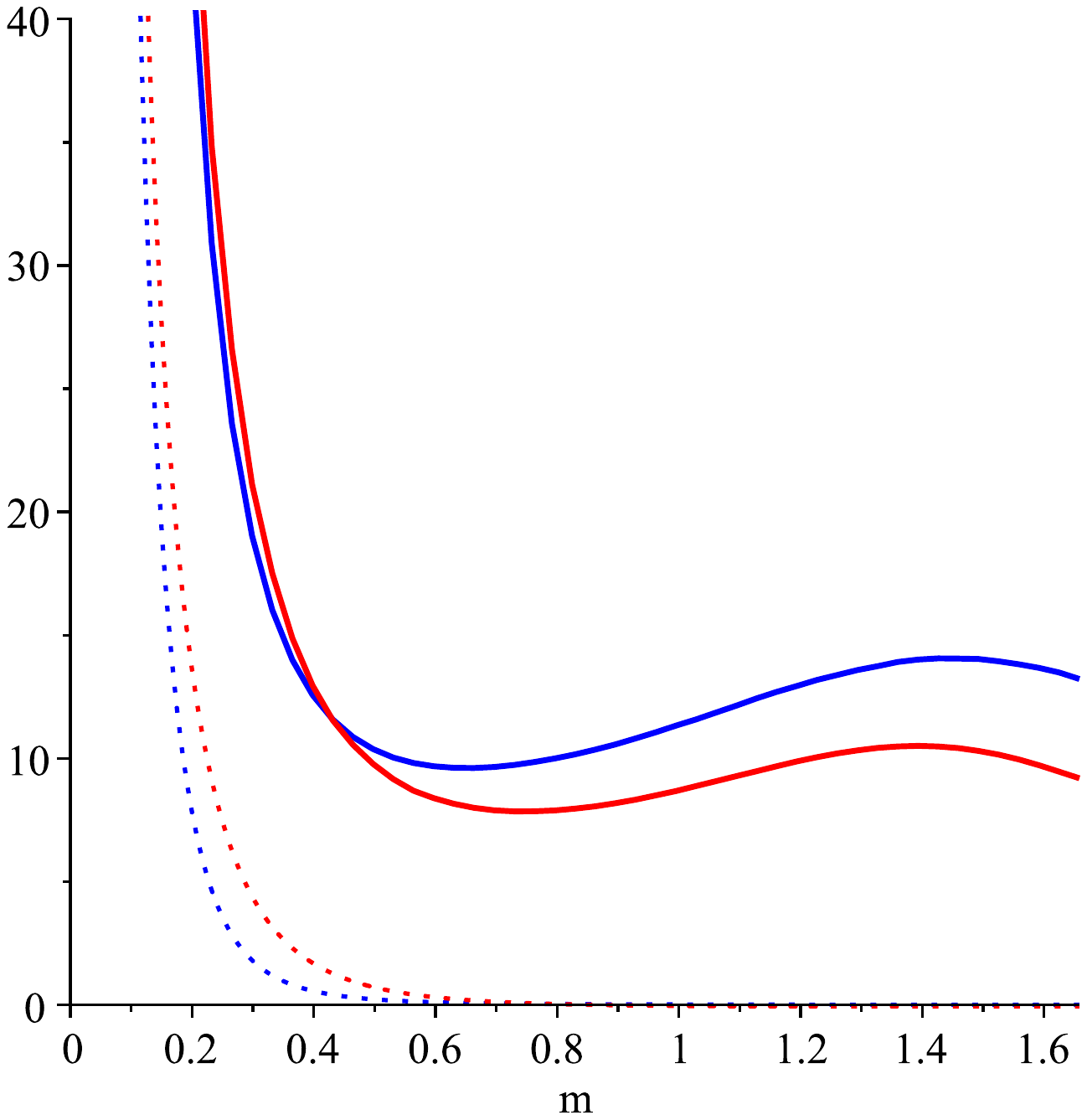}
%q=.5-v4.1-NumB=50-Blow=.33e-1-Bhigh=1.658-omega_low=0-omega_high=2.2-nrange=1000
\vspace{-6.5cm}
\subcaption{\scriptsize $\scriptstyle q=0.5$}
\end{subfigure}

\vspace{-5cm}

\begin{subfigure}[t]{0.45\textwidth}
\hspace{-2.5cm}\includegraphics[scale=0.5]{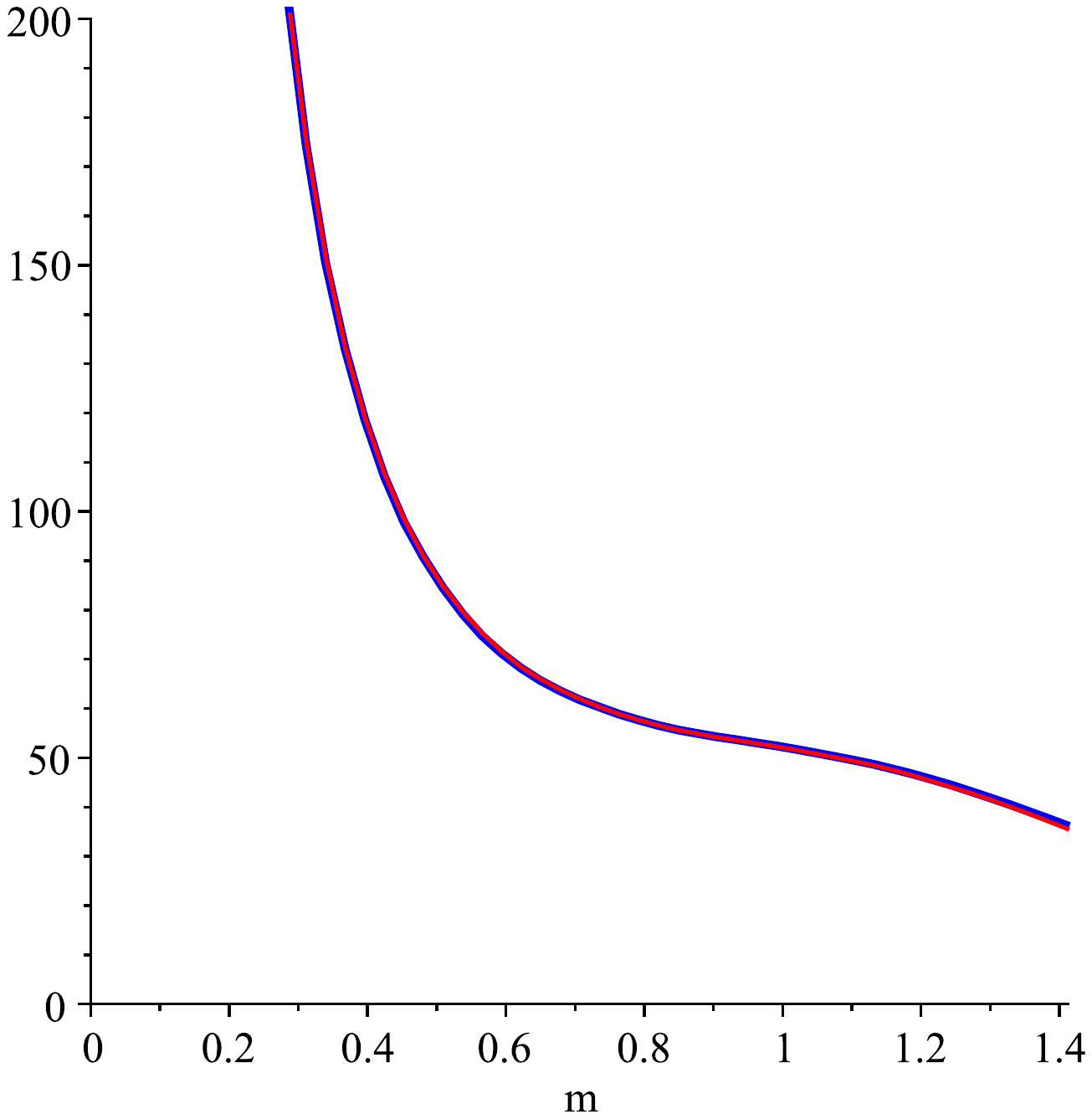}
% AdS-CFT-sigma-equation-v4.6: data/q=1-v4.1/ OmegaRange=1.1,  NumB=50, nrange=1000, Bhigh=sqrt(3-q^2);
\vspace{-6.5cm}
\subcaption{\scriptsize $\scriptstyle q=1$}
\end{subfigure}
\begin{subfigure}[t]{0.5\textwidth}
\hspace{-1.3cm}\includegraphics[scale=0.5]{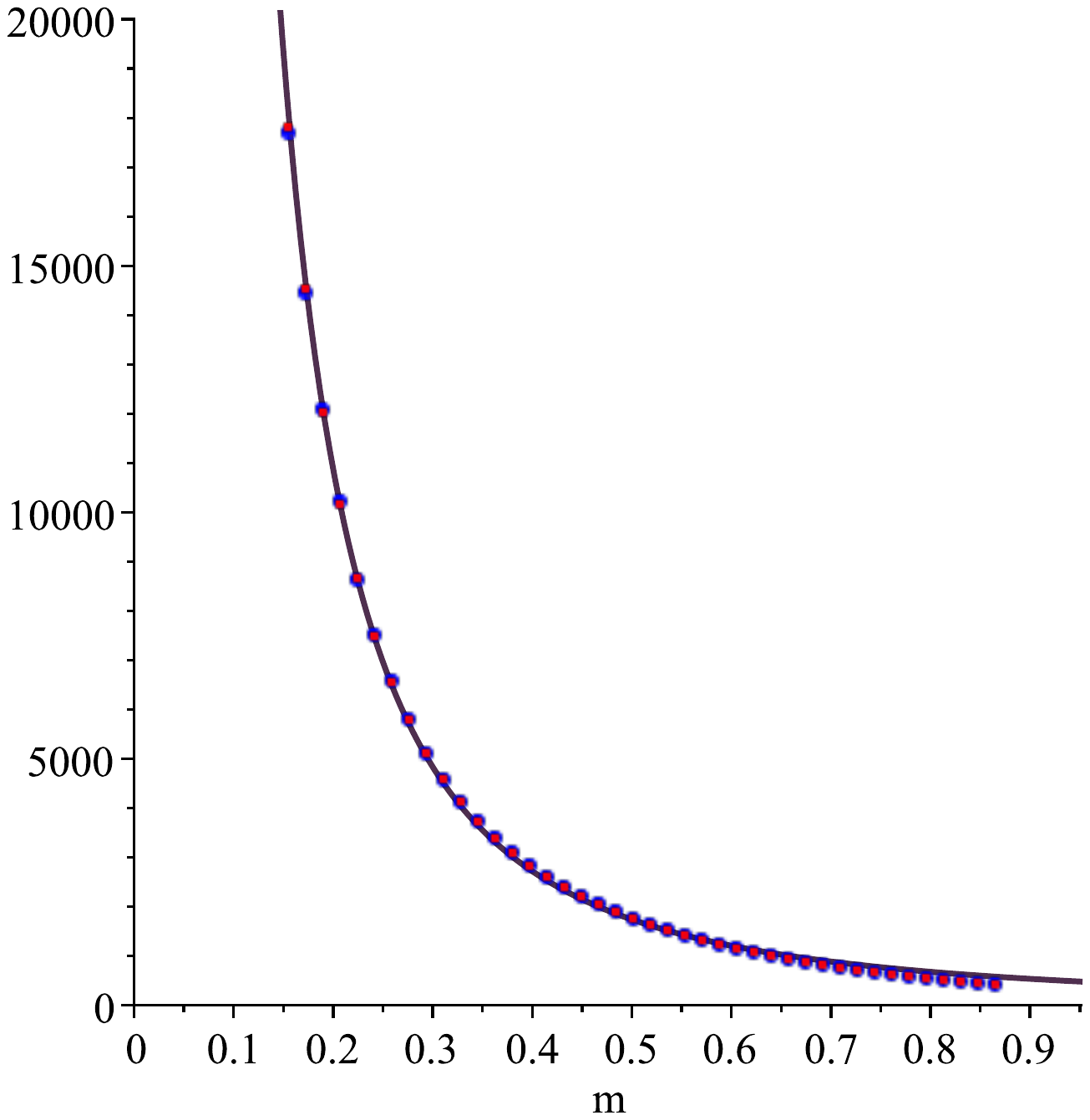}
%q=1.5-v4.1-NumB=50-Blow=.17e-1-Bhigh=.866-omega_low=0-omega_high=.6-nrange=1000
\vspace{-6.5cm}
\subcaption{\scriptsize $\scriptstyle  q=1.5$}
\end{subfigure}
\vspace{-5cm}
\caption{\scriptsize $\Gamma^{-1}$ for the Ohmic conductivity (red) and Hall conductivity (blue) at fixed $q$ as a function of $m$. For clarity the data are represented as continuous lines in figures (a), (b) and (c) and as dots in figure (d).
 The approximations (\ref{eq:Pade-sigma-xx})
 and (\ref{eq:Pade-sigma-xy})  are shown in the dotted curves in (a) and (b).
For $q=1.5$ in (d) the fit $\scriptstyle \Gamma^{-1} =\frac{435}{ m^2}$ is shown in grey.}
\label{fig:Gamma}
\end{figure}

\subsubsection{The ${\cal Q}$ factor at fixed $q$ as a function of $m$.}
The ${\cal Q}$-factor for the resonances is shown as a function of $m$
at fixed $q$ in Fig.\,\ref{fig:Q-h}. For small $q$ the system is heavily damped for filling factor 
$\nu \approx 1$,
as anticipated in equation (\ref{eq:Q-factor-approx}), as $q$ increases the damping is reduced but is greatest when $\nu$ is just below $1$ (once $q$ is greater than  $\sqrt{\frac{3}{2}}=1.225$, $\nu$ can no longer attain the value $1$). 

\begin{figure}
\begin{subfigure}[t]{0.45\textwidth}
\hspace{-2cm}\includegraphics[scale=0.5]{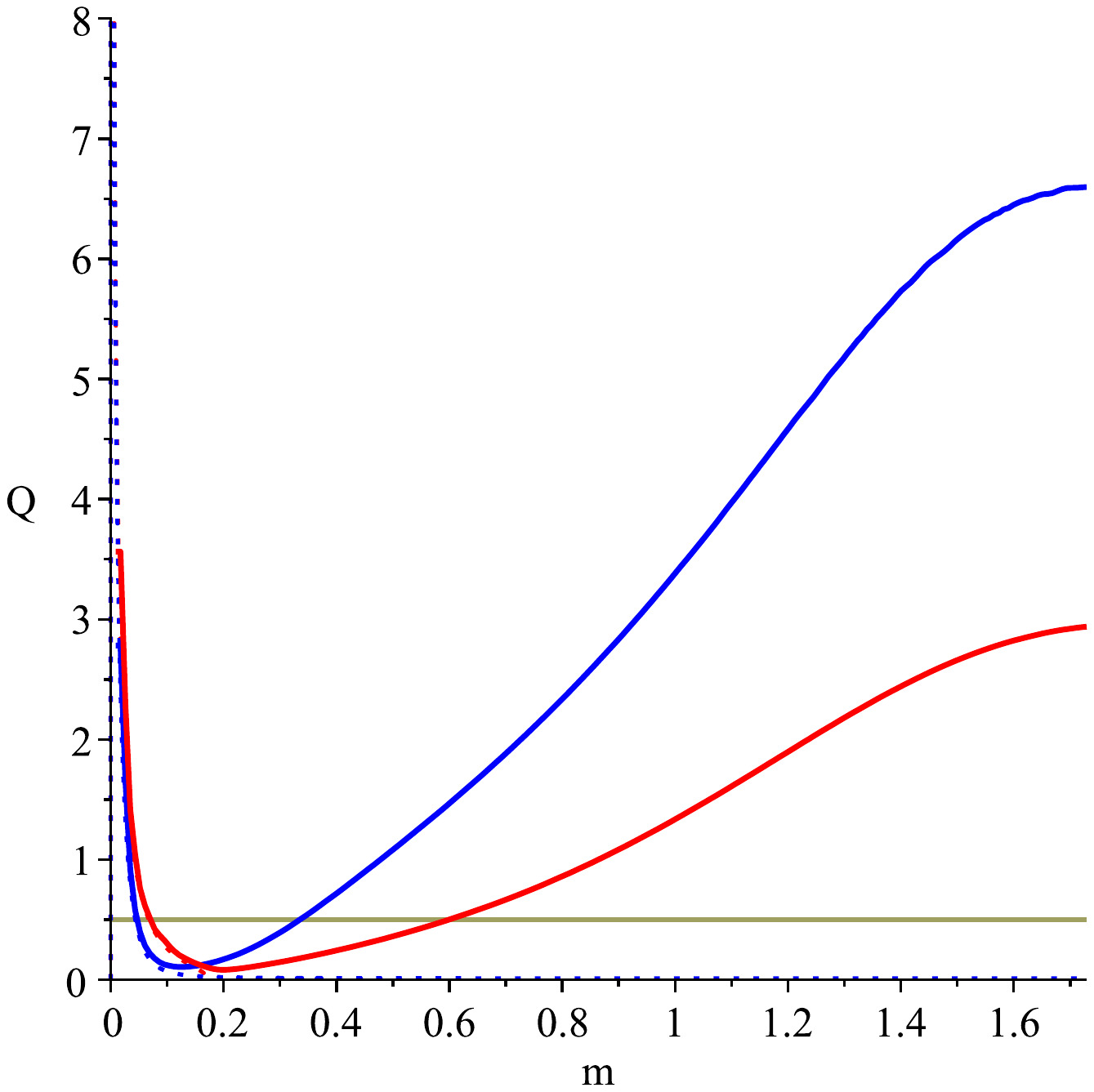}
% AdS-CFT-sigma-equation-v4.6: data v4.40 OmegaRange=1.6,  NumB=200, nrange=2000;
\vspace{-6.5cm}
\subcaption{ \scriptsize $\scriptstyle q=0.1$, the horizontal line is critical damping.}
\end{subfigure}
\begin{subfigure}[t]{0.45\textwidth}
\hspace{-1cm}\includegraphics[scale=0.5]{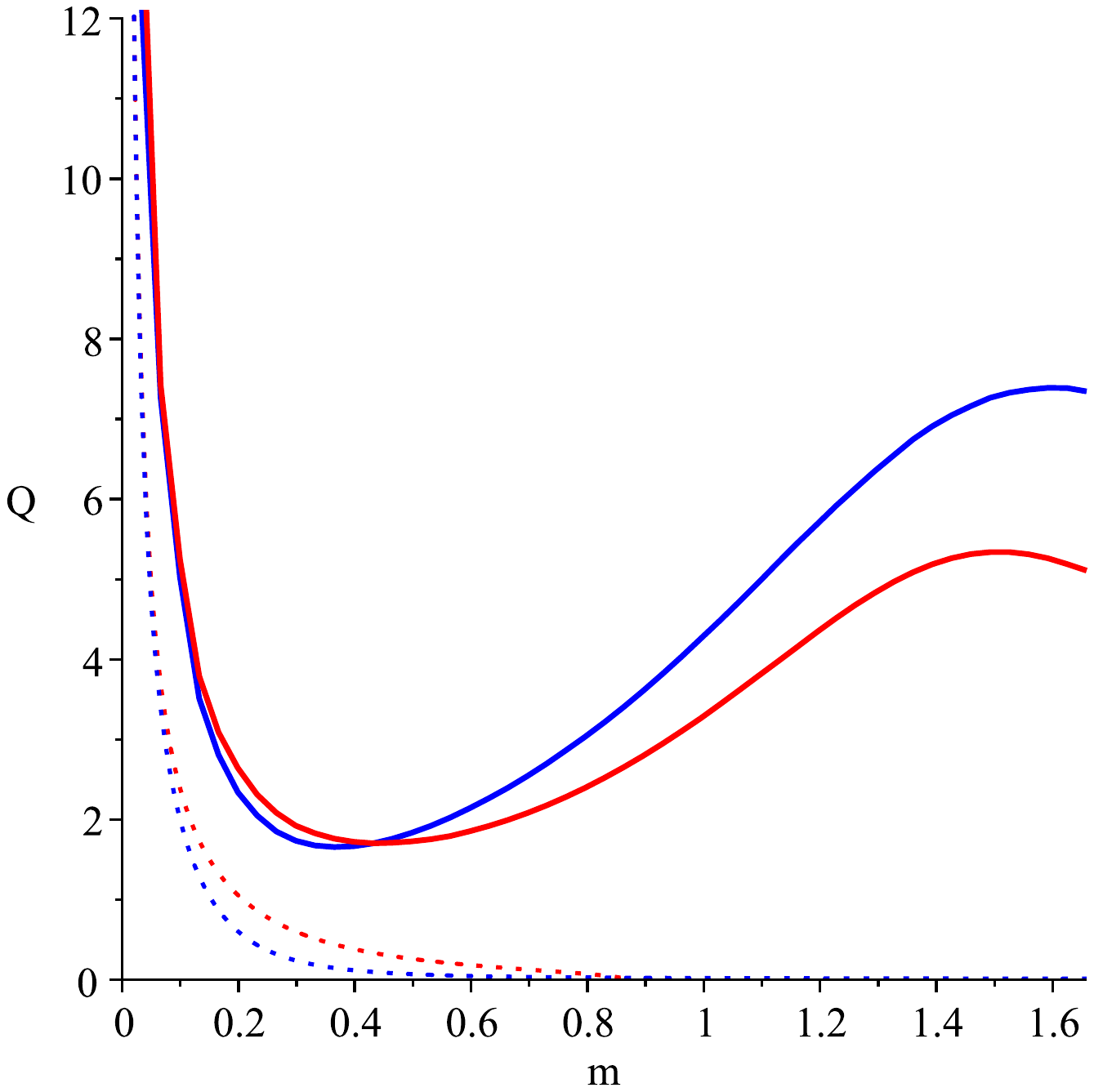}
% AdS-CFT-sigma-equation-v4.6.2: data/q=.5-v4.1-NumB=50-Bhigh=1.658-OmegaRange=2.2-nrange=1000
\vspace{-6.5cm}
\subcaption{$\scriptstyle q=0.5$}
\end{subfigure}

\vspace{-5cm}

\centering
\begin{subfigure}[t]{0.45\textwidth}
\hspace{-2.5cm}\includegraphics[scale=0.5]{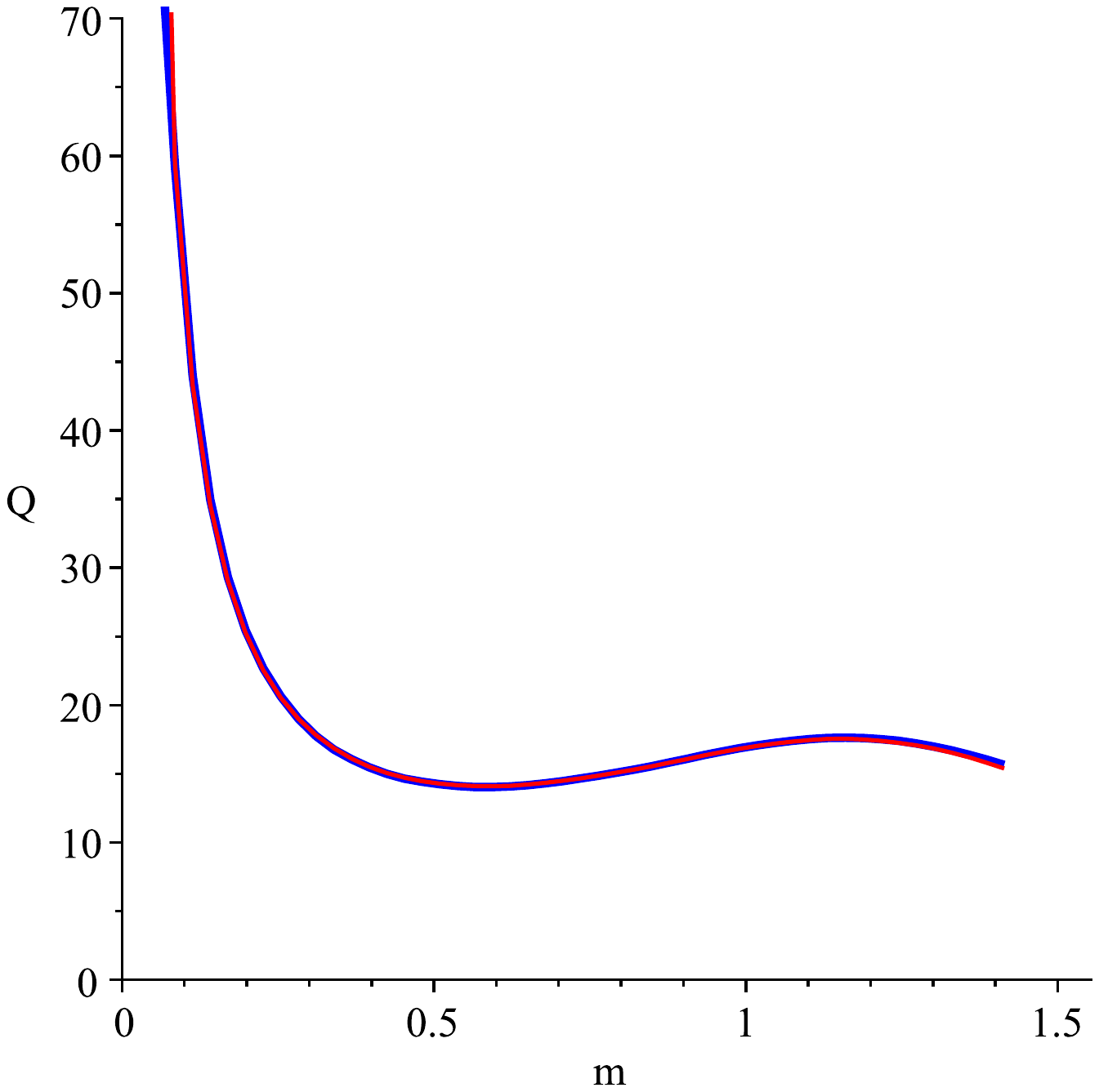}
% AdS-CFT-sigma-equation-v4.6: "data/q=1-v4.1/ OmegaRange=1.1,  NumB=50, nrange=1000, Bhigh=sqrt(3-q^2);
\vspace{-6.5cm}
\subcaption{$\scriptstyle  q=1$}
\end{subfigure}
\vspace{-6cm}
\begin{subfigure}[t]{0.45\textwidth}
\hspace{-1.5cm}\includegraphics[scale=0.5]{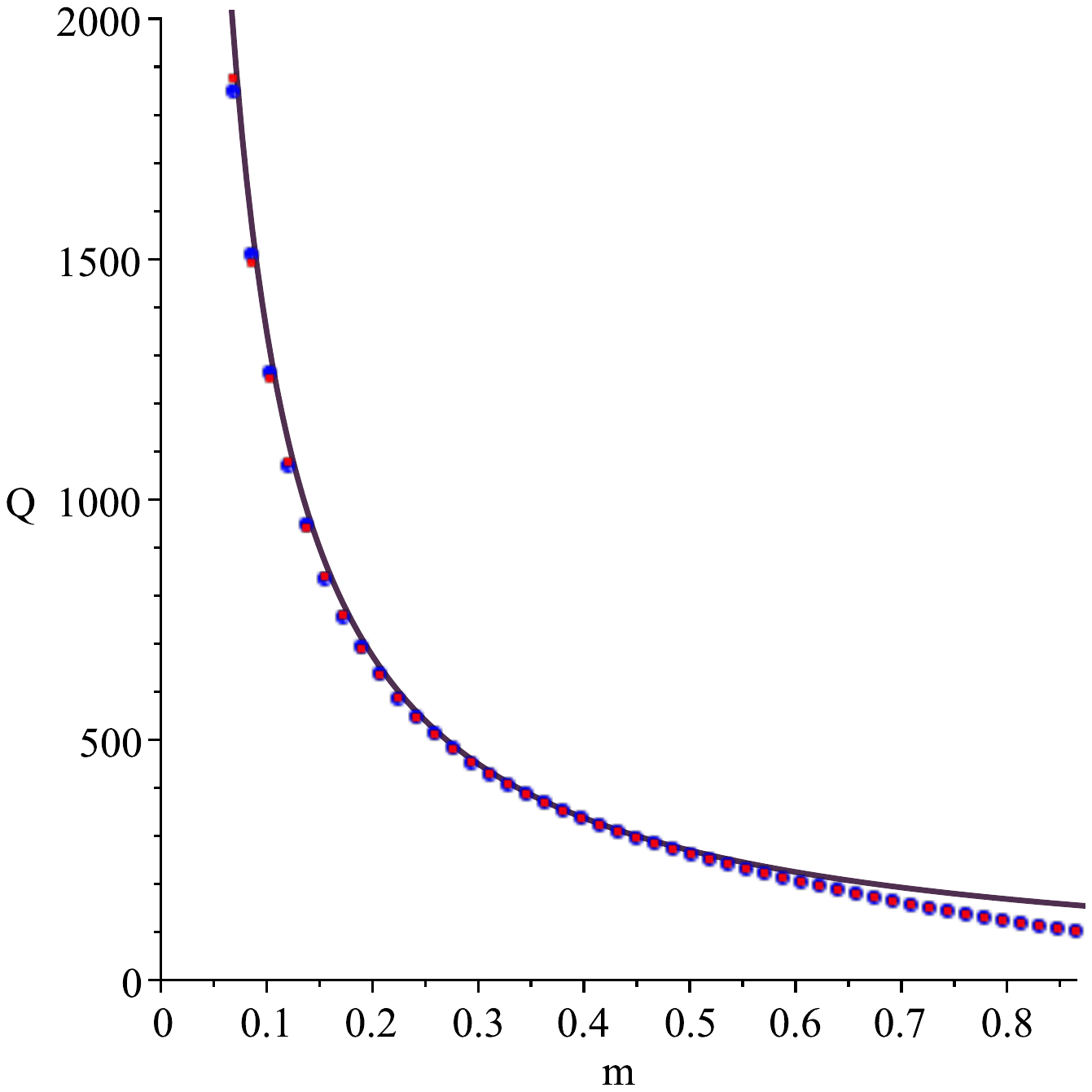}
%q=1.5-v4.1-NumB=50-Blow=.17e-1-Bhigh=.866-omega_low=0-omega_high=.6-nrange=1000
\vspace{-6.5cm}
\subcaption{\scriptsize $\scriptstyle  q=1.5$}
\end{subfigure}
\vspace{1cm}
\caption{\scriptsize The ${\cal Q}$-factor for the resonance of the Ohmic conductivity (red) and Hall conductivity (blue) at fixed $q$. For clarity the data are represented as continuous lines in figures (a), (b) and (c) and as solid circles in figure (d).
 The approximations (\ref{eq:Pade-sigma-xx})
 and (\ref{eq:Pade-sigma-xy}) are shown in the dotted curves in (a) and (b) and the horizontal line in (a) is critical damping.
 In (d) the hyperbolic fit $\scriptstyle {\cal Q}=90\frac{q}{m}$ is shown in grey.}
\label{fig:Q-h}
\end{figure}

\subsubsection{Details of the resonances for $q=0.1.$}
In Fig.\,\ref{fig:small-qm} more detailed characteristics of the resonance are shown for $q=0.1$ and low $m$, $0<m<0.2$.
When $\sigma_0$ in (\ref{eq:sigma-Drude}) is real the peak in $Re\bigl(\sigma(\wo))\bigr)$  coincides with a zero in $Im\bigl(\sigma(\wo)\bigr)$ and this relation is obeyed well for $q\ge 0.5$
but it is does not apply in the approximation (\ref{eq:sigma_xx_pole_approx}) 
and (\ref{eq:sigma_xy_pole_approx}) and numerically it is not true for $q=0.1$. The notion of a resonance frequency is ambiguous for low ${\cal Q}$ and, as described above, the graphs in Fig.\ref{fig:small-qm} use the maximum of $Re\bigr(\sigma(\wo)\bigr)$ to define $\wo_0$ rather than the zero of $Im\bigr(\sigma(\wo)\bigr)$.

\begin{figure}
%  \vspace{-1cm}
\begin{subfigure}[t]{0.45\textwidth}
\hspace{-2cm}\includegraphics[scale=0.5]{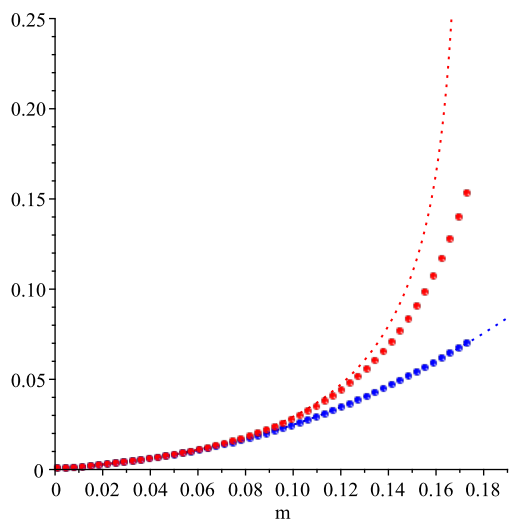}\hspace{1cm}
% AdS-CFT-sigma-equation-v4.6: 
% data/q=.1-v4.6.4-NumB=50-Blow=.1e-2-Bhigh=.173-omega_low=.1e-2-omega_high=.5-range=10000/
\vspace{-6.5cm}
\subcaption{\scriptsize The resonance frequencies as a function of $m$.}
\end{subfigure}
\vspace{-7cm}
\begin{subfigure}[t]{0.45\textwidth}
\hspace{-0.5cm}\includegraphics[scale=0.5]{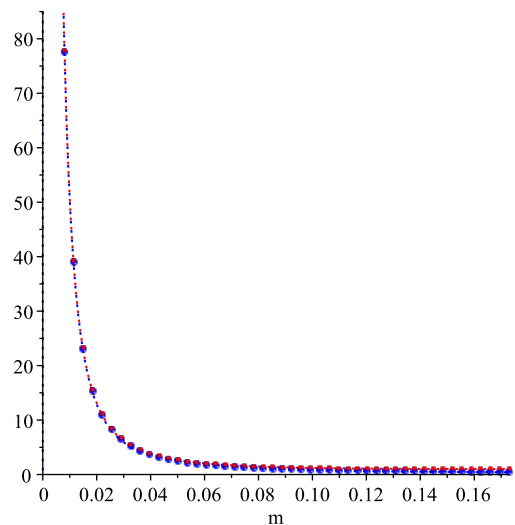}
% AdS-CFT-sigma-equation-v4.6: 
% data/q=.1-v4.6.4-NumB=50-Blow=.1e-2-Bhigh=.173-omega_low=.1e-2-omega_high=.5-nrange=10000/ReSigmaxx(h)-q=.1-b=1.data
\vspace{-6.5cm}
\subcaption{\scriptsize The height of the resonance peaks.}
\end{subfigure}

\vspace{1.5cm}

\centering
\begin{subfigure}[t]{0.45\textwidth}
\hspace{-2cm}\includegraphics[scale=0.5]{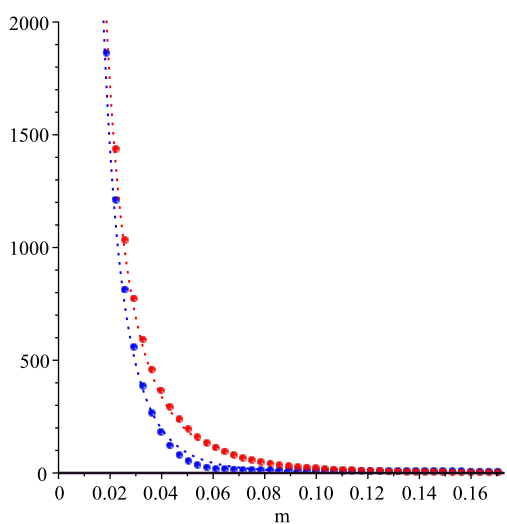}
% AdS-CFT-sigma-equation-v4.6: "data/q=1-v4.1/ OmegaRange=1.1,  NumB=50, nrange=1000, Bhigh=sqrt(3-q^2);
\vspace{-6.5cm}
\subcaption{\scriptsize The inverse damping factors $\scriptstyle \Gamma^{-1}$
for the Ohmic and Hall conductivities.}
\end{subfigure}
\qquad
\vspace{-6cm}
\begin{subfigure}[t]{0.45\textwidth}
\hspace{-1.5cm}\includegraphics[scale=0.5]{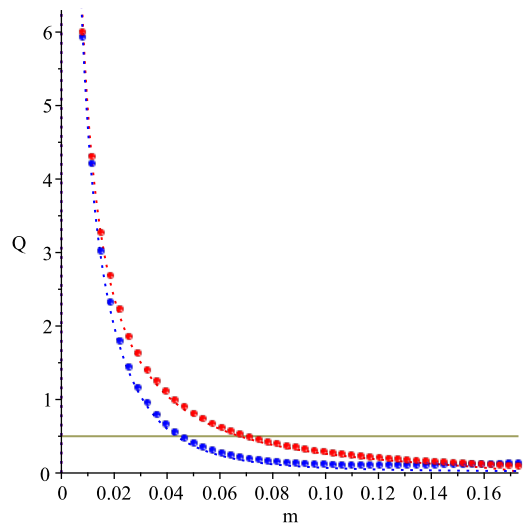}
% AdS-CFT-sigma-equation-v4.6.1: data v4.61 OmegaRange=0.6,  NumB=200, nrange=500;
\vspace{-6.5cm}
\subcaption{\scriptsize The $\scriptstyle {\cal Q}$-factor for the Ohmic and Hall
conductivities. The horizontal line is critical damping.}
\end{subfigure}
\vspace{1.5cm}
\caption{\scriptsize Further details of the resonance at $q=0.1$ at low values of $m$. For clarity the data are represented as solid circles in all figures.
 The approximations (\ref{eq:sigma_xx_pole_approx}) 
and (\ref{eq:sigma_xy_pole_approx}) are shown as dotted lines.}
\label{fig:small-qm}
\end{figure}

An instructive way of viewing the conductivities is to plot the real parts of the conductivities against the imaginary parts. At fixed $q$ this can be done using a parametric plot with $\nu = \frac{q}{m}$ as a parameter. To keep the temperature positive $m$ is restricted to the range
$0\le |m| < \sqrt{ 3 - q^2}$, so there is a lower bound on $|\nu|$ when $q$ is fixed, namely $\frac{|q|}{\sqrt{3 - q^2}}\le |\nu|$. The range of $|\nu|$ is therefore greater for smaller $q$ and this kind of plot is more useful for smaller $q$ values than for larger ones.\footnote{The alternative strategy of fixing $m$, and varying $\nu$ by varying $q$, imposes an upper bound on $|\nu|$.}
In Fig.\,\ref{fig:parametric} the real parts of the Ohmic and Hall conductivities are plotted against their respective imaginary parts for $q=0.1$ ($0.058\le \nu < \infty$) for six different frequencies. 
The Ohmic conductivity is red and the Hall conductivity is blue: for $\sigma^{x x}$ the    vertical axis is $Re(\sigma^{x x})$ and the horizontal axis is $Im(\sigma^{x x})$, for
$\sigma^{x y}$ the vertical axis is $-Im(\sigma^{x y})$ and the horizontal axis
is $Re(\sigma^{x y})$. The dotted curves are numerical data, the red and blue solid lines are the approximations (\ref{eq:sigma_xx_pole_approx}) and (\ref{eq:sigma_xy_pole_approx})
respectively. The Ohmic conductivities are very close to being circular arcs,
particularly for the larger values of $\wo$, $0.64$ and $1.12$.
The numerical data are well approximated by circles in the complex
Ohmic conductivity plane
\[ \sigma^{x x}(\wo) = \frac{ a_0 - i b_0 (\wo-\wo_0)}{1 - i \tau (\wo - \wo_0)
  },\]
which is a generalisation of the Drude form
\beq \sigma(\wo) = \frac{\sigma_0}{1 - i \tau (\omega - \omega_0)},
\label{eq:cyclotron-Drude}\eeq
with $\omega_0$ monotonic in $\nu$.

Similar plots for the Hall conductivity
from experimental data on Ga/As heterojunctions at THz frequencies, were given in \cite{DSSMSP}.

\begin{figure}
\vspace{-3cm}
\begin{subfigure}{0.8\textwidth}
\hspace{-1cm}
\hbox{\begin{minipage}[t]{0.45\textwidth}
%    \begin{overpic}[unit=1mm,scale=0.4,grid]{fig-11a.pdf}
    \begin{overpic}[unit=1mm,scale=0.4]{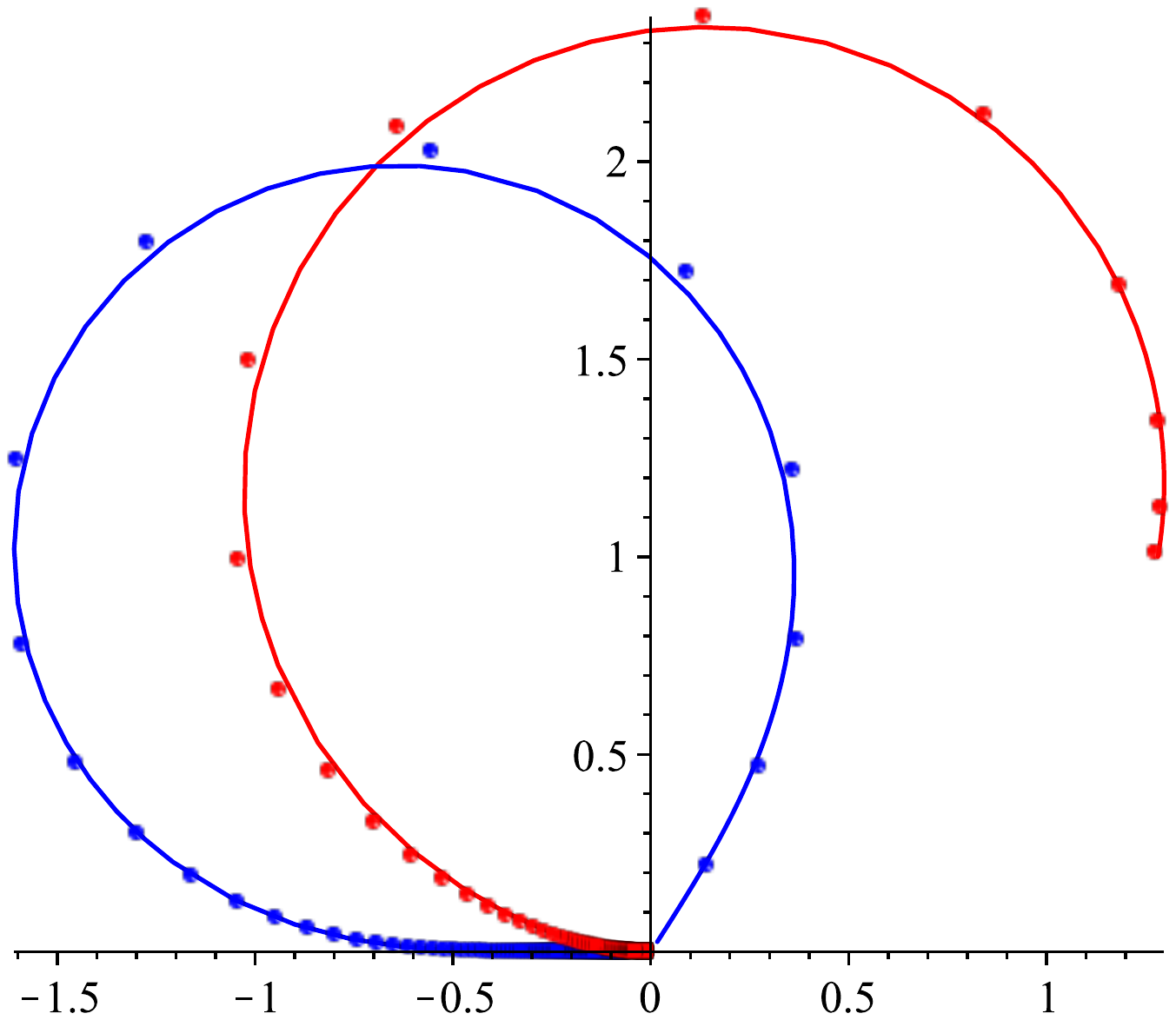}
   %   \put(50,64){$\wo = 0.01$}
      \put(51,82){$\searrow \nu $}
      \put(57,62){$\nwarrow $}
      \put(59,59){$\scriptstyle \nu \rightarrow \infty$}
      \put(37,50){$\swarrow $}
      \put(41,52){$\scriptstyle \nu = 0.058$}
    \end{overpic}
    \vspace{-5cm}
    \subcaption{$\scriptstyle \wo = 0.01$}
\end{minipage} 
\hspace{2.5cm}
\begin{minipage}[t]{0.45\textwidth}
  \begin{overpic}[unit=1mm,scale=0.4]{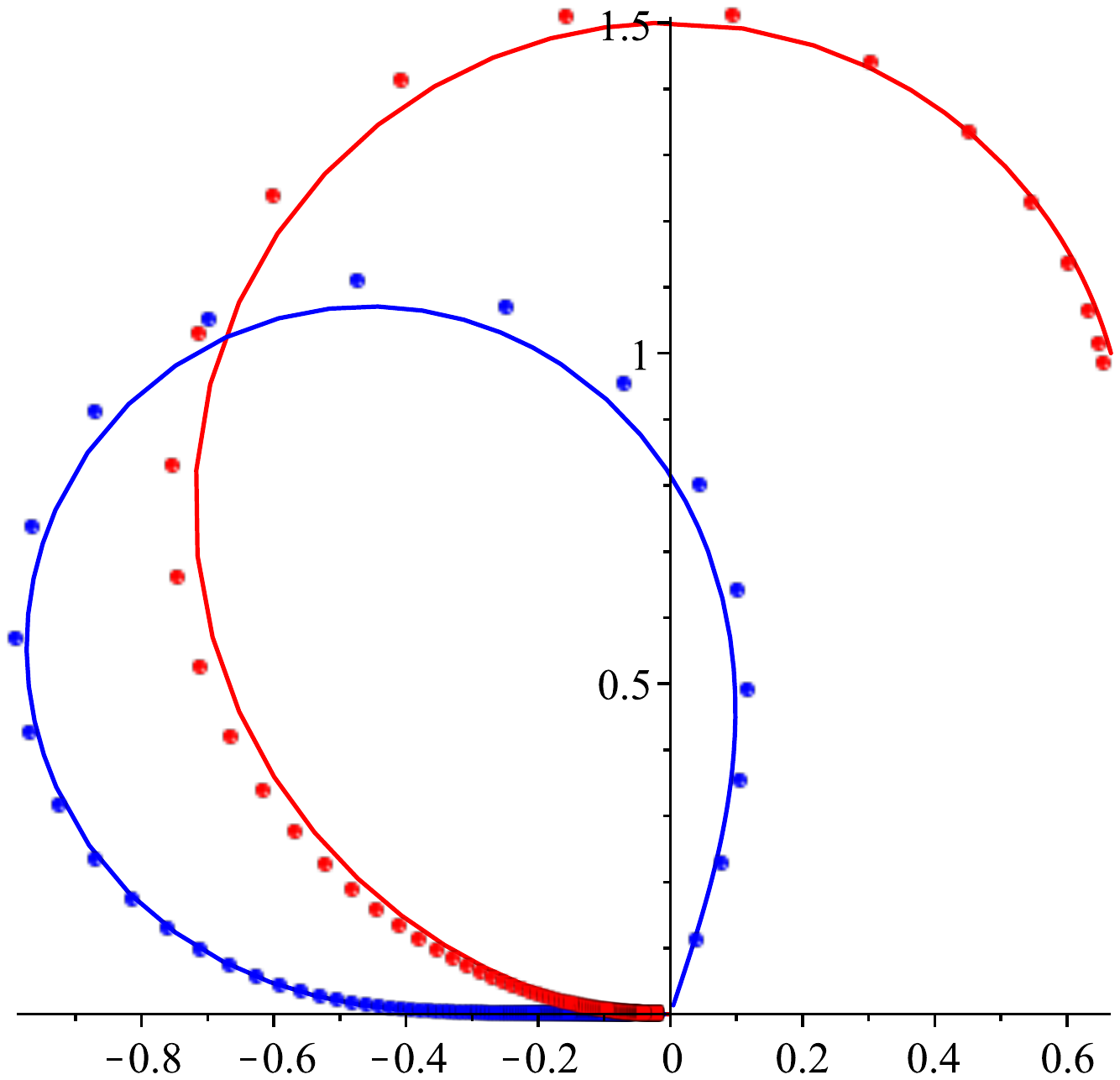}
   %  \put(50,64){$\wo = 0.02$}
\end{overpic}
\vspace{-5cm}
\subcaption{$\scriptstyle \wo = 0.02$}
\end{minipage}}
\end{subfigure}

\vspace{-2cm}

\centering
\begin{subfigure}[t]{0.8\textwidth}
\vspace{-2cm}
\hspace{-2.5cm}
\hbox{\begin{minipage}[t]{0.45\textwidth}
  %  \centering
 \begin{overpic}[unit=1mm,scale=0.4]{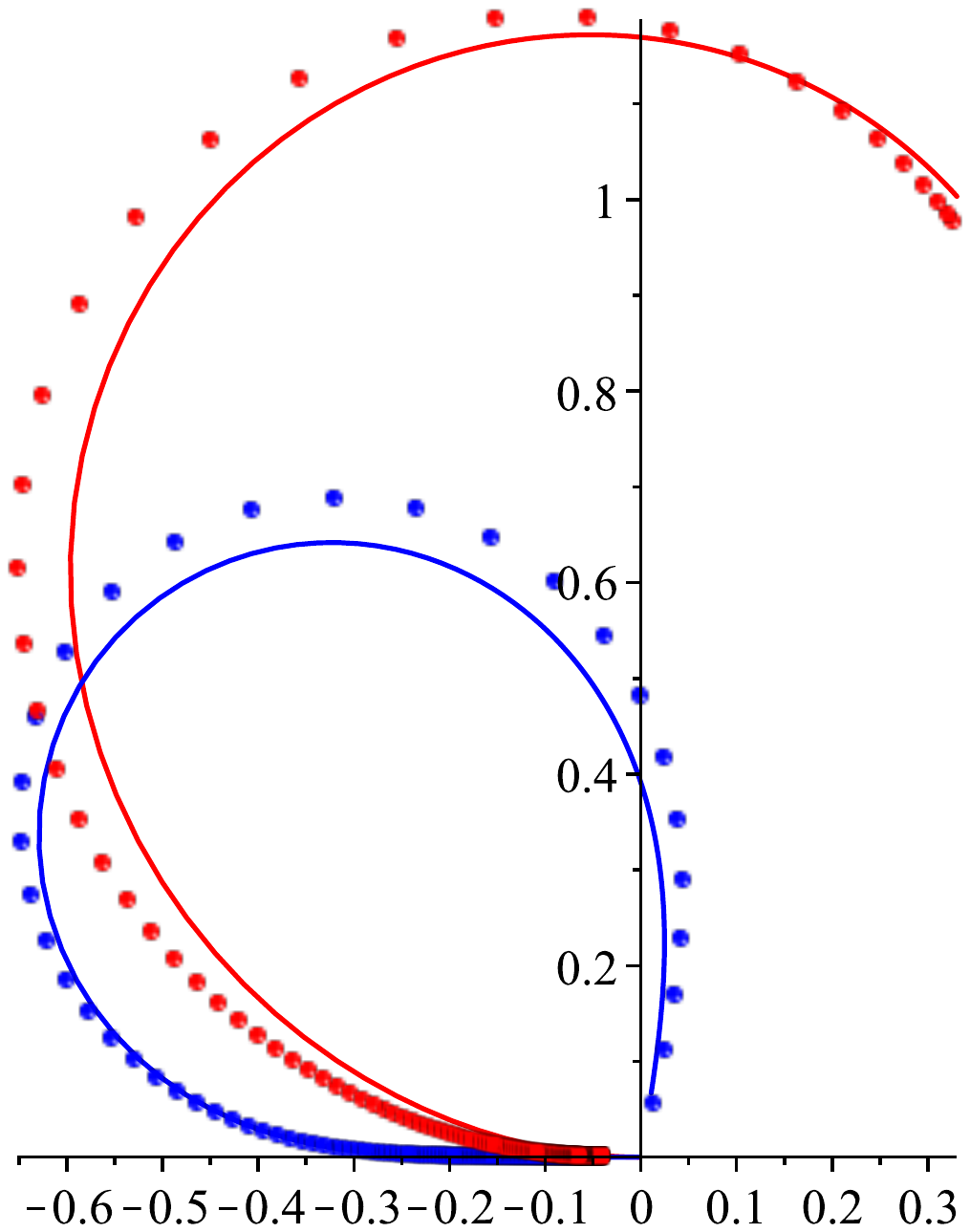}
%  \put(50,80){$\wo = 0.04$}
 \end{overpic}
 \vspace{-5cm}
 \subcaption{$\scriptstyle \wo = 0.04$}
\end{minipage}
\hspace{2cm}
\begin{minipage}[t]{0.45\textwidth}
    \begin{overpic}[unit=1mm,scale=0.4]{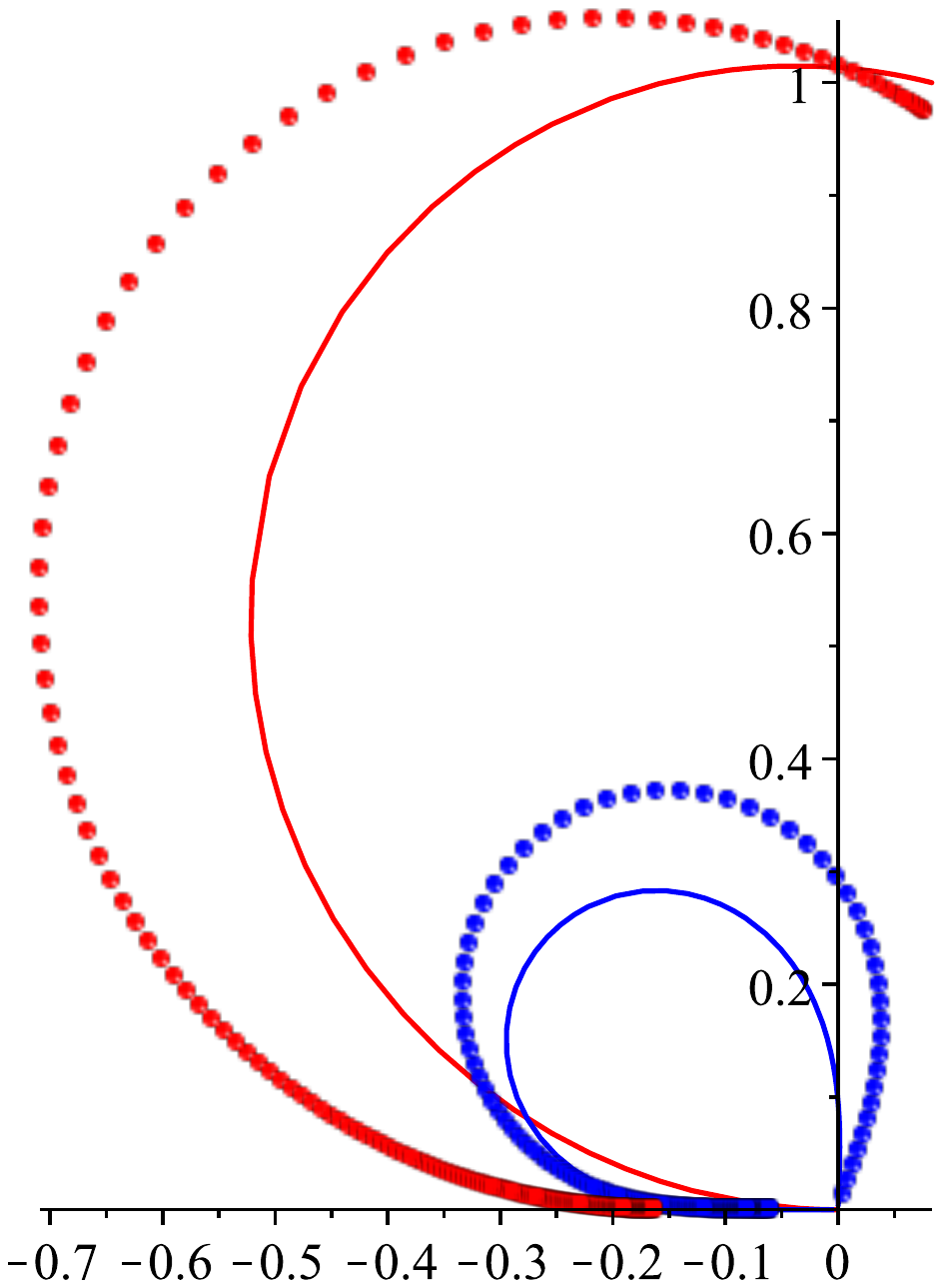}
    %  \put(55,80){$\wo=0.16$}
    \end{overpic}
\vspace{-5cm}
    \subcaption{$\scriptstyle \wo=0.16$}
\end{minipage}}
\vspace{-5cm}
\end{subfigure}

\vspace{3cm}

\begin{subfigure}[t]{0.8\textwidth}
\vspace{-2cm}
\hskip -2cm 
\hbox{\begin{minipage}[t]{0.45\textwidth}
    \centering
      \begin{overpic}[unit=1mm,scale=0.4]{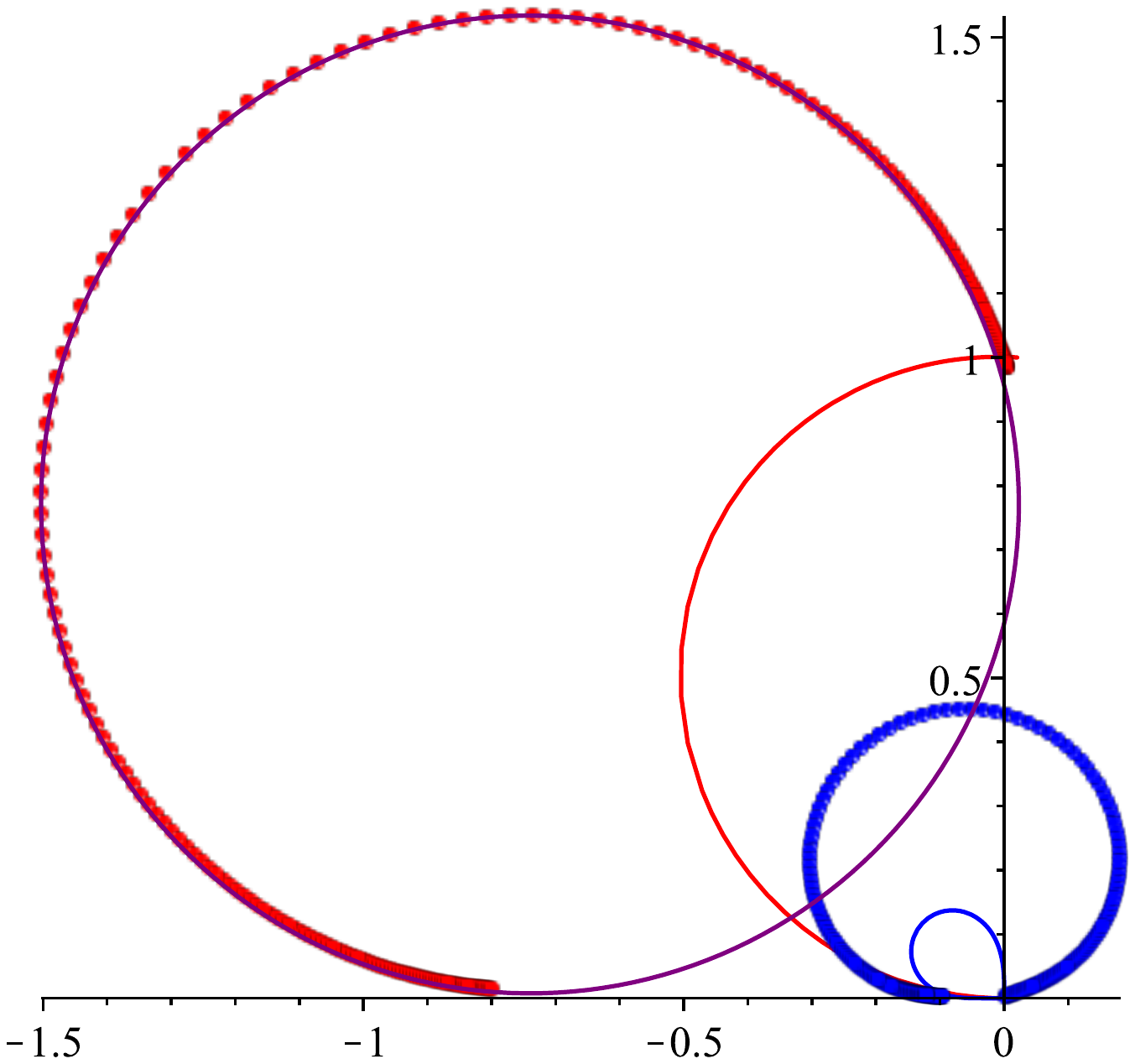}
 % \put(25,85){$\wo=0.64$}
      \end{overpic}
      \vspace{-5cm}
      \subcaption{$\scriptstyle \wo=0.64$}
\end{minipage}
\hspace{2.5cm}
\begin{minipage}[t]{0.45\textwidth}
  \begin{overpic}[unit=1mm,scale=0.4]{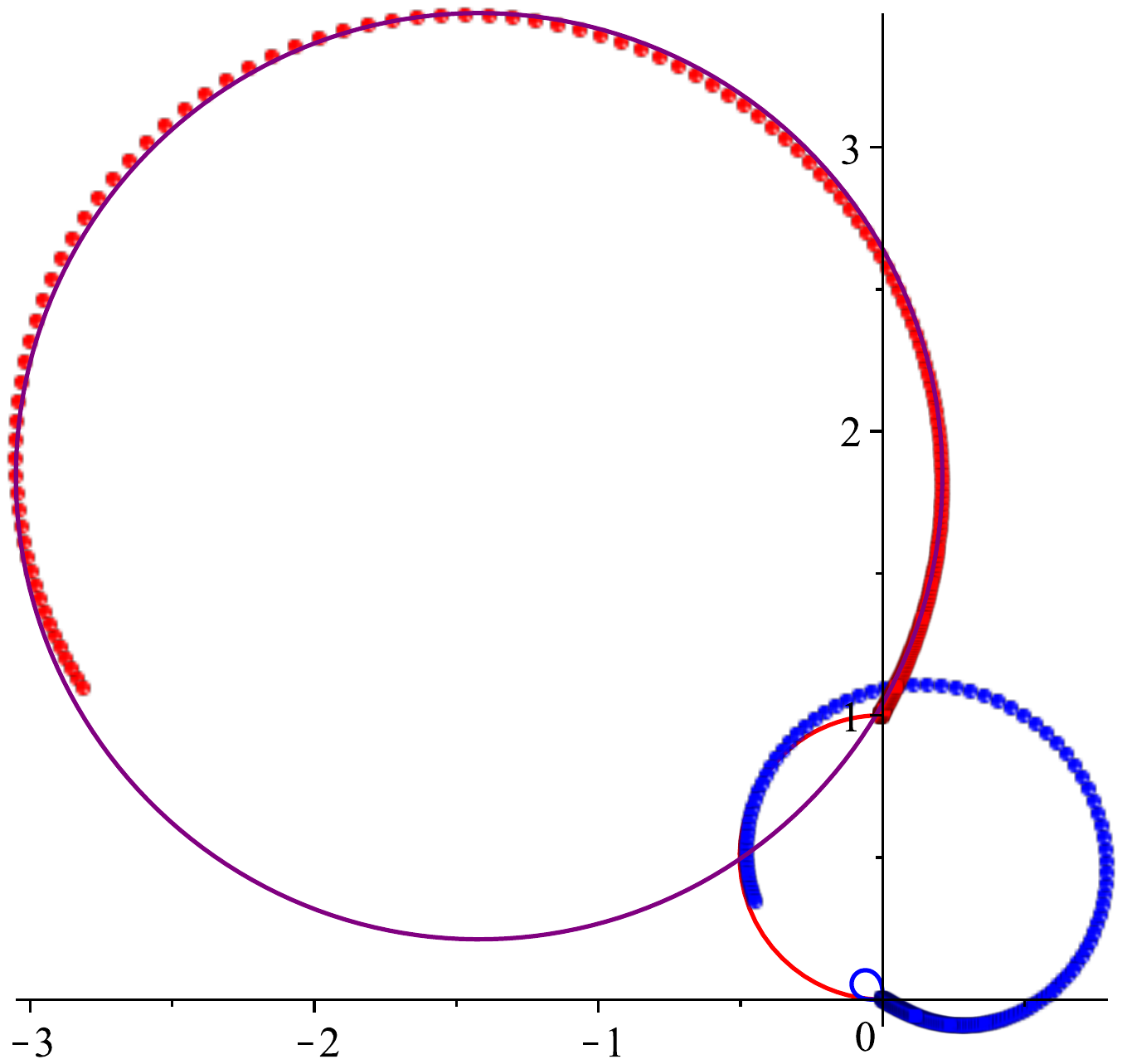}
%       \put(15,70){$\swarrow $}
%      \put(21,74){$\scriptstyle \nu =0.058 $}
       \put(13,64){$\swarrow $}
      \put(17,67){$\scriptstyle \nu =0.058 $}
\end{overpic}
\vspace{-5cm}
\subcaption{$\scriptstyle \wo = 1.12$}
\end{minipage}}
\end{subfigure}
\vspace{-4cm}
\caption{\scriptsize The real parts of the conductivities plotted against the imaginary parts
  for $q=0.1$ at various frequencies (Ohmic conductivity in red, Hall conductivity in blue). For the Ohmic conductivity the vertical axis is $Re\bigl(\sigma^{x x}(\wo)\bigr)$
  and the horizontal axis is  $Im\bigl(\sigma^{x x}(\wo)\bigr)$, for the Hall conductivity the
  vertical axis is  $-Im\bigl(\sigma^{x y}(\wo)\bigr)$ and the horizontal axis is  $Re\bigl(\sigma^{x y}(\wo)\bigr)$. The dots are numerical data, red and blue solid lines are the approximation (\ref{eq:sigma_xx_pole_approx}) and (\ref{eq:sigma_xy_pole_approx}).
  For the higher values of $\wo$ ($0.64$ and $1.12$)  the purple curves are fits to the Ohmic conductivity using a perfect circle.  Increasing $\nu$ is
clockwise in the figures.}
  \label{fig:parametric}
\end{figure}

\subsubsection{Details of the resonances for $q=1.5$}
In Fig.\,\ref{fig:large-q-small-m} the resonance parameters at $q=1.5$ are studied
for low $m$, $0<m<0.2$ (again with the Ohmic conductivity in red and the Hall
conductivity in blue).  The numerical data in the four figures
fit remarkably well with the analytic forms
\beq \wo_0=0.41 q m, \quad \Gamma = 0.0023\, m^2,  \quad \sigma_0
=90\frac{q^2}{m^2},  \quad {\cal Q}= 90\left| \frac{q}{m}\right|,
\label{eq:large-q-analytic-fit}\eeq
 for both conductivities, 
at least for $m$ in the lower part of its allowed range.
This is the same functional from as (\ref{eq:omega_0-linear})
but with different co-efficients.
Indeed the four numbers in (\ref{eq:large-q-analytic-fit}) can be obtained 
from only three parameters in the analytic expression 
\beq
\sigma^{x x} = \frac{\wo + i a q^2- b q m}{\wo + b  q m+ i c m^2 }\label{eq:sigma-Pade-renormailsed}
\eeq
with $a=0.21$, $b=-0.41$ and $c=0.0023 $, with errors of $O\left(\frac{m}{q}\right)^2$,
which also fits Fig.\,\ref{fig:Gammasigma0}(c) with $Re\bigl((\sigma^{x x}(\wo)\bigr) \Gamma^{x x}
=\bigl|Im\bigl((\sigma^{x y}(\wo)\bigr)\bigr| \Gamma^{x y}= 0.46=0.21 q^2$.
This suggests that perhaps it may be possible 
to justify some similar analytic approximation when $|q|$ is close to its maximum value of $\sqrt{3}$, but we have not found an analytic argument for this.
\begin{figure}
%\vspace{-2.5cm}
\begin{subfigure}[t]{0.45\textwidth}
\hspace{-2cm}\includegraphics[scale=0.5]{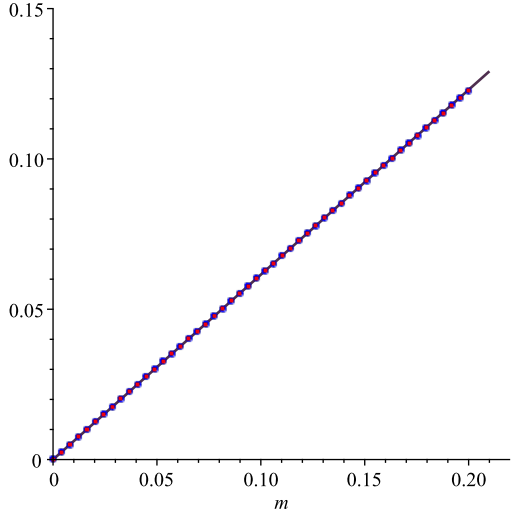}
% AdS-CFT-sigma-equation-v4.6.4: data/q=1.5-v4.6.4-NumB=50-Blow=0-Bhigh=.200-omega_low=0-omega_high=.6-nrange=2000
\qquad
\vspace{-6.5cm}
\subcaption{\scriptsize The resonance frequencies for the Ohmic (red) and Hall conductivities (blue) at $\scriptstyle q = 1.5$ for $\scriptstyle 0<m<0.2$.
The linear fit $\scriptstyle \wo_0 = 0.41 q m$ is shown in grey.}
\end{subfigure}
\qquad
\begin{subfigure}[t]{0.45\textwidth}
\hspace{-1.5cm}\includegraphics[scale=0.5]{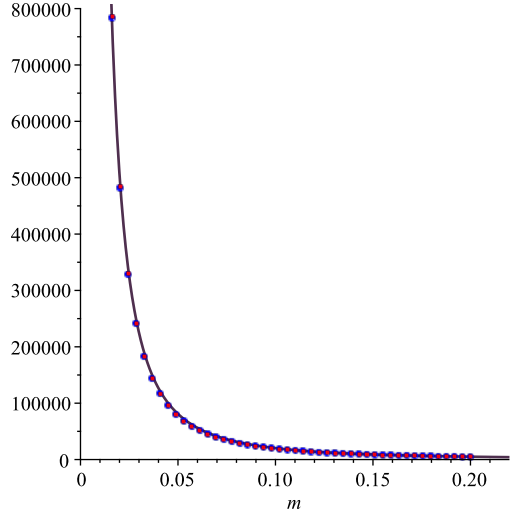}
% AdS-CFT-sigma-equation-v4.6.4: data/q=1.5-v4.6.4-NumB=50-Blow=0-Bhigh=.200-omega_low=0-omega_high=.6-nrange=2000
\vspace{-6.5cm}
\subcaption{\scriptsize The maxima of $\scriptstyle Re\bigl(\sigma^{x x}(\wo)\bigr)$
  (red) and $\scriptstyle -Im\bigl(\sigma^{x y}(\wo)\bigr)$ (blue)
conductivities, overlaying the analytic fit $\scriptstyle 90 \frac{q^2}{m^2}$.}
\end{subfigure}

\vspace{-4.5cm}

\begin{subfigure}[t]{0.45\textwidth}
\hspace{-2cm}\includegraphics[scale=0.5]{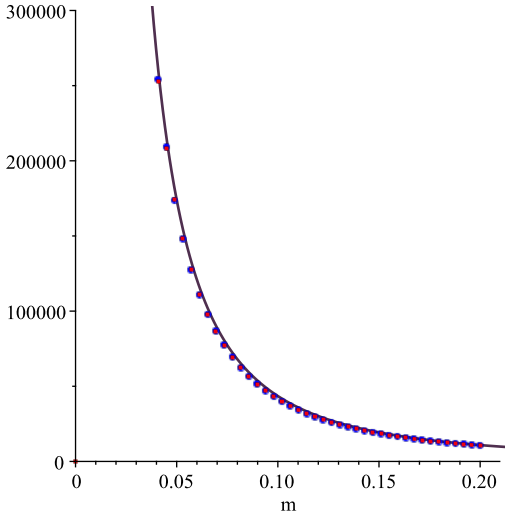}
% AdS-CFT-sigma-equation-v4.6.4: data/q=1.5-v4.6.4-NumB=50-Blow=0-Bhigh=.200-omega_low=.1e-2-omega_high=.6-nrange=2000
\vspace{-6.5cm}
\subcaption{\scriptsize The inverse widths $\scriptstyle \Gamma^{-1}$ for the Ohmic (red) and 
Hall (blue) conductivities at $\scriptstyle q = 1.5$ for $\scriptstyle 0<m<0.2$.
The fit $\scriptstyle \Gamma^{-1} = \frac{435}{m^2}$ is shown in grey.}
\end{subfigure}
\qquad
\begin{subfigure}[t]{0.45\textwidth}
\hspace{-1.5cm}\includegraphics[scale=0.5]{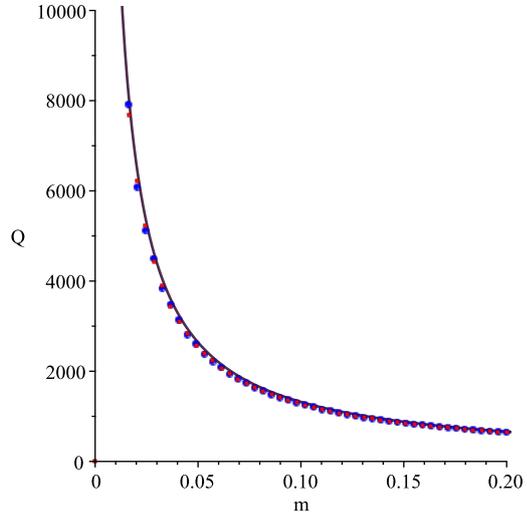}
% AdS-CFT-sigma-equation-v4.6.4: data/q=1.5-v4.6.4-NumB=50-Blow=0-Bhigh=.200-omega_low=0-omega_high=.6-nrange=2000
\vspace{-6.5cm}
\subcaption{\scriptsize The $\scriptstyle {\cal Q}$-factor for the Ohmic (red) and Hall (blue) conductivities at $\scriptstyle q = 1.5$ for $\scriptstyle 0<m<0.2$.
The hyperbolic fit $\scriptstyle {\cal Q}=90\frac{q}{m}$ is shown in grey.}
\end{subfigure}
\vspace{-4cm}
\caption{\scriptsize More details of the resonance at $q=1.5$ for $0<m<0.2$. The data are represented as solid circles in all figures.
  The analytic fit (\ref{eq:large-q-analytic-fit}) is shown in grey.}
\label{fig:large-q-small-m}
\end{figure}

\subsubsection{The residues at
  $q=0.5$, $q=1$ and $q=1.5$ as functions of $m$}
Inspection of Figs.\,\ref{fig:peaks} and \ref{fig:Gamma} shows a possible correlation
between $\sigma_0$ and $\Gamma^{-1}$, they are strikingly similar.
In Fig.\,\ref{fig:Gammasigma0} the product  $\Gamma_{x x} Re(\sigma_0^{ x x})$ (red) 
and $\Gamma_{x y} \bigl|Im(\sigma_0^{x y})\bigr|$ (blue) are plotted for $q=0.5$, $1$ and $1.5$.
For a well-defined resonance of the form (\ref{eq:sigma-Drude})
this the residue of the pole at $\wo_* = \wo_0 -i\Gamma$.
For $q=0.5$ the plots for $\sigma^{x x}$ and $\sigma^{x y}$ are rather different but for $q=1$ and $q=1.5$ they exhibit a remarkable correlation. Indeed for $q=1.5$ the residue 
appears to be a constant, independent of $m$,  suggestive of the Drude -from
(\ref{eq:cyclotron-Drude}) with $\sigma_0$ and $\Gamma$ independent of $m$ and
$\wo_0$ linear in $m$.

\begin{figure}
\centering
\begin{subfigure}[t]{0.45\textwidth}
\hspace{-1.5cm}\includegraphics[scale=0.5]{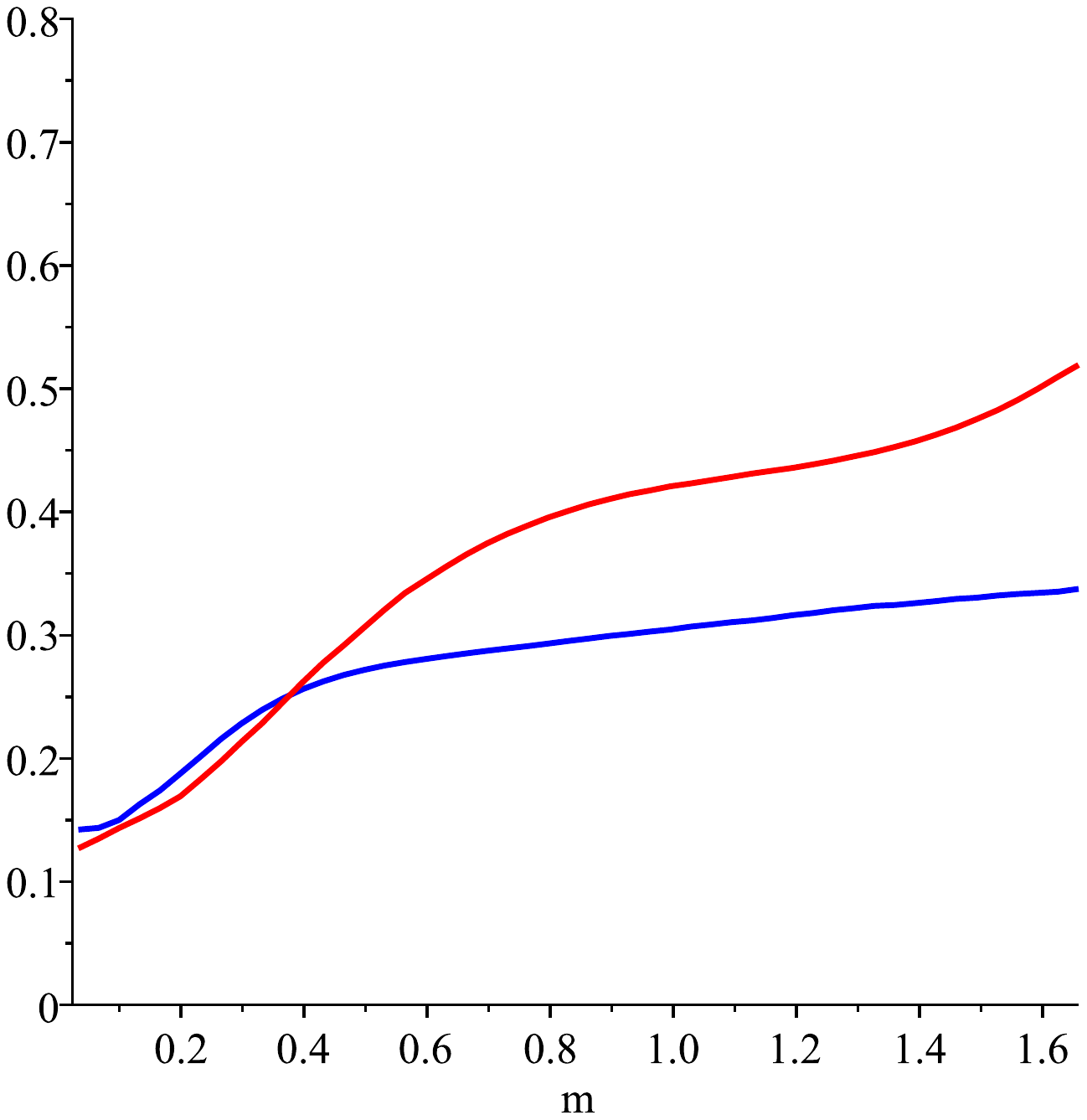}
% AdS-CFT-sigma-equation-v4.6.1: data v4.61 OmegaRange=0.6,  NumB=200, nrange=500;
\vspace{-6.5cm}
\subcaption{\scriptsize $\scriptstyle q=0.5$: there is little correlation between the Ohmic and Hall conductivities.}
\end{subfigure}

\vspace{-5cm}

\centering
\begin{subfigure}[t]{0.45\textwidth}
\hspace{-2cm}\includegraphics[scale=0.5]{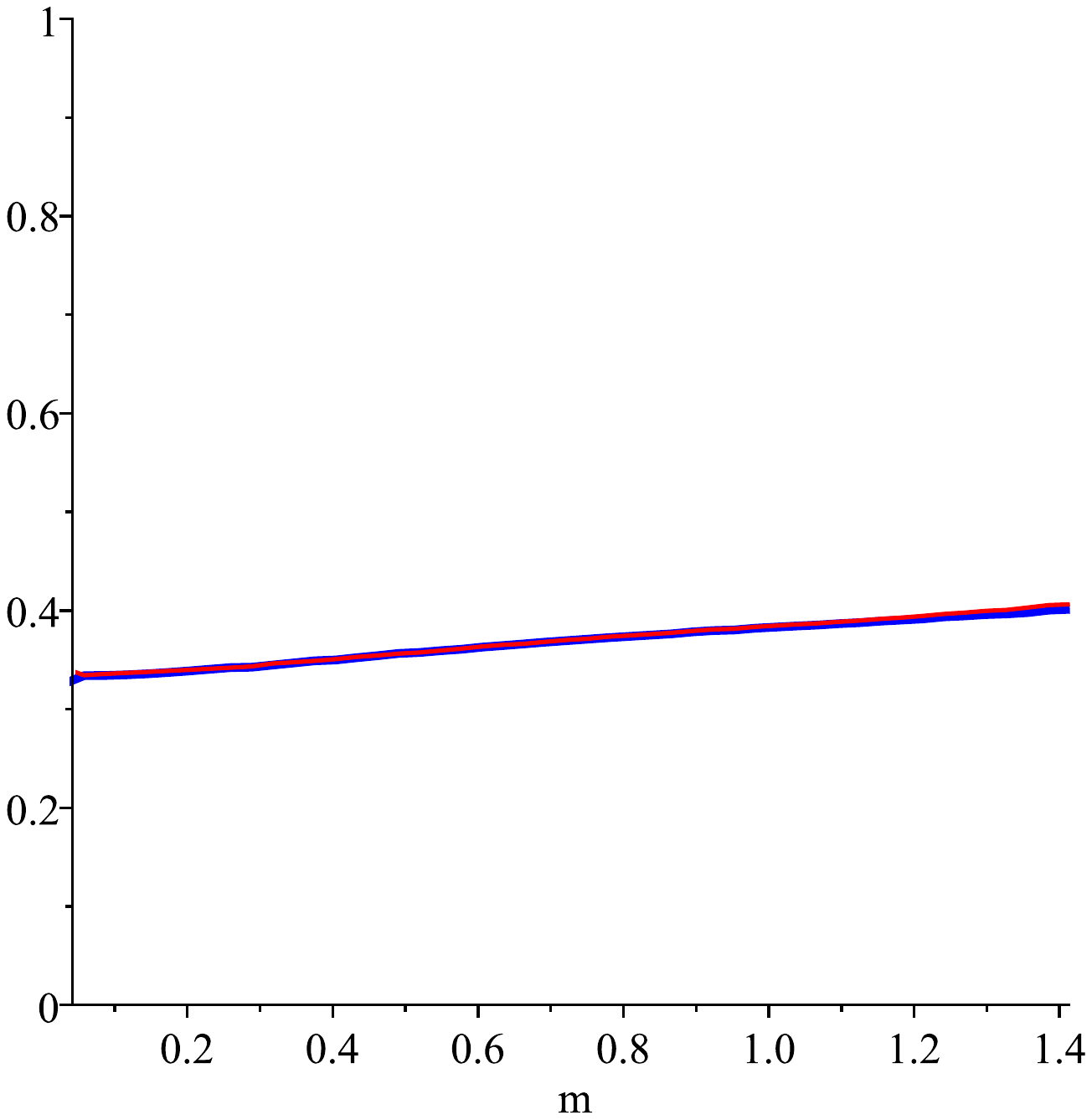}
\vspace{-6.5cm}
\subcaption{\scriptsize $\scriptstyle q=1$: the plots for the Ohmic and Hall conductivities are
nearly linear and almost indistinguishable.}
\end{subfigure}
\centering
\qquad
\vspace{-5cm}
\begin{subfigure}[t]{0.45\textwidth}
\hspace{-1.5cm}\includegraphics[scale=0.5]{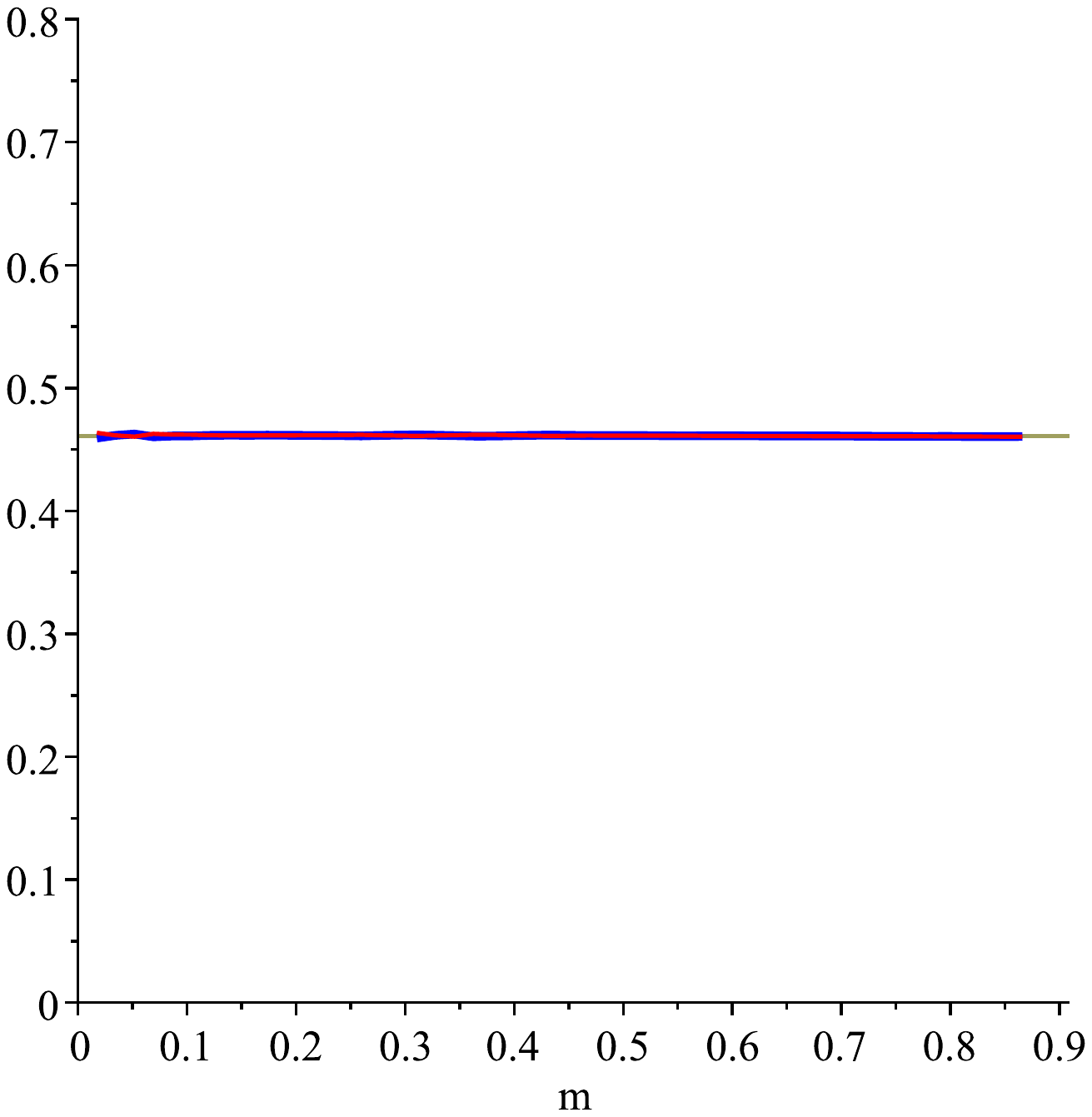}
% AdS-CFT-sigma-equation-v4.6.1: data v4.1 OmegaRange=0.6,  NumB=50, nrange=1000;
\vspace{-6.5cm}
\subcaption{\scriptsize $\scriptstyle q=1.5$: the resonance widths are inversely proportional to the peak amplitudes, independently of $\scriptstyle m$, and the product is constant at $\scriptstyle 0.46$ (shown in grey).}
\end{subfigure}
\vspace{1cm}
\caption{\scriptsize Numerical plots of the products $Re(\sigma^{xx}_0 )\Gamma_{xx}$ and $\Bigl|Im(\sigma^{xy}_0 )\Bigr|\Gamma_{xy}$ at
the resonance peaks as a function of $m$. The values for $\sigma^{x x}$ (red) and $\sigma^{x y}$ (blue) are superimposed.}
\label{fig:Gammasigma0}
\end{figure}

\subsubsection{Resonance properties as functions of temperature.}

To get an intuitive understanding of the physics it is better to transform from the
geometric parameters $q$, $m$ and $z_h$ to physical parameters, such as charge density $\rho$,
magnetic field $B$ and temperature $T$, \cite{Hartnoll+Herzog}. We have
\[ \nu = \frac{q}{m} = \frac{\tilde q}{\tilde m}= \frac{\rho}{B}
\qquad \hbox{and} \qquad
T =\frac{\hbar }{4\pi}  \sqrt{\frac{B}{L}} \left(\frac{\kappa^2}{2}\right)^{\frac 1 4} \frac{ \bigl( 3-(1+\nu^2)m^2\bigl)}{\sqrt{|m|}}.\]
Fixing $\nu$, $B$ and $L$ gives a monotonic relation between $T$ and $m$.
Fig.\,\ref{fig:QT} shows the peak conductivities at resonance and ${\cal Q}$-factors as a function of the reduced temperature
\[T_{Red}= \frac{3-(1+\nu^2)m^2}{4 \pi \sqrt{|m|}}\]
for three different values of $\nu$.
We see that the conductivity decreases
and the damping increases as $T$ increases, which is what one would expect from electron-phonon scattering. 

\begin{figure}
\begin{subfigure}{0.8\textwidth}
  \hskip -0.4cm 
\hbox{\begin{minipage}[t]{0.45\textwidth}
    \centering
\includegraphics[scale=0.4]{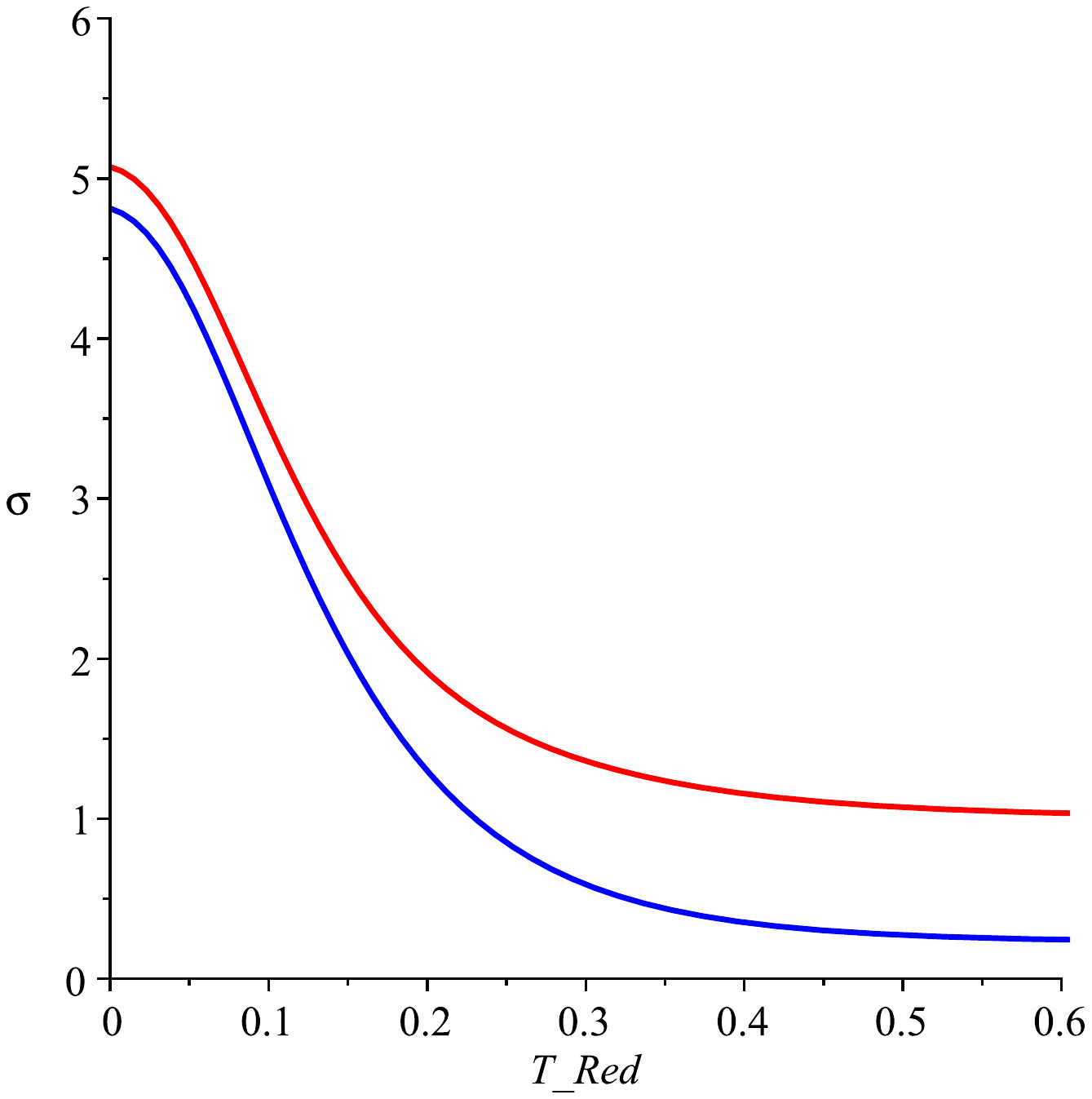}
\end{minipage} 
\hspace{4cm}
\hskip -3cm
\begin{minipage}[t]{0.45\textwidth}
  \begin{overpic}[unit=1mm,scale=0.4 ]{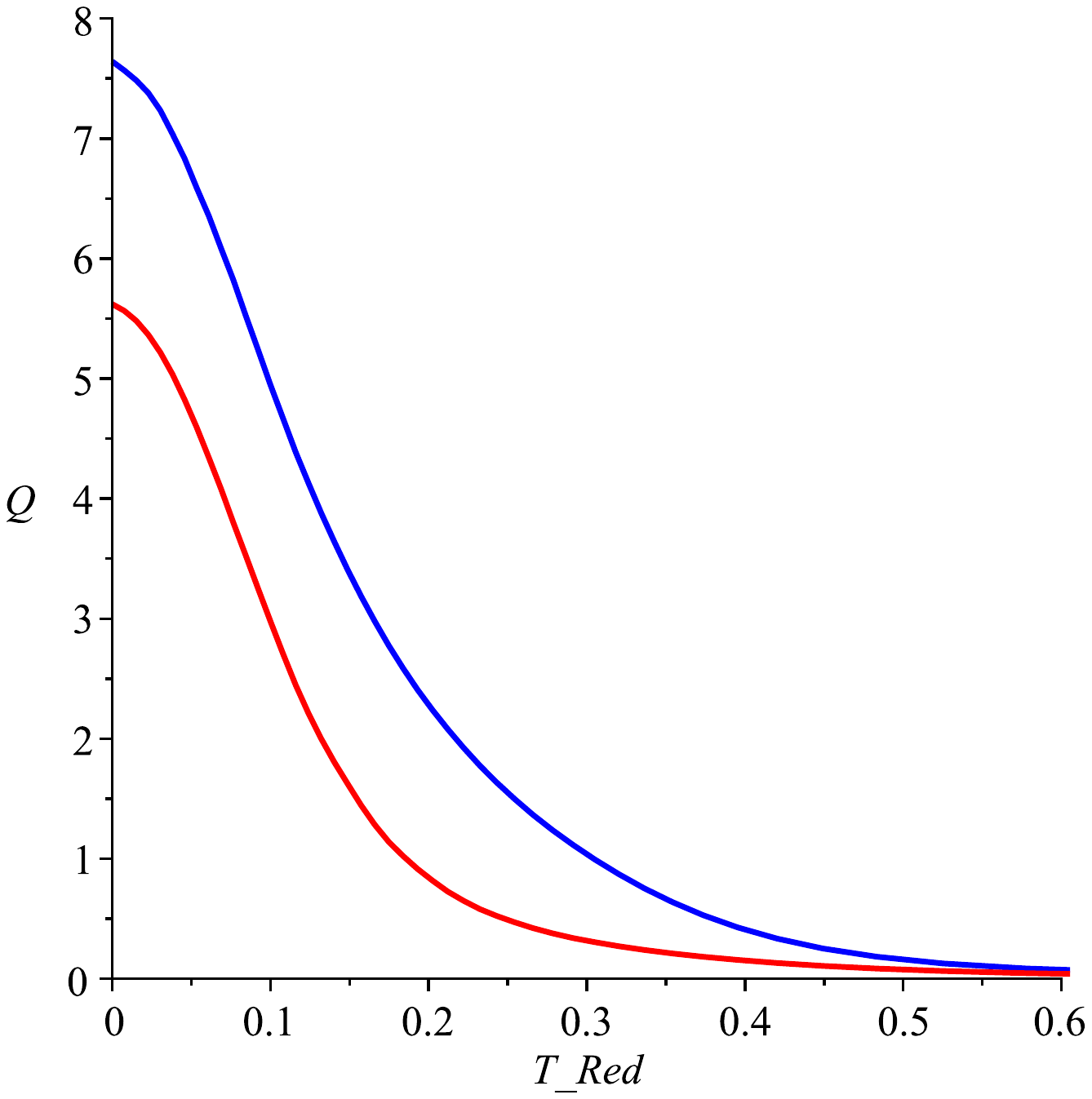}
  \put(40,85){$\nu = \frac{1}{3}$}
\end{overpic}
%nu=1-NumB=36-Blow=.34e-1-Bhigh=1.225-omega_low=0-omega_high=2.5-nrange=1000
\end{minipage}}
\vspace{-5cm}
\end{subfigure}

\vspace{2cm}

\centering
\begin{subfigure}[t]{0.8\textwidth}
\vspace{-2cm}
\hskip -2cm 
\hbox{\begin{minipage}[t]{0.45\textwidth}
    \centering
\includegraphics[scale=0.4]{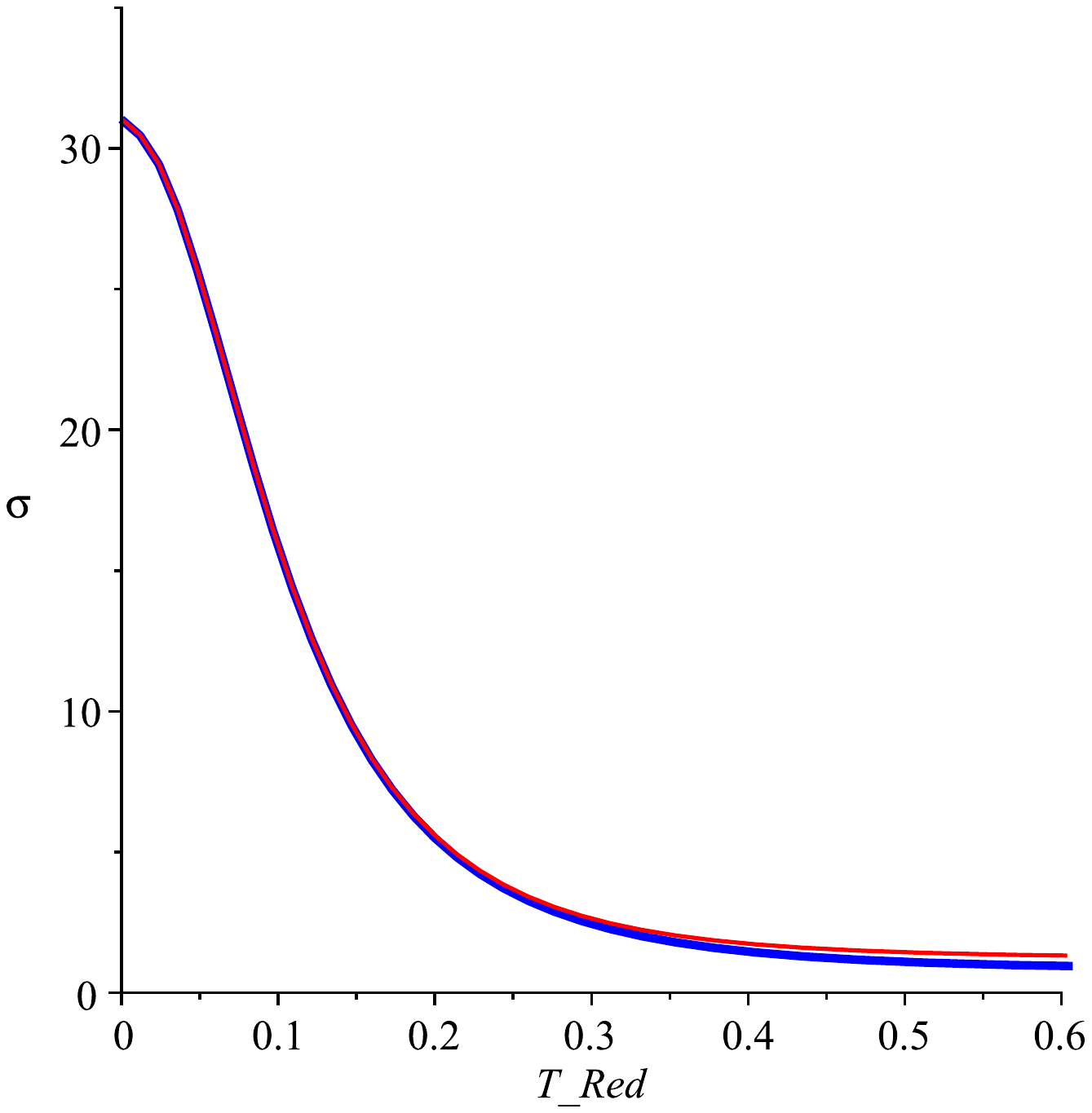}
%nu=1-NumB=36-Blow=.34e-1-Bhigh=1.225-omega_low=0-omega_high=2.5-nrange=1000
\end{minipage}
\hspace{4cm}
\hskip -3cm
\begin{minipage}[t]{0.45\textwidth}
    \begin{overpic}[unit=1mm,scale=0.4]{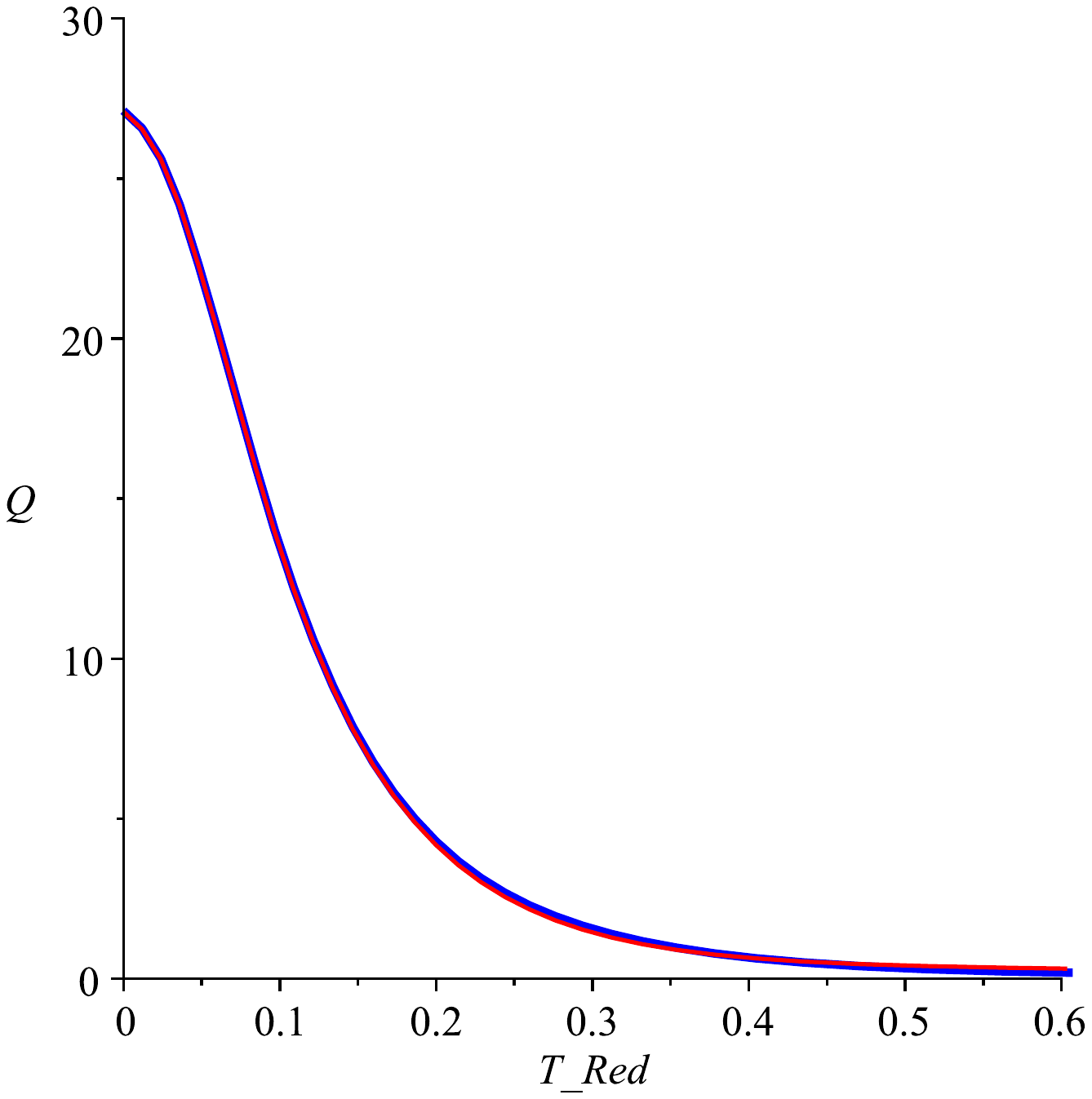}
  \put(40,85){$\nu = 1$}
\end{overpic}
%nu=2-v4.6.4-NumB=50-Blow=.1e-1-Bhigh=.775-omega_low=.1e-1-omega_high=.5-nrange=1000"
\end{minipage}}
\vspace{-5cm}
\end{subfigure}

\vspace{2cm}

\begin{subfigure}[t]{0.8\textwidth}
\vspace{-2cm}
\hskip -2cm 
\hbox{\begin{minipage}[t]{0.45\textwidth}
\centering
\includegraphics[scale=0.4]{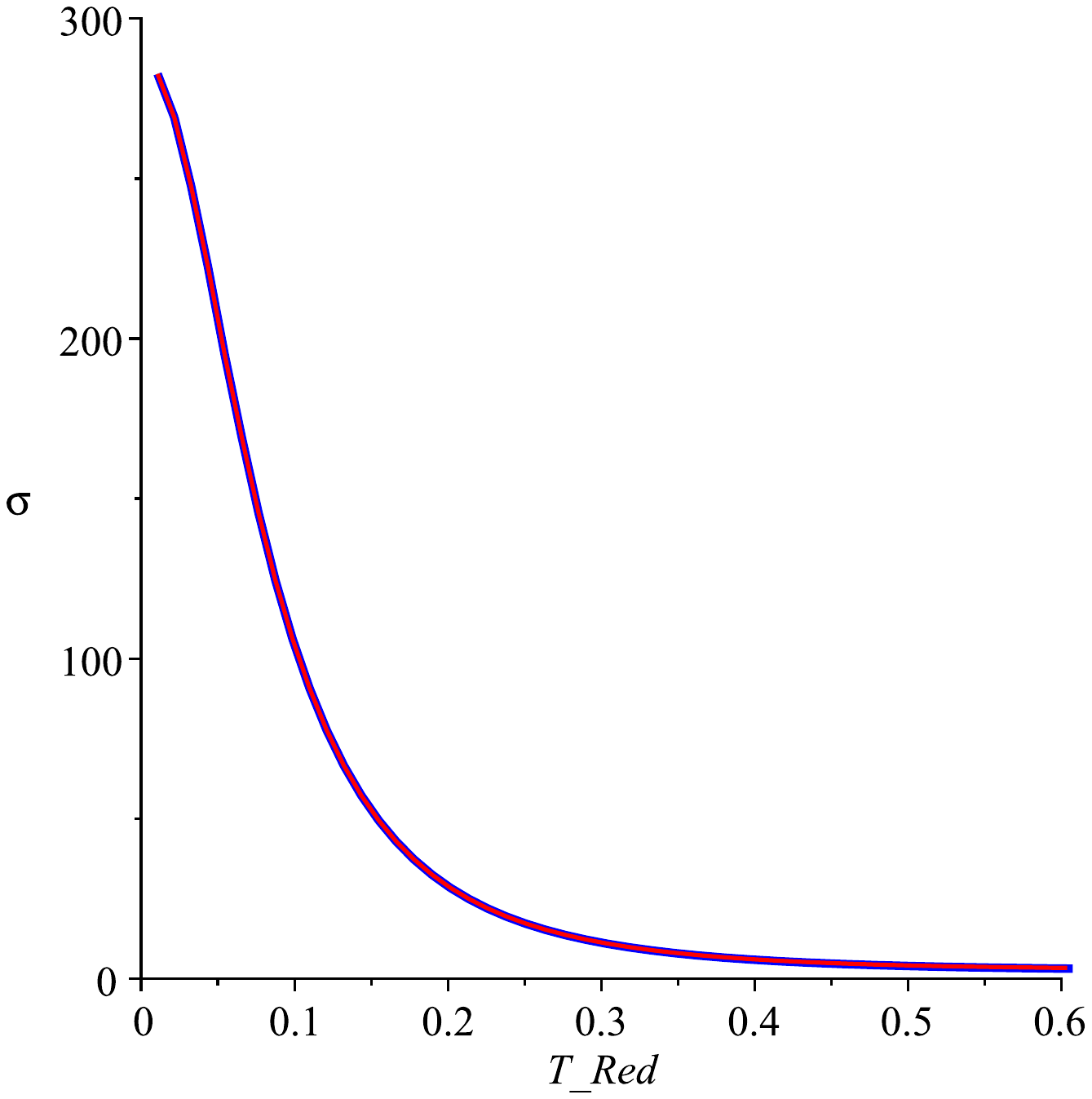}
\end{minipage}
\hspace{1cm}
\begin{minipage}[t]{0.45\textwidth}
    \begin{overpic}[unit=1mm,scale=0.4]{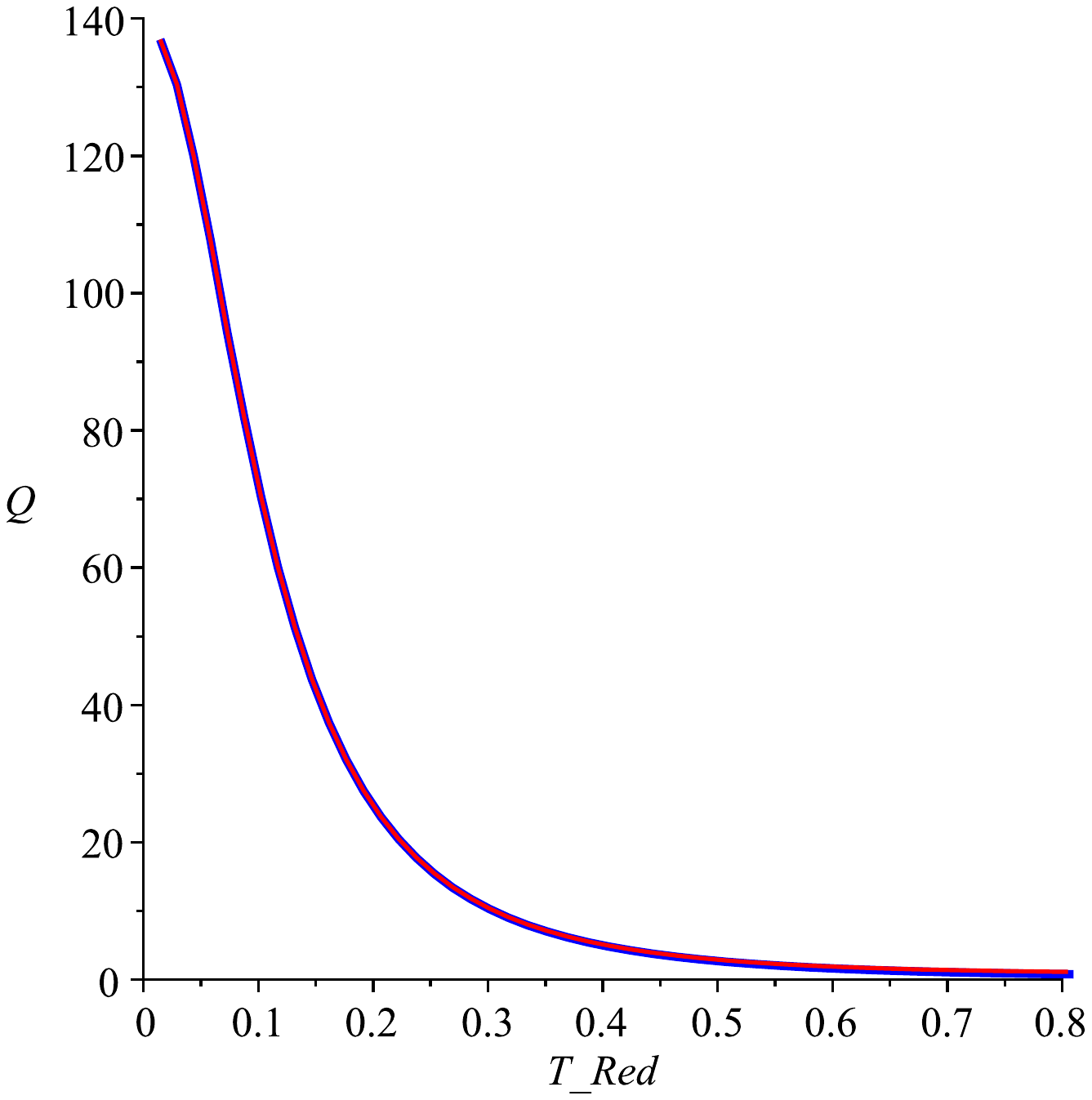}
  \put(40,85){$\nu = 2$}
\end{overpic}
\vspace{-7cm}
\end{minipage}}
\end{subfigure}
\vspace{-4cm}
\caption{\scriptsize Numerical plots of the peak conductivities at resonance (left-hand figures) and their associated
  ${\cal Q}$-factors (right-hand figures) as a function of dimensionless temperature $T_{Red}$ at various filling factors.}
% nu=1/3: q=.547722557505167-v4.4.4-NumB=49-Blow=.33e-1-Bhigh=1.643-omega_low=.2e-2-omega_high=2.5-nrange=1000
\label{fig:QT}
\end{figure}

\subsubsection{Zero temperature conductivities as functions of the filling factor $\nu$.}
To investigate the low temperature regime the conductivities are plotted
in Fig.\,\ref{fig:T=0} for various values of $\nu$ with $q^2+m^2=3$, corresponding to $T=0$ regardless of the value of $z_h$.  The resonance becomes sharper as $|\nu|$ increases, {\it i.e.}
as the charge density increases at fixed $B$ or as $B$ decreases at fixed charge density.
The DC Ohmic conductivity has the $\delta$-function behaviour of
a superconductor as $B\rightarrow 0$ at fixed $\rho$.
There is significant damping at
smaller values of $|\nu|$ but the resonance persists all the way down to $\nu=0$,
where the ${\cal Q}$-factor is finally reduced to $3.15$.  Thus increasing $B$ at fixed $\rho$ damps the
system from a superconductor to an ordinary conductor but it is not clear
what the physical origin of this damping is.
It may be indicative of a strengthening magnetic field
destroying superconductivity but this merits further investigation.
\begin{figure}
\vspace{-3cm}
\begin{subfigure}{0.8\textwidth}
\hskip -0.6cm 
\hbox{\begin{minipage}[t]{0.45\textwidth}
\centering
\begin{overpic}[unit=1mm,scale=0.4]{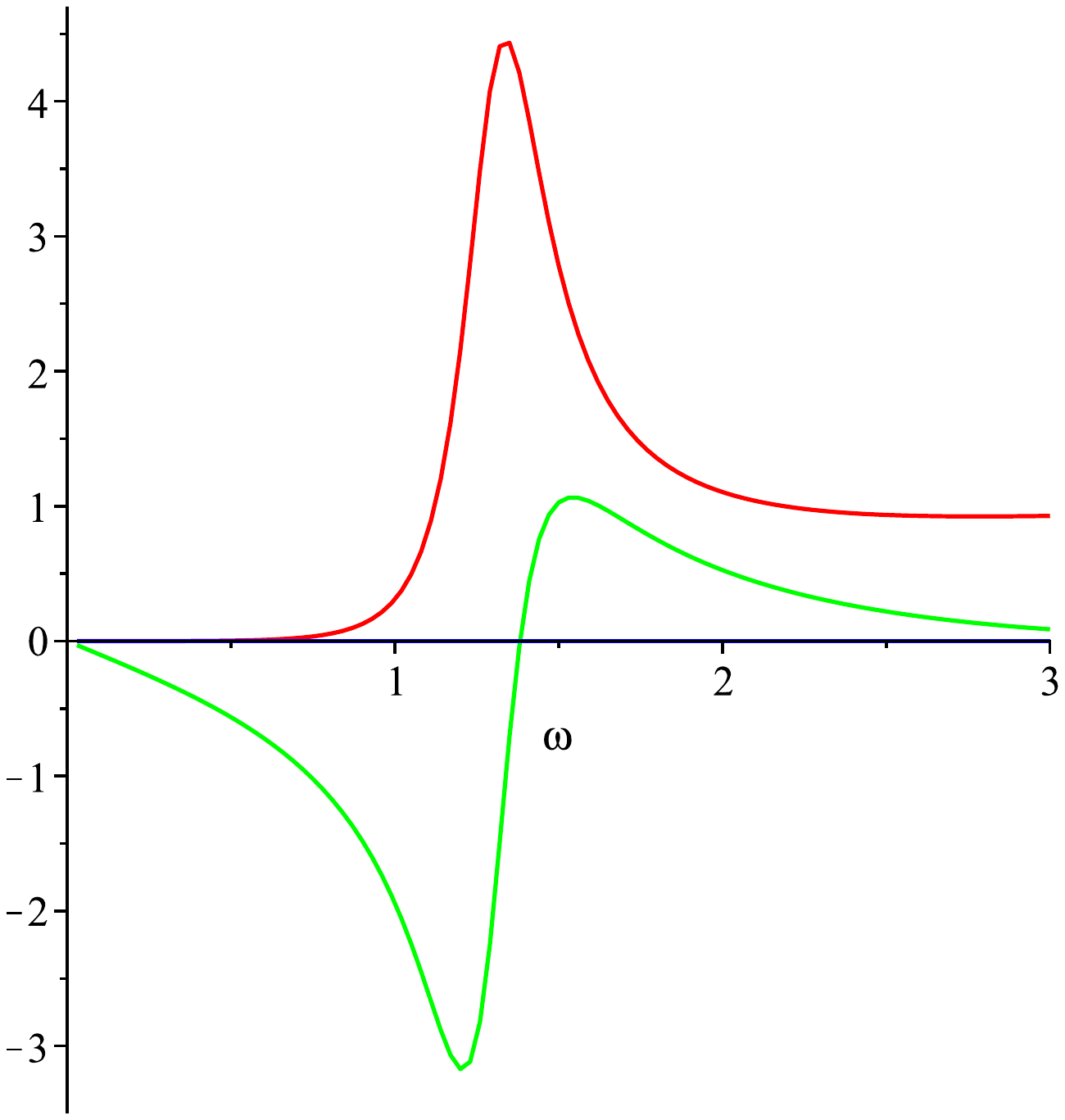}
\put(45,85){$\nu =0 $}
\end{overpic}
\end{minipage} 
\hspace{2cm}
\begin{minipage}[t]{0.45\textwidth}
\begin{overpic}[unit=1mm,scale=0.4]{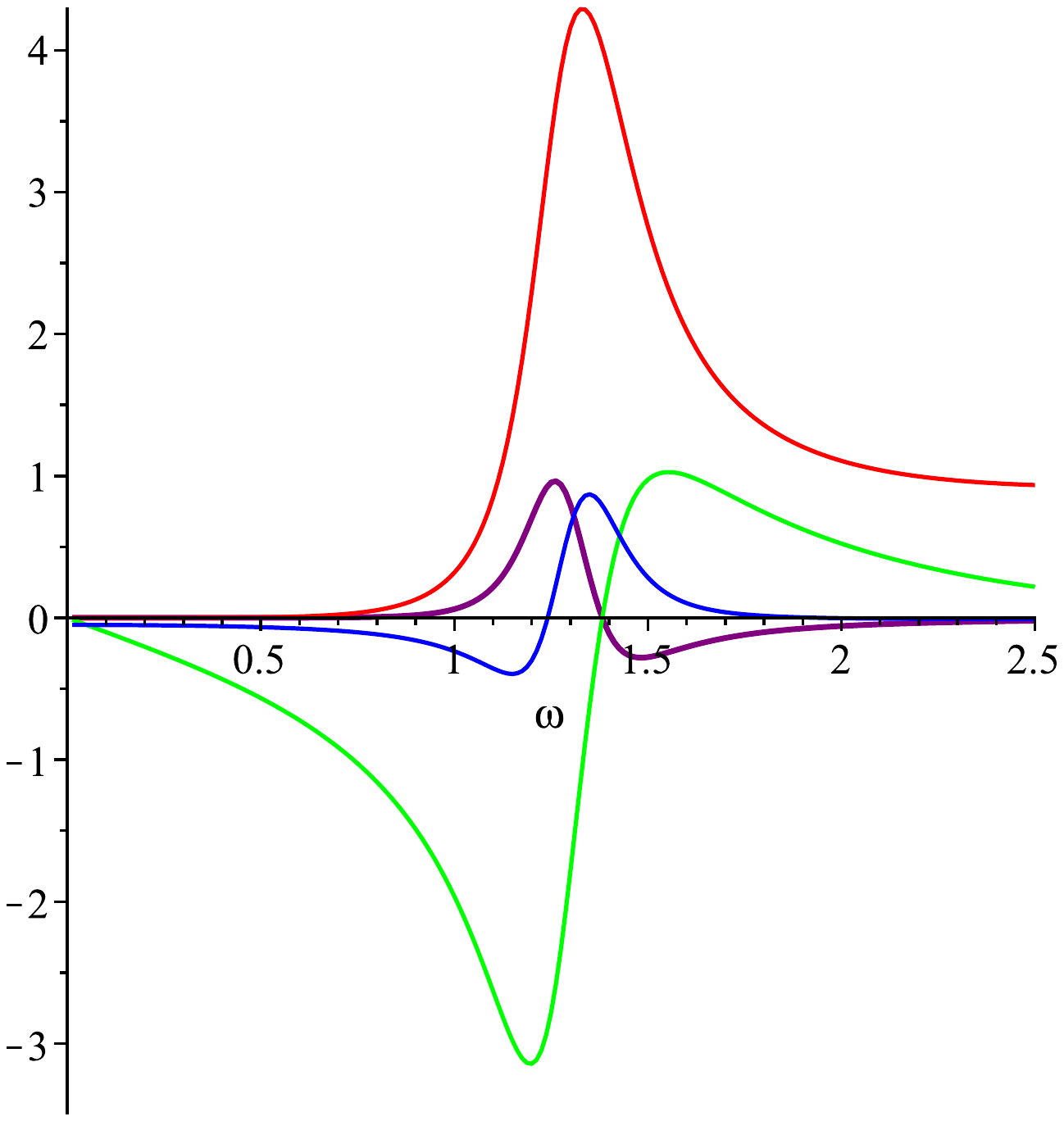}
  \put(45,85){$\nu = 0.05$}
\end{overpic}
\end{minipage}}
\end{subfigure}

\vspace{-3.5cm}

\centering
\begin{subfigure}[t]{0.8\textwidth}
\vspace{-2cm}
\hskip -2cm 
\hbox{
  \begin{minipage}[t]{0.45\textwidth}
    \centering
    \begin{overpic}[unit=1mm,scale=0.4]{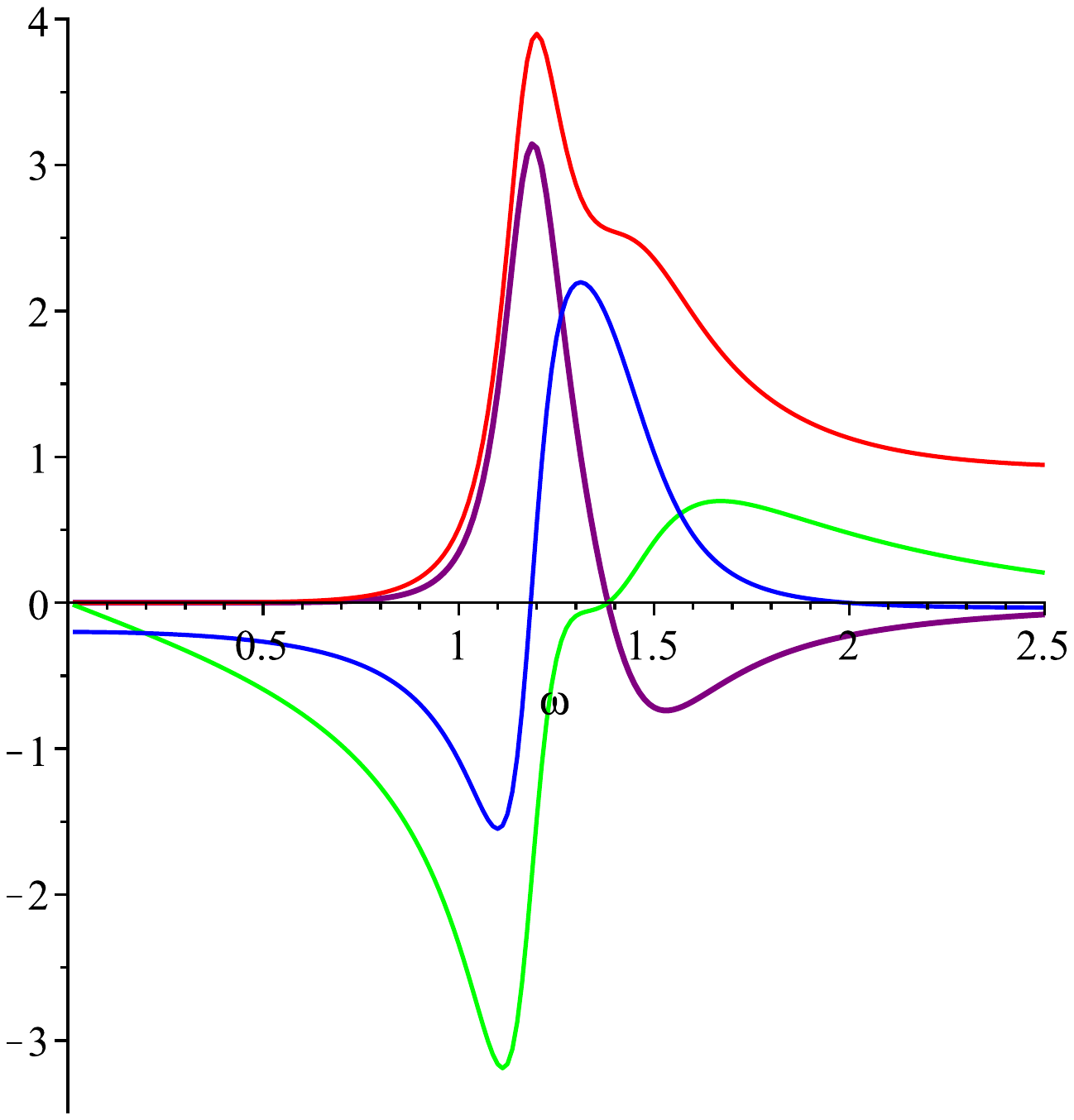}
  \put(45,85){$\nu = 0.2$}
\end{overpic}
\end{minipage}
\hspace{2cm}
\begin{minipage}[t]{0.45\textwidth}
\begin{overpic}[unit=1mm,scale=0.4]{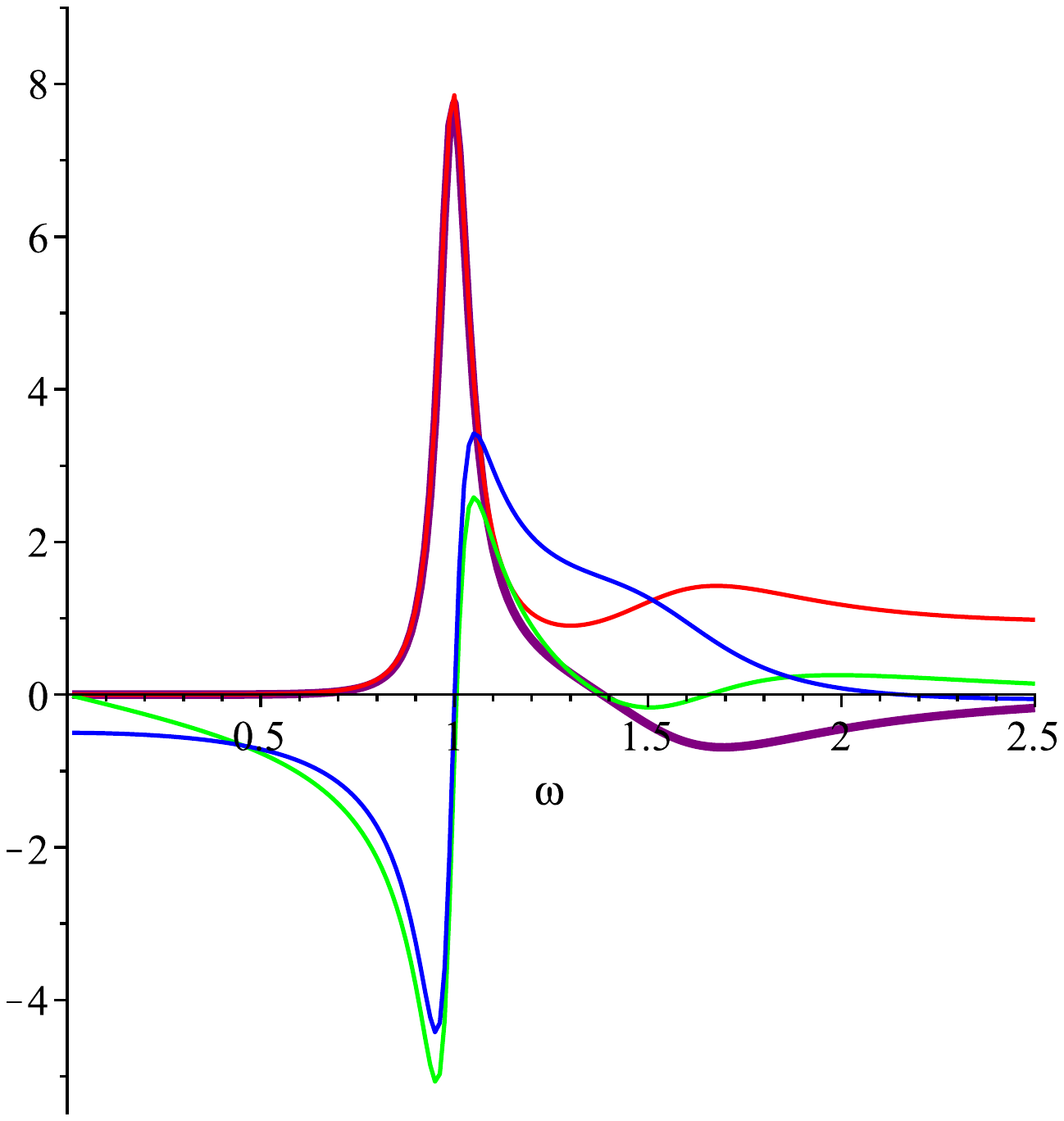}
  \put(45,85){$\nu = 0.5$}
\end{overpic}
\end{minipage}}
\end{subfigure}

\vspace{-3.5cm}

\begin{subfigure}[t]{0.8\textwidth}
\vspace{-2cm}
\hskip -2cm 
\hbox{\begin{minipage}[t]{0.45\textwidth}
    \centering
    \begin{overpic}[unit=1mm,scale=0.4]{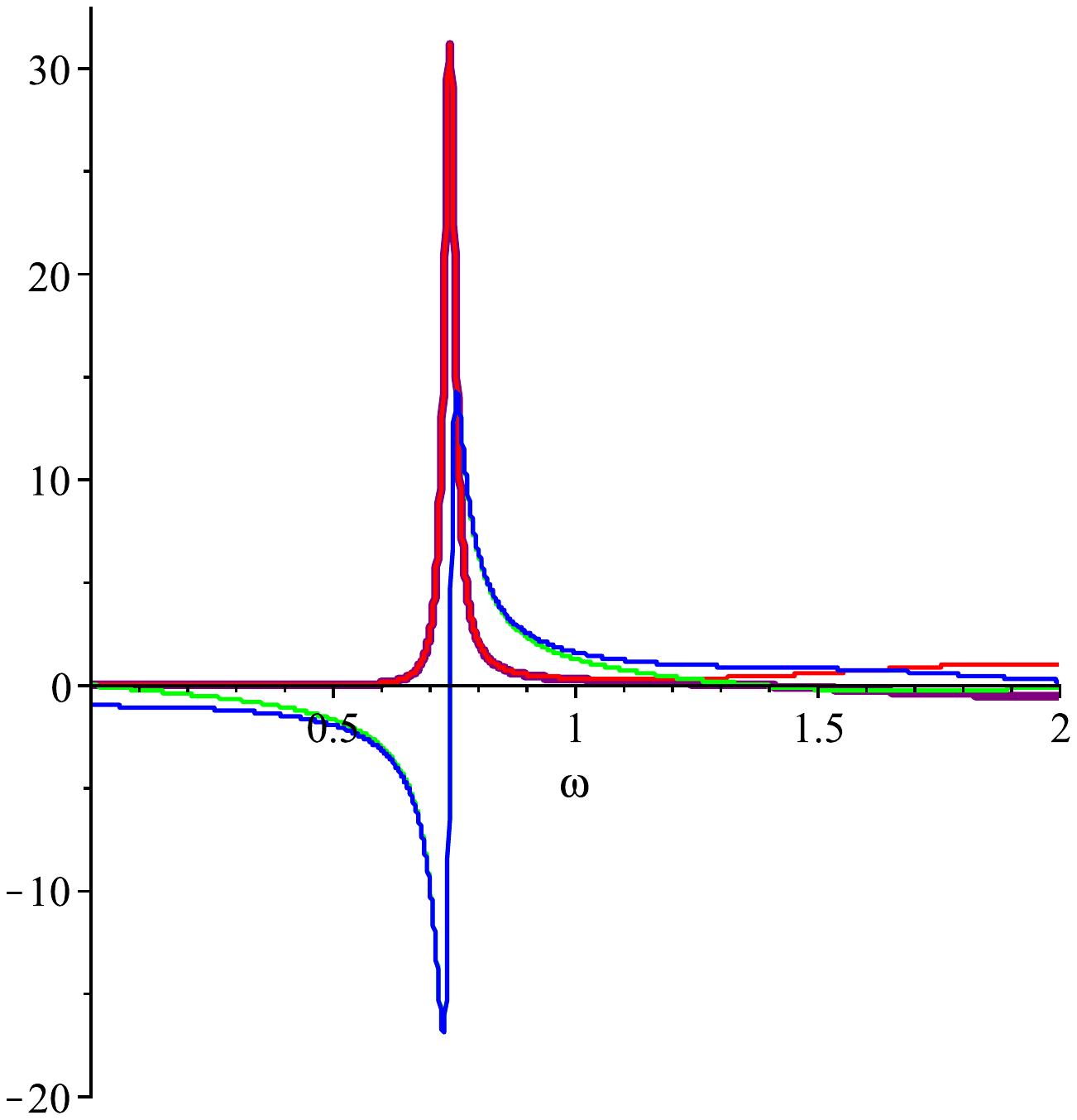}
      \put(45,85){$\nu = 1$}
\end{overpic}
\end{minipage}
\hspace{1.9cm}
\begin{minipage}[t]{0.45\textwidth}
      \begin{overpic}[unit=1mm,scale=0.4]{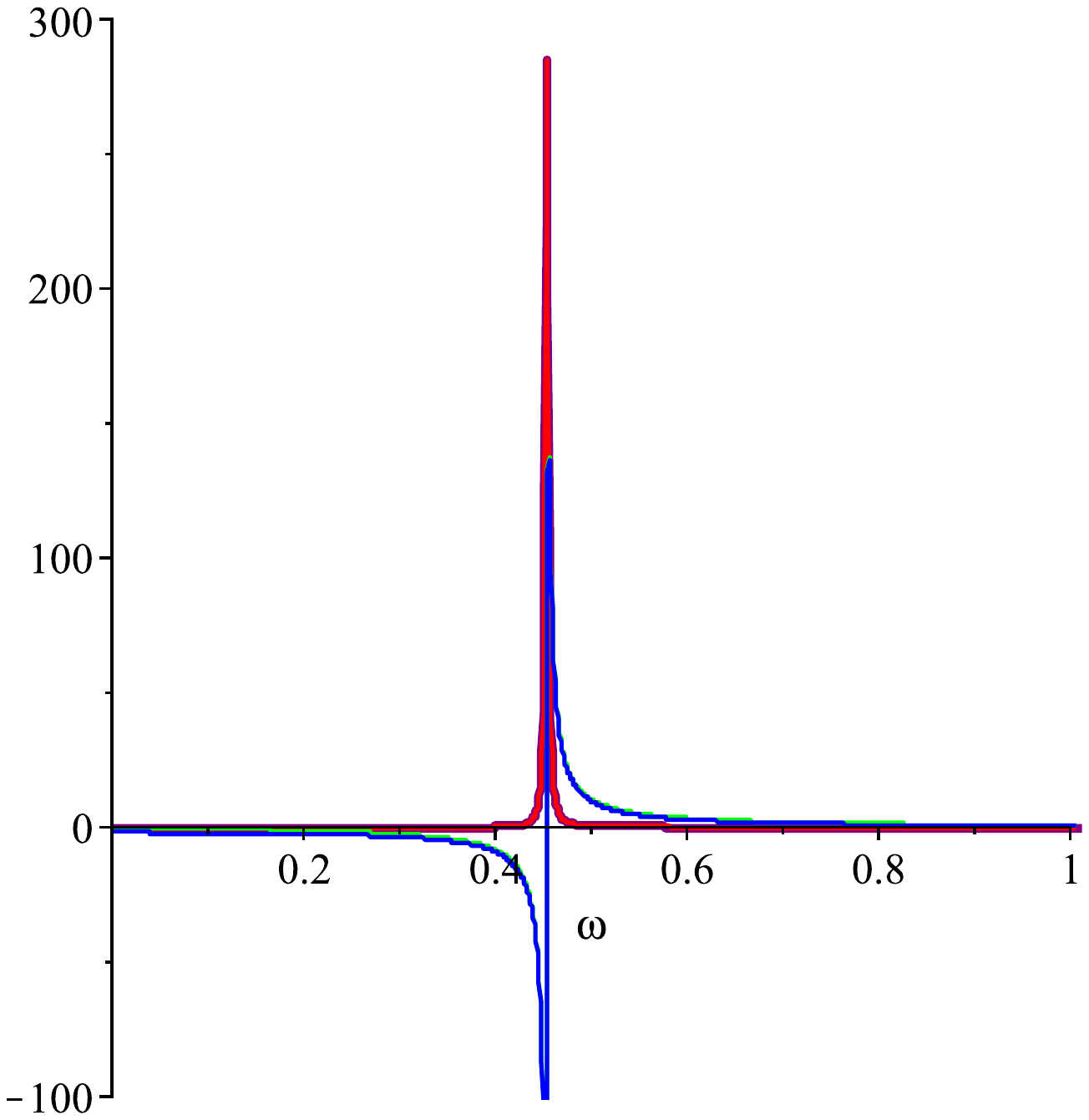}
  \put(45,85){$\nu = 2$}
\end{overpic}
\vspace{-7cm}
\end{minipage}}
\end{subfigure}

\vspace{-3.5cm}

\begin{subfigure}[t]{0.8\textwidth}
\vspace{-2cm}
\hskip -2.1cm 
\hbox{\begin{minipage}[t]{0.45\textwidth}
    \centering
        \begin{overpic}[unit=1mm,scale=0.4]{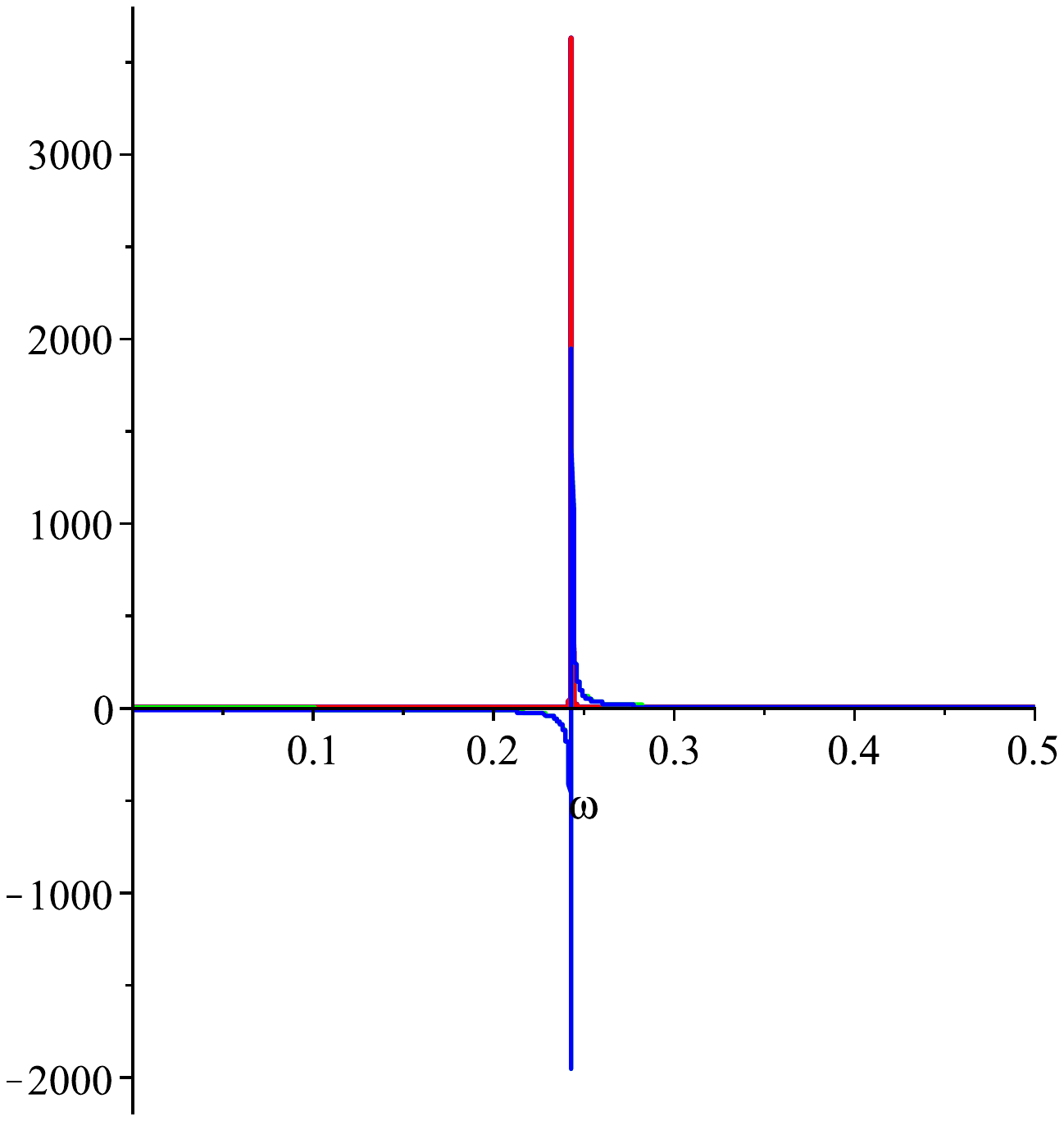}
          \put(45,85){$\nu = 4$}
\end{overpic}
\end{minipage}
\hspace{2.3cm}
\begin{minipage}[t]{0.45\textwidth}
      \begin{overpic}[unit=1mm,scale=0.4]{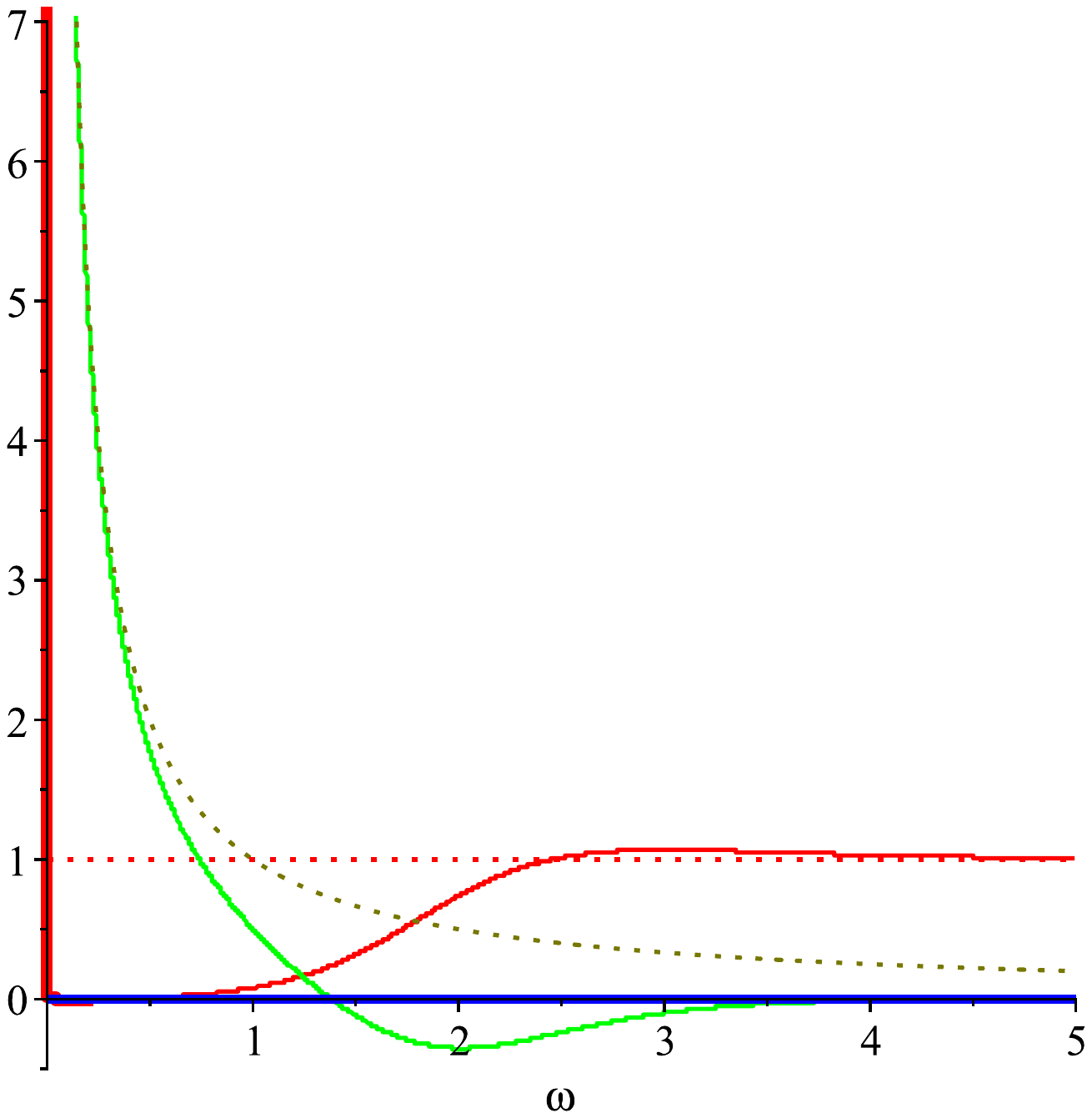}
        \put(45,85){$\nu \rightarrow \infty$}
\end{overpic}
\vspace{-7cm}
\end{minipage}}
\end{subfigure}
\vspace{-5cm}
\caption{\scriptsize Numerical plots of the conductivities at $T=0$ for various values of $\nu$.
The colours are: $Re\bigl( \sigma^{x x}(\wo) \bigr)$, red;  $Im\bigl( \sigma^{x x}(\wo) \bigr)$ Ohmic conductivity, green;  $-Im\bigl( \sigma^{x y}(\wo) \bigr)$, purple;  $Re\bigl( \sigma^{x y}(\wo) \bigr)$, blue. There is a $\delta$-function singularity in $Re\bigl(\sigma^{x x}(\wo)\bigr)$ at $\wo=0$ for the case $\nu\rightarrow\infty$,}
\label{fig:T=0}
\end{figure}

\newpage

\section{Conclusions\label{sec:Conclusions}}

 Equation (\ref{eq:dsigma-du}) is a renormalisation group equation for the conductivity 
of a two-dimensional system whose solutions can be used to extract AC and DC Ohmic and Hall
conductivities, in the presence of a magnetic field. The infra-red conductivities at the event horizon are
input as boundary conditions and are fixed by the singular nature of the differential equation there. Both the real and the imaginary parts of the Ohmic and Hall  conductivities can be obtained and the cyclotron resonance peaks and corresponding ${\cal Q}$-factors show qualitatively reasonable behaviour as functions of temperature and
charge density at fixed magnetic field.  As $q^2$ approaches $3$ the resonance peaks
of the Ohmic and Hall
conductivities become very nearly equal and the ${\cal Q}$-factor increases at small $m$ 
giving very sharp resonance peaks with high conductivities, tending to superconductor behaviour as $m\rightarrow 0$.  On the other hand the
resonance forms of the Ohmic and Hall conductivities are somewhat different when $q$ and $m$ are both low, for $q\approx  m \approx 0.1$  the
${\cal Q}$-factor is so small that the cyclotron peaks can hardly
be called resonances.

The RG equation has a simple truncation (\ref{eq:dsigma-du-truncated})
when $q^2 \approx m^2 \approx \wo \ll 1$ for which analytic solutions exist, corresponding to the approximate expressions for
the conductivity found in \cite{Hartnoll+Herzog}. Near extremality for the black hole, when
$q^2$ approaches $3$, numerical results indicate that a similar analytic form to (\ref{eq:dsigma-du-truncated}), but with different co-efficients, is again a good approximation.

A notable  feature of the numerical analysis presented above,
particularly evident in Fig. \ref{fig:large-q-small-m}, is the regularity of the resonance
peaks for $q\ge 1$. The characteristics of the resonances for the Ohmic and Hall conductivities are almost identical for $q>1$ but differ substantially for $q<1$.
When $q >1$ the cyclotron frequency shows a linear dependence on the magnetic charge $m$,  at least for small $m$, Fig.\,\ref{fig:omega_0-versus-m}(d), as it does for $q$ and $m$ small, but with a reduced slope as if the effective mass of the charged particle has increased, or its effective charge decreased.
Within numerical accuracy the residue becomes independent of $m$.

The present analysis only produces one peak in the dissipative Ohmic conductivity,
there is no sign of Shubnikov-de Haas oscillations or the quantum Hall effect.
This may be expected as there is no fermionic matter in the bulk, but there is no charged matter at all in the bulk --- it is  interpreted a classical effective theory for the electro-magnetic field after charged matter is integrated out.
The quantum Hall effect has been related to an emergent $Sl(2,{\bf R})$ symmetry in the infra-red \cite{Lutken+Ross1, Lutken+Ross2,SIGMA-Review,Lutken-Review}
and, while the bulk theory used here does enjoy an electromagnetic duality symmetry, giving rise to $S$-duality of the conductivity $\sigma_\pm \rightarrow -1/\sigma_\pm$
as pointed out in \cite{Hartnoll+Herzog}, this does not extend to $Sl(2,{\bf R})$ symmetry in
the bulk unless the bulk action is changed.  A bulk theory with near horizon $Sl(2,{\bf R})$
electro-magnetic duality was studied in \cite{GIKPTW}, and in a subsequent paper \cite{BD-in-preparation} we shall extend these ideas to study a different action which has full  $Sl(2,{\bf R})$
symmetry in the bulk and exhibits features that show strong parallels with
the experimental situation with the integer and fractional quantum Hall effects.

\appendix

\section{The piezoelectric effect\label{app:piezoelectric}}

Non-relativistically the piezoelectric effect occurs when a distortion of
a medium generates an electric induction \cite{LL-ECoCM},
\beq D^\alpha = \epsilon^{\alpha \beta} E_\beta + \gamma^{\alpha,\beta\gamma} S_{\beta\gamma}\label{eq:piezoelectric}\eeq
where $S_{\beta \gamma}= \nabla_\beta v_\gamma + \nabla_\gamma v_\beta$ is
the strain tensor generated by a displacement $v_\alpha$ and $\gamma^{\alpha,\beta\gamma}$ is the piezoelectric tensor.
As emphasised in \cite{LL-ECoCM} the definition the electric permittivity $\epsilon^{\alpha\beta}$ in such situations is somewhat conventional, one could equally use the stress tensor on the right-hand side, rather than the strain tensor, and define $\hat \epsilon^{\alpha\beta}$ and $\hat \gamma^{\alpha,\beta\gamma}$ using
\[  D^\alpha = \hat \epsilon^{\alpha \beta} E_\beta + \hat \gamma^{\alpha,\beta\gamma} T_{\beta\gamma}\]
and $\epsilon^{\alpha \beta}$ and $\hat \epsilon^{\alpha \beta}$ would in general be different.

For our purposes the first definition is the more appropriate.
Consider for example a distortion of the metric components due to a
diffeomorphism generated by a vector field $v_\alpha$,
$\delta g_{\alpha \beta}=\nabla_\alpha v_\beta + \nabla_\beta v_\alpha$, clearly
$\delta g_{\alpha\beta}$ plays the role of the stress tensor.
Now consider the case where the electric field and electric induction are time dependent infinitesimal variations with an oscillatory
phase, $\delta D^\alpha =e^{-i\omega t} \delta \widetilde D^\alpha$, and $
\delta E_\alpha= e^{-i\omega t} \delta \widetilde E_\alpha$,
and the diffeomorphism is due to a vibration $e^{-i\omega t} \delta \tilde v_\alpha$ with $\delta \tilde v_\alpha$ a constant displacement.  Then $\delta g_{t \alpha} = -i\omega \delta v_\alpha$ and differentiating equation (\ref{eq:piezoelectric}), then
using Maxwell's equations with zero magnetic field, implies
\[ \delta j^\alpha = \delta \dot D^\alpha = -i \omega(\epsilon^{\alpha\beta} \delta E_\beta  +  \gamma^{\alpha, t \beta} \delta v_\beta).\] 
It is perfectly reasonable that a vibration with oscillating charges generates a current. In a conductor the electric
permittivity has a pole at $\omega=0$, $\epsilon^{\alpha\beta}(\omega)=\frac{i \sigma^{\alpha\beta}}{\omega}$, and the residue gives the conductivity
\[ \delta J^\alpha = \sigma ^{\alpha\beta} \delta E_\beta  -i\omega \gamma^{\alpha, t \beta} \delta v_\beta.\]
If
\[\gamma^{\alpha,t \beta} = -\frac{i \tilde q}{\omega} g^{\alpha\beta}
=-\frac{i \tilde q  u^2}{\omega} \delta^{\alpha\beta} =\frac{i \rho(u)}{\omega} \delta^{\alpha\beta},\]
with $\rho(u)=- u^2 \tilde q$ the charge density (\ref{eq:rho}) and
$\delta v_\alpha = \frac{\delta G_\alpha}{u^2}$.
The current is
\[ \delta J^\alpha = \sigma ^{\alpha\beta} \delta E_\beta  + \rho\,
  \delta^{\alpha\beta} v_\beta  = \sigma ^{\alpha\beta} \delta E_\beta  -\tilde q \, \delta G_\alpha.\] 
where the second term is a contribution to the current generated by the forced oscillation, completely independent of any microscopic properties of the material, apart from the charge density (see equation (\ref{eq:J-delta-G})).
It would not be appropriate to include it in the conductivity tensor $\sigma^{\alpha \beta}$.

\section{The RG equation as a second order linear equation \label{app:2nd-order-RG}}

Equation (\ref{eq:dsigma-du}) is a Riccati equation and, as is well known, it
can easily be recast as a second order, linear, homogeneous equation. 
Let $\Sigma = \wo \sigma_+$ and write (\ref{eq:dsigma-du}) as
\beq \Sigma'(u)=  g_2  \Sigma^2    + g_1 \Sigma + g_0\label{eq:dSigma-du}\eeq
 with
\begin{align*}
g_2 & =  \frac{4 u^2 m^2}{\wo^2} - \frac{1}{f},\\
g_1 & = \frac{8 u^2 q m}{\wo},\\
g_0 & = 4 u^2 q^2 - \frac{\wo^2}{f}.
\end{align*}
Now define the function $X(u)$ via
\beq X' + g_2 \Sigma X=0,  \label{app:X-def}\eeq 
then equation (\ref{eq:dSigma-du}) for $\Sigma$ is
equivalent to 
\begin{align}
X'' - \left(g_1 + \frac{g_2'}{g_2} \right) X' + g_2 g_0 X =  0  & \hskip 20pt \Rightarrow  \label{eq:D2X}\\
\wo^2 f^2(4 u^2 f m^2-\wo^2) X''
-\wo f \Bigl\{ 8 u^2 f q m (4 u^2 f m^2 -\wo^2) + &\wo(8 u f^2 m^2 + \wo^2 f')\Bigr\}X'\nonumber \\
&  \kern -80pt+ (4 u^2 f q^2 - \wo^2)(4 u^2 f m^2 -\wo^2)^2 X  = 0  \nonumber 
\end{align}
for $X$.
This is not the same as the second order equation for ${\cal E}_+$ analysed in
\cite{Hartnoll+Herzog}, but it has the same physical content.

%\listoffigures

\end{document}